\documentclass[default,iicol]{sn-jnl}

\usepackage{soul}

\usepackage{subfigure}
\usepackage{graphicx}
\usepackage[T1]{fontenc}

\newcommand\phcms{photons~cm$^{-2}$~s$^{-1}$}

\newcommand\apj{{Astrophys.\ J.\ }}%
\newcommand\apjl{{Astrophys.\ J.\ }}%
\newcommand\apjs{{ApJS\ }}%
\newcommand\aap{{Astron. Astrophys.\ }}%
\newcommand\mnras{{Mon.\ Not.\ R.\ Astron.\ Soc.\ }}%
\newcommand\nat{{Nature\ }}%
\newcommand\araa{{Annu.\ Rev.\ Astron.\ Astrophys.\ }}%
\newcommand\pasj{{PASJ\ }}%
\newcommand\prl{{Phys.~Rev.~Lett.}}%
\newcommand\prd{{Phys.~Rev.~D}}%
\newcommand\baas{{BAAS}}%
\newcommand\aj{{AJ}}%
\newcommand\jcap{{J. Cosmology Astropart. Phys.}}%

\jyear{2023}%

\theoremstyle{thmstyleone}%
\theoremstyle{thmstyletwo}%

\theoremstyle{thmstylethree}%

\begin{document}

\title[MASS]{MeV Astrophysical Spectroscopic Surveyor (MASS): A Compton Telescope Mission Concept}


\author[1]{\fnm{Jiahuan} \sur{Zhu}}
\equalcont{These authors contributed equally to this work.}
\author[2]{\fnm{Xutao} \sur{Zheng}}
\equalcont{These authors contributed equally to this work.}
\author*[1]{\fnm{Hua} \sur{Feng}}\email{hfeng@tsinghua.edu.cn}
\author*[2]{\fnm{Ming} \sur{Zeng}}\email{zengming@tsinghua.edu.cn}
\author[3]{\fnm{Chien-You} \sur{Huang}}
\author[3]{\fnm{Jr-Yue} \sur{Hsiang}}
\author[3]{\fnm{Hsiang-Kuang} \sur{Chang}}
\author[1]{\fnm{Hong} \sur{Li}}
\author[2]{\fnm{Hao} \sur{Chang}}
\author[2]{\fnm{Xiaofan} \sur{Pan}}
\author[2]{\fnm{Ge} \sur{Ma}}
\author[1]{\fnm{Qiong} \sur{Wu}}
\author[2]{\fnm{Yulan} \sur{Li}}


\author[4]{\fnm{Xuening} \sur{Bai}}
\author[5]{\fnm{Mingyu} \sur{Ge}}
\author[6]{\fnm{Long} \sur{Ji}}
\author[7]{\fnm{Jian} \sur{Li}}
\author[8]{\fnm{Yangping} \sur{Shen}}
\author[9]{\fnm{Wei} \sur{Wang}}
\author[5]{\fnm{Xilu} \sur{Wang}}
\author[10,11,12]{\fnm{Binbin} \sur{Zhang}}
\author[13]{\fnm{Jin} \sur{Zhang}}

\affil*[1]{\orgdiv{Department of Astronomy}, \orgname{Tsinghua University}, \orgaddress{\city{Beijing}, \postcode{100084}, \country{China}}}

\affil*[2]{\orgdiv{Department of Engineering Physics}, \orgname{Tsinghua University}, \orgaddress{\city{Beijing}, \postcode{100084}, \country{China}}}

\affil[3]{\orgdiv{Institute of Astronomy}, \orgname{National Tsing Hua University}, \orgaddress{\city{Hsinchu}, \postcode{300044}, \state{Taiwan}}}

\affil[4]{\orgdiv{Institute for Advanced Study}, \orgname{Tsinghua University}, \orgaddress{\city{Beijing}, \postcode{100084}, \country{China}}}

\affil[5]{\orgdiv{Key Laboratory of Particle Astrophysics}, \orgname{Institute of High Energy Physics}, \orgaddress{\city{Beijing}, \postcode{100049}, \country{China}}}

\affil[6]{\orgdiv{School of Physics and Astronomy}, \orgname{Sun Yat-Sen University}, \orgaddress{\city{Zhuhai}, \postcode{519082}, \country{China}}}

\affil[7]{\orgdiv{Hefei National Laboratory for Physical Sciences at the Microscale}, \orgname{University of Science and Technology of China}, \orgaddress{\city{Hefei}, \postcode{230026}, \country{China}}}

\affil[8]{\orgdiv{China Institute of Atomic Energy}, \orgaddress{\city{Beijing}, \postcode{102413}, \country{China}}}

\affil[9]{\orgdiv{School of Physics and Technology}, \orgname{Wuhan University}, \orgaddress{\city{Wuhan}, \postcode{430072}, \country{China}}}

\affil[10]{\orgdiv{School of Astronomy and Space Science}, \orgname{Nanjing University}, \orgaddress{\city{Nanjing}, \postcode{210093}, \country{China}}}

\affil[11]{\orgdiv{Key Laboratory of Modern Astronomy and Astrophysics (Nanjing University)}, \orgname{Ministry of Education}, \orgaddress{\city{Nanjing}, \postcode{210093}, \country{China}}}

\affil[12]{\orgdiv{Purple Mountain Observatory}, \orgname{Chinese Academy of Sciences}, \orgaddress{\city{Nanjing}, \postcode{210023}, \country{China}}}

\affil[13]{\orgdiv{School of Physics}, \orgname{Beijing Institute of Technology}, \orgaddress{\city{Beijing}, \postcode{100081}, \country{China}}}


\abstract{
We propose a future mission concept, the MeV Astrophysical Spectroscopic Surveyor (MASS), which is a large area Compton telescope using 3D position sensitive cadmium zinc telluride (CZT) detectors optimized for emission line detection. The payload consists of two layers of CZT detectors in a misaligned chessboard layout, with a total geometric area of 4096~cm$^2$ for on-axis observations. The detectors can be operated at room-temperature with an energy resolution of 0.6\% at 0.662~MeV. The in-orbit background is estimated with a mass model. At energies around 1~MeV, a line sensitivity of about $10^{-5}$~\phcms\ can be obtained with a 1~Ms observation. The main science objectives of MASS include nucleosynthesis in astrophysics and high energy astrophysics related to compact objects and transient sources. The payload CZT detectors weigh roughly 40~kg, suggesting that it can be integrated into a micro- or mini-satellite. We have constructed a pathfinder, named as MASS-Cube, to have a direct test of the technique with 4 detector units in space in the near future.
}




\maketitle

\section{Introduction}
\label{sec:intro}

Gamma-ray astronomy in the MeV energy range is an important window for the study of stellar evolution, supernova explosion, neutron star merger, and in particular, nucleosynthesis in these systems, and also a useful probe to constrain the physics of compact objects and even planetary formation.  In the MeV band, a unique feature is the presence of multiple strong astrophysical emission lines, including the positron-electron annihilation line (0.511~MeV) and radioactive decay lines from radioisotopes such as $^{26}$Al (1.809~MeV), $^{44}$Ti (1.157~MeV), $^{56}$Co (0.847 \& 1.238~MeV), and $^{60}$Fe (1.173 \& 1.332~MeV). Compton imaging is so far the best practical means for MeV detection in astronomy~\citep{Kierans2022}. A Compton telescope with a fine spectral resolution in the MeV band may address the following science questions. 

The gamma-ray lines observed from supernovae will provide direct or unique information regarding the total mass of the synthesized nickel, its spatial distribution, explosion energy, etc., facilitating our understandings of the supernova progenitors and how they explode~\cite{Maoz2014,Timmes2019}. A precise measurement of the $^{56}$Ni mass can help improve the Phillips relation for type Ia supernovae~\cite{Phillips1993,Phillips1999}. In the Galactic plane or starburst regions, the optical or infrared confirmation of a Ia supernova may take more than 2 weeks, while gamma-rays can be detected within days if there exists surface $^{56}$Ni and this detection can uniquely probe the ejecta structure~\cite{Wang2019}. The radioisotopes such as $^{44}$Ti from supernova remnants can provide key information for the explosion asymmetry and mechanism, as they are synthesized in the innermost region of core-collapse supernovae~\cite{The1998,Wongwathanarat2017}. So far, MeV observations of supernovae~\cite{Matz1988,Churazov2015} or supernova remnants~\cite{Renaud2006,The1990,Tsygankov2016} are rare. A larger sample is needed. 

The formation and evolution of elements, especially heavy elements, has been a mystery for centuries~\cite{Burbidge1957}. This process is driven by the star formation rate and nucleosynthesis event rate, including novae, supernovae, and kilonovae~\cite{Timmes2019}. The MeV signal can possibly distinguish the light and heavy r-process elements, and confirm whether kilonovae are the main r-process sites in the universe~\cite{Wang2020}. Measurements of radionuclides like $^{26}$Al, $^{60}$Fe, $^{53}$Mn, and $^{182}$Hf can directly trace the flow of ejecta from the nucleosynthesis sites into the ambient interstellar medium, and their mapping in the galaxy can provide information about the stellar evolution~\cite{Diehl2006,Wang2007,Wu2019}. 

The 0.511~MeV line map can probe not only individual positron sites, but also the structure and kinematics of the galaxy~\cite{Knoedlseder2005,Siegert2016a,Fuller2019}.
The origin of the positrons that produce the 0.511~MeV emission at the Galactic center has been a long standing puzzle in astrophysics~\cite{Prantzos2011}. In particular, the bright bulge component could be either due to astrophysical origins~\cite{Totani2006,Alexis2014,Bartels2018} or exotic explanations that involve the dark matter or primordial black holes~\cite{Pospelov2007,Hooper2008,Khalil2008,Keith2021,Cai2021}. The key to solving this problem is to determine the morphology of the diffuse 0.511~MeV emission in the Galactic center, and find similar signatures in satellite galaxies~\cite{Siegert2016b}.

The matter composition of relativistic jets produced by accreting compact objects is largely unknown~\cite{Margon1979,Liu2015}. If the jets are composed of position-electron pair plasma~\cite{Begelman1984,Reynolds1996}, one may expect the appearance of a narrow 0.511~MeV emission line when the jets hit ambient mediums. However, a search in the radio galaxy 3C 120 yielded a non-detection~\cite{Marscher2007}. The line could also be relativistically broadened and shifted, and manifest itself as a broad bump in the MeV spectrum~\cite{Zhang2010}. A possible positron annihilation signature has been observed during an outburst of the microquasar V404 Cygni~\cite{Siegert2016}, likely produced by positions created via $\gamma$-$\gamma$ interaction. However, there is controversy that it could be an instrument effect~\cite{Roques2016}. Recently, a possible emission line around 10~MeV was observed from the brightest ever gamma-ray burst (GRB) 221009A, and interpreted as blue-shifted positron annihilation emission~\citep{EdvigeRavasio2023}. This implies that bright transient gamma-ray sources could be important targets to search for positron annihilation lines. 

For accreting pulsars, the neutron capture line at 2.223~MeV is theoretically predicted and may be revealed as a broad, blueshifted line if produced in a hot accretion flow, a redshifted line if produced on the neutron star surface, or narrow lines on the surface of the companion~\cite{Reina1974,Brecher1980,Agaronian1984,Bildsten1993,Jean2001,Guessoum2002}. However, current observations only provided upper limits~\cite{Bildsten1993,Boggs2006,Caliskan2009}. MeV observations with high line sensitivity will supply tighter constraints or bring new insights for these objects.

$^{26}$Al serves as an important heating source for planetary differentiation~\cite{Jacobsen2008} and ionization of the solar nebula~\cite{Turner2009}. There is strong evidence on the presence of $^{26}$Al and other short-lived radionuclides at the very beginning of the solar system formation~\cite{Lee1976}. It has long been debated where $^{26}$Al is originated, from nearby supernovae or massive stars~\cite{Gounelle2008,Gaidos2009,Gritschneder2012}. The $^{26}$Al emission line has been detected with COMPTEL and INTEGRAL toward the nearby Upper Sco association of massive and young stars~\cite{Diehl2010}. Detection in other star-forming regions~\citep{Reiter2020} may constrain the underlying production channels~\citep{Forbes2021}. For example, as massive stellar winds do not produce $^{60}$Fe, detection or a stringent limit on the $^{60}$Fe line emission can distinguish the two scenarios~\cite{Wang2020}. 

A Compton telescope offers the capability of gamma-ray polarimetry for free. So far, MeV gamma-ray polarization measurements are only available for two persistent sources, the Crab nebula/pulsar and Cygnus X-1~\cite{Zdziarski2014,Vadawale2018,Chauvin2018,Chattopadhyay2023}. Less certain polarization measurements were reported for V404 Cygni~\cite{Laurent2016} and for several GRBs~\cite{Chattopadhyay2022}. A phase-resolved gamma-ray polarimetry can help distinguish different pulsar models~\cite{Takata2007,Petri2013,Harding2017}. However, the current measurements for the Crab pulsar obtained with different instruments do not agree with each other~\cite{Vadawale2018,Chauvin2018}.  A high degree of polarization is seen in Cygnus X-1 in 230--370 keV, suggestive of a jet origin with a magnetic field stronger than that estimated assuming equipartition~\cite{Zdziarski2014}. Polarization measurements of the GRB prompt emission may shed light on the magnetic environment of GRBs.  

In this paper, we propose a future mission concept, the MeV Astrophysical Spectroscopic Surveyor (MASS) based on 3D position-sensitive CdZnTe (CZT) detectors~\cite{Barrett1995,He1996,He1997,He1999,Du2001,Zhang2006,Zhang2007,Kim2012,Yang2020} for Compton imaging in the energy range of 0.2--5~MeV optimized for emission line detection. The detector works at room-temperature and has a fine spectral resolution of 0.6\% at 0.662~MeV. MASS will be a successor to the NASA mission Compton Spectrometer and Imager (COSI) that is scheduled to launch around 2027~\citep{Tomsick2021}. MASS has an energy resolution similar to that of COSI, but an effective area at least one order of magnitude larger according to the current design. We will perform a flight test of the MASS detector with a CubeSat in the near future, named as MASS-Cube. In the following, we will introduce the design of MASS and MASS-Cube, and the expected performance based on simulations and laboratory measurements.

\section{Mission concept}

\subsection{The CZT detector}
\label{sec:czt}

The unit CZT detector, manufactured by Redlen Technologies Inc.\footnote{\url{https://www.redlen.com/}}, has a geometry of $2 \times 2 \times1.5$~cm$^3$. The anode is pixelated into an array of $11 \times 11$ pixels, with a size of $1.72 \times 1.72$~mm for each pixel, directly coupled to an ASIC chip for pixel readout. The pixel size also determines the spatial resolution along $x$ and $y$. When any anode pixel is triggered, charges at all 121 pixels are measured independently. The whole cathode is read out by a single charge-sensitive preamplifier and the waveform is digitized at a sample rate of 40~MHz~\cite{Yang2020}. The waveform from 6.4~$\mu$s before the trigger and 3.6~$\mu$s after the trigger will be saved. The anode trigger is assigned with a timestamp based on a 200 MHz system clock. The reconstructed interaction time from the waveform has an accuracy better than 0.2~$\mu$s. The cathode has no readout chip or circuit board and is used as the entrance window for gamma-rays.

\begin{figure}
\centering
\includegraphics[width=0.7\columnwidth]{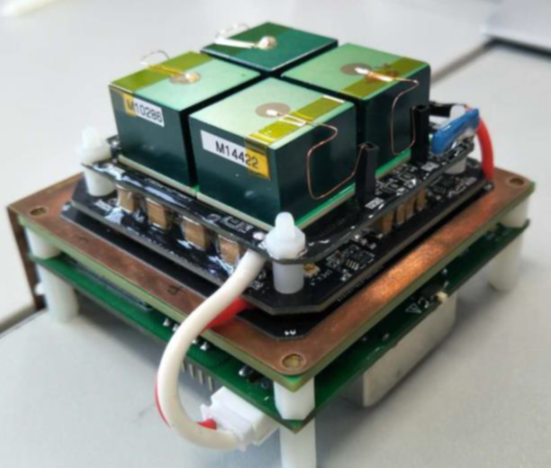}
\caption{Picture of a prototype detector module.}
\label{fig:prototype}
\end{figure}

\begin{figure}
\centering
\includegraphics[width=0.7\columnwidth]{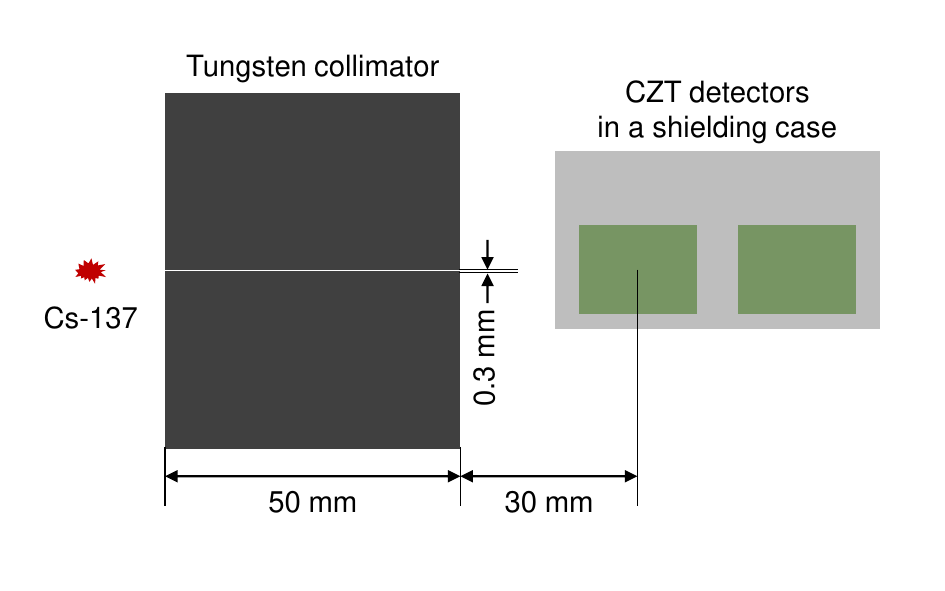}
\caption{Schematic drawing of the experiment setup for the depth calibration.}
\label{fig:experiment_setup}
\end{figure}

When a gamma-ray interacts with the detector, electron and hole pairs are created and drift toward the anode and cathode, respectively. The $x$ and $y$ positions of the interaction point are directly obtained by the location of the pixel that collects the electrons. The relative amount of charges measured from the anode and cathode is a function of the $z$ location where electrons and holes are created~\cite{He1996,He1997}. Therefore, the $z$ position of the interaction point can be inferred by the ratio of charges measured on the cathode to those on the anode (C/A). If there are multiple hits, i.e., due to energy deposits in a single CZT detector within the readout time or charge sharing among neighboring pixels, this technique no longer works. The charge drift time depends on the drift distance along the $z$ direction. Therefore, in the case of multiple hits in a single detector unit, the $z$ positions of different interaction points can be reconstructed by the charge drift times encoded in the readout waveform~\cite{Zhang2006,Kim2012,Yang2020}. An example of the cathode waveform resulted from a two-pixel energy deposit is illustrated in Figure~\ref{fig:depth}. We note that events with multiple hits in a single pixel will be discarded due to large uncertainties on depth reconstruction. 

\subsection{Detector performance}
\label{sec:performance}

\begin{figure}
\centering
\includegraphics[width=0.7\columnwidth]{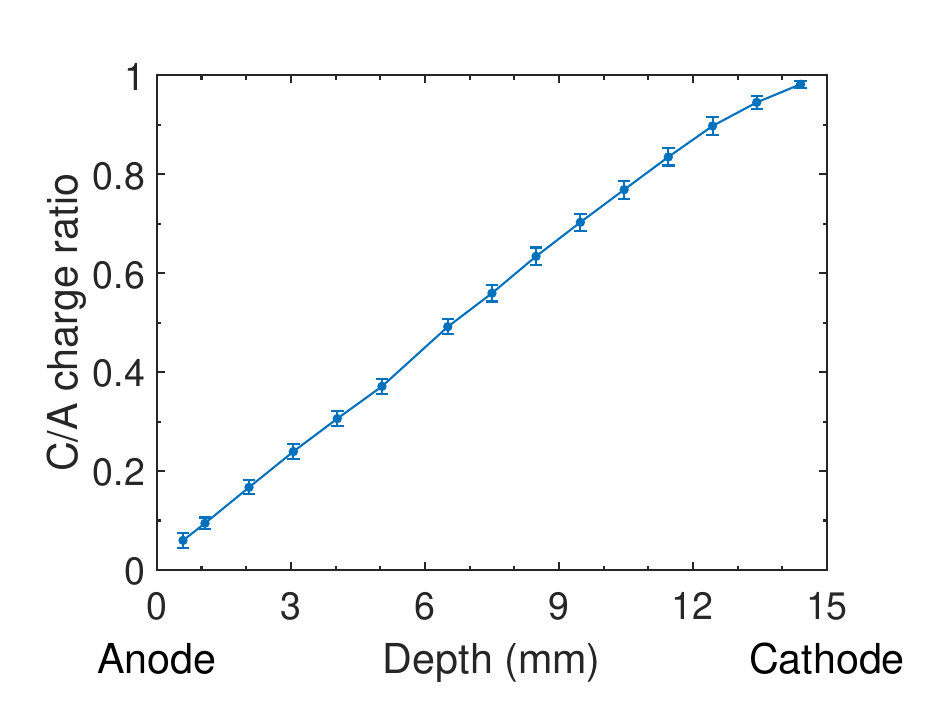}
\includegraphics[width=0.7\columnwidth]{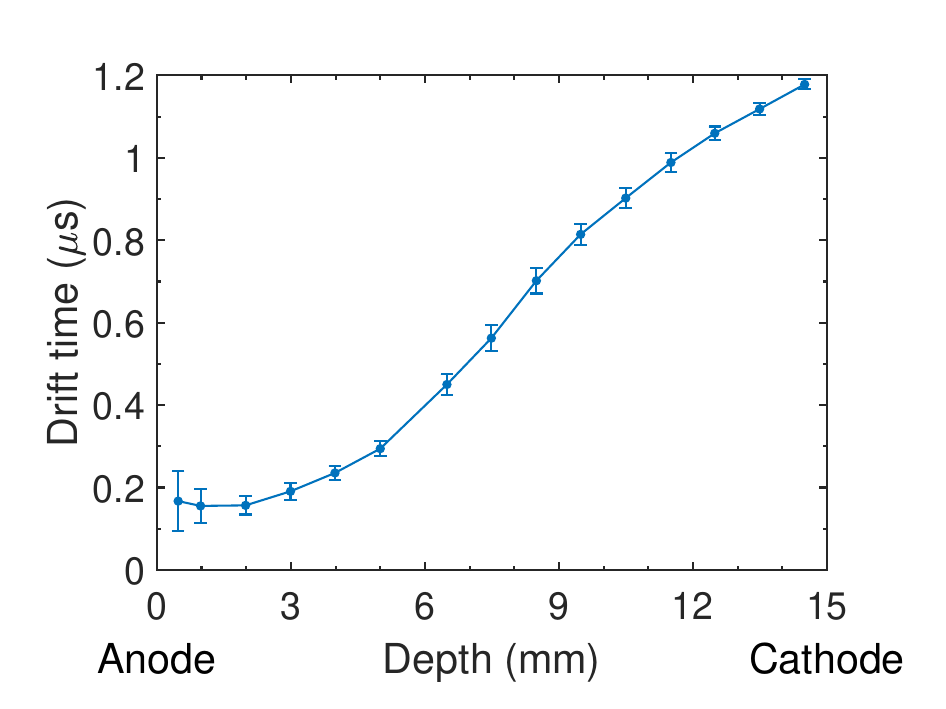}
\includegraphics[width=0.7\columnwidth]{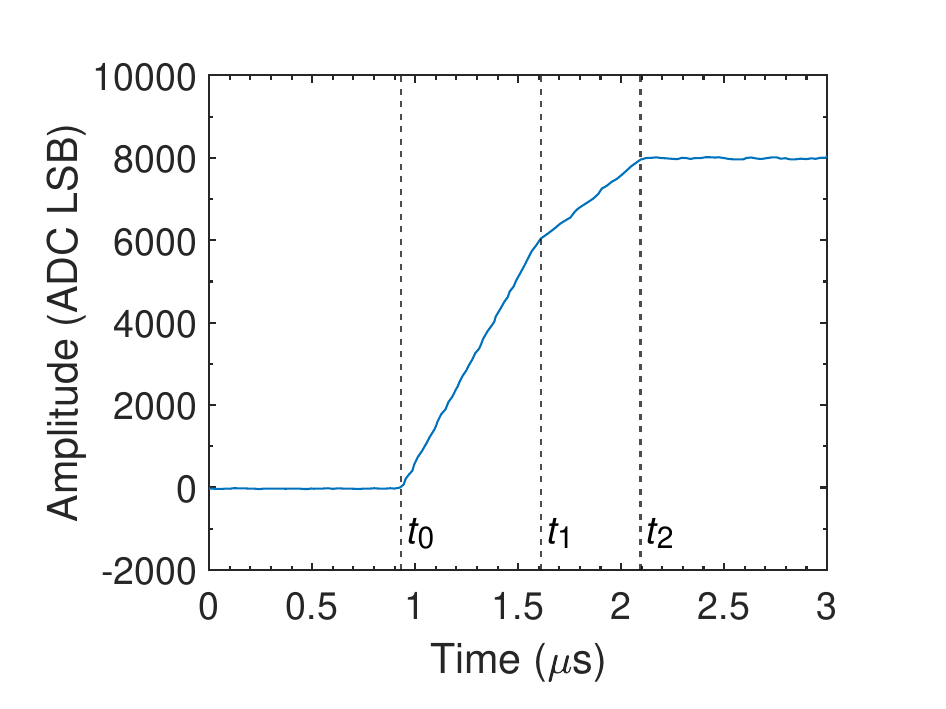}
\caption{C/A charge ratio (left) and electron drift time (middle) as a function of energy deposit depth ($z$) in the CZT detector. The right panel illustrates a waveform measured from the cathode resulted from a two-pixel event. $t_0$ is the time of interactions, and $t_1$ and $t_2$ are the times when the charges stop drifting in the first and second pixel, respectively.}
\label{fig:depth}
\end{figure}

A prototype of the detector module (Fig.~\ref{fig:prototype}) was built to evaluate its performance~\cite{Yang2020}. We used a $^{137}$Cs radioactive source (0.662~MeV gamma-rays) collimated with a 0.3~mm wide slit to illuminate the detector at different depths (see Fig.~\ref{fig:experiment_setup}). The C/A charge ratio and drift time as a function of detector depth are measured and shown in Fig.~\ref{fig:depth}. The drift time vs.\ depth shows a larger degree of nonlinearity and uncertainty than the C/A charge ratio, because the drift velocity of electrons is sensitive to the imperfection of the crystal and nonuniformity of the electric field, as explained in Ref.~\cite{Kim2012}.

\begin{figure}
\centering
\includegraphics[width=0.7\columnwidth]{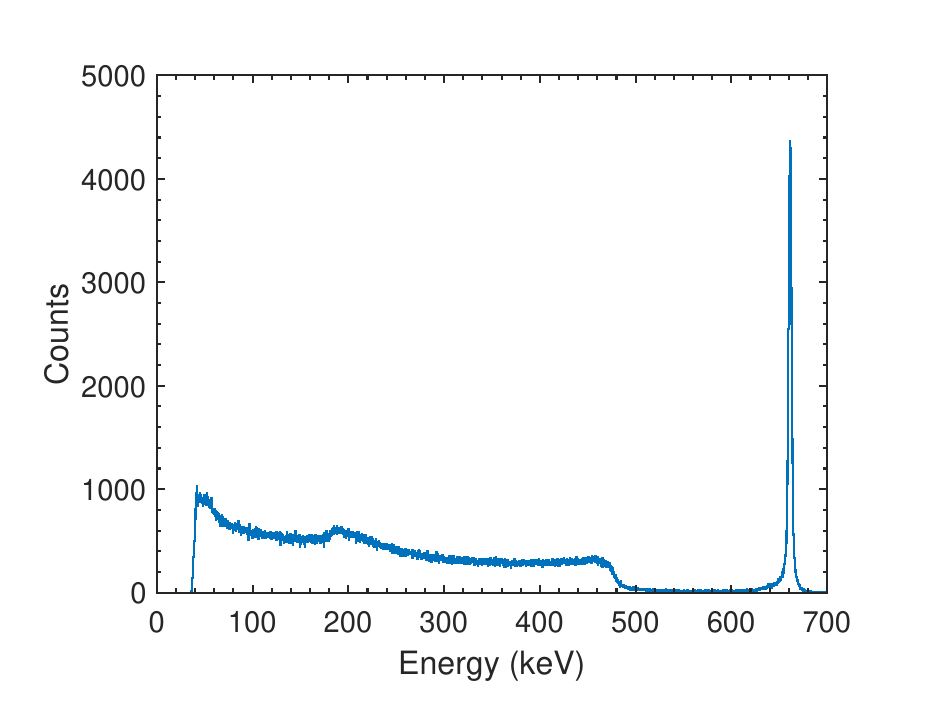}
\caption{Energy spectrum measured with $^{137}$Cs using events summed over all pixels from the four CZT units shown in Fig.~\ref{fig:prototype}. The full energy peak has an FWHM of 0.6\% at 0.662~MeV.}
\label{fig:spec}
\end{figure}

These relations are used for finding the $z$ positions of the interaction points, and the typical FWHM uncertainty is 0.5 mm and 0.57 mm, respectively, for the C/A charge ratio and drift time technique~\cite{He1996,He1997,Kim2012,Yang2020}. These relations can also be used to correct the charge loss during drift and help improve the energy resolution. As a result, the energy resolution at 0.662~MeV is measured to be 0.6\% FWHM (see the spectrum in Fig.~\ref{fig:spec}) after charge loss correction, using events averaged over all pixels of the four detectors shown in Fig.~\ref{fig:prototype}. 

\section{The MASS and MASS-Cube payloads}

\begin{figure}
\centering
\includegraphics[width=\columnwidth]{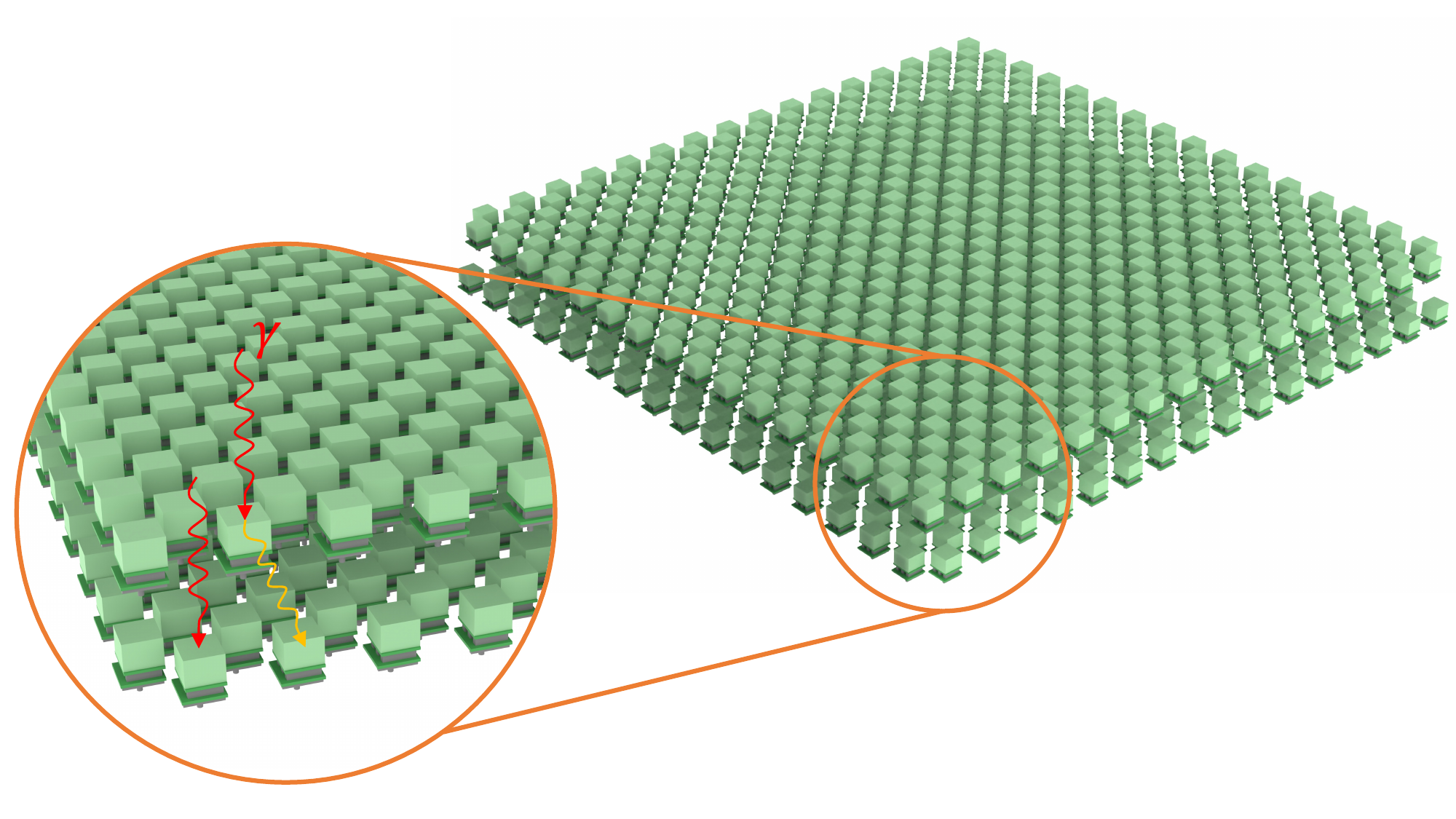}
\caption{Conceptional design of the MASS payload, containing two layers of CZT detectors, with $32 \times 16$ detectors in each layer. The detectors in each layer have a chessboard layout, with a horizontal shift so that there is no blocking for the detectors at the bottom from a top view. The red arrows indicate that on-axis gamma-rays can hit both the top and bottom layers directly. Detectors in the bottom layer can also absorb some of the secondary gamma-rays scattered from the top layer. Active shielding with BGO is not shown.}
\label{fig:mass}
\end{figure}

The MASS payload is an assembly of $32 \times 32 = 1024$ CZT detectors mounted in two layers. Each layer has $32 \times 16$ detectors in a chessboard layout. The spacing between detectors (from center to center) is 2.4~cm along the $x$ or $y$ direction. The bottom layer is 6.5~cm (also from center to center) below the top layer and has a horizontal shift such that all the detectors in the bottom layer are vertically aligned with the open space in the top layer, which means that detectors at the bottom are completely visible from the top. A schematic drawing of the MASS CZT detectors is shown in Fig.~\ref{fig:mass}. The instrument has a field of view of about 2$\pi$. Here as a conceptual design, the support structure is not shown, but will not block the front side of the instrument.

Such a format gives rise to a total detection area of 4096~cm$^2$ for an on-axis source. We refer to events that deposit all of their energies in the CZT detectors as full-energy events, and those with at least two hits as Compton events. Due to the large volume of the unit detector ($2 \times 2 \times 1.5$~cm$^3$), at an energy of 1~MeV, our Geant4 simulations reveal that 61\% of the full-energy Compton events deposit all of their energies within a single detector unit. Thus, the current layout can maximize the effective area for on-axis observations. Also at the same energy, 34\% of the full-energy Compton events are detected by two unit detectors; among them, 18\% are detected by detectors in the same layer and 16\% are detected by both layers. Events detected with both layers, with a small fraction though, provide a better spatial resolution due to a longer baseline for Compton imaging. 

In order to suppress the background, we consider to add Bismuth Germanate (BGO) scintillators surrounding the CZTs as active shielding. There are several possible schemes to mount the BGO detectors. First of all, it is perhaps not a good idea to have anti-coincident detectors in between the CZT detectors, such that Compton events across different CZT detectors will be reduced. Second, as the CZT detectors are almost mounted in a planar geometry, BGOs around the four sides will have a small chance to actively shield CZTs near the center, unless they are very tall. Therefore, a large format of BGO detectors beneath the CZT detectors could be a reasonable solution if anti-coincidence is needed, such that the albedo background can be effectively suppressed. On the other hand, a Compton telescope itself has some degree of capability of vetoing background events. Taking into account the high density and large mass of BGOs, whether or not adding active shielding is a question that deserves further in-depth studies. In the following of this paper, we just assume a planar BGO detector with a thickness of 0.8~cm right beneath the CZT detectors for anti-coincidence. Such a design indicates a total mass of 39~kg for CZTs and 59~kg for BGOs.  The payload CZT envelope is estimated to have a geometry of about $80 \times 80 \times 10$~cm including detectors, electronics, and structures. Such a mass and geometry indicate that the Compton telescope could be integrated into a micro-satellite ($<200$~kg), or at least a mini-satellite even if BGO vetos are needed. 

\begin{figure}
\centering
\includegraphics[width=0.5\columnwidth]{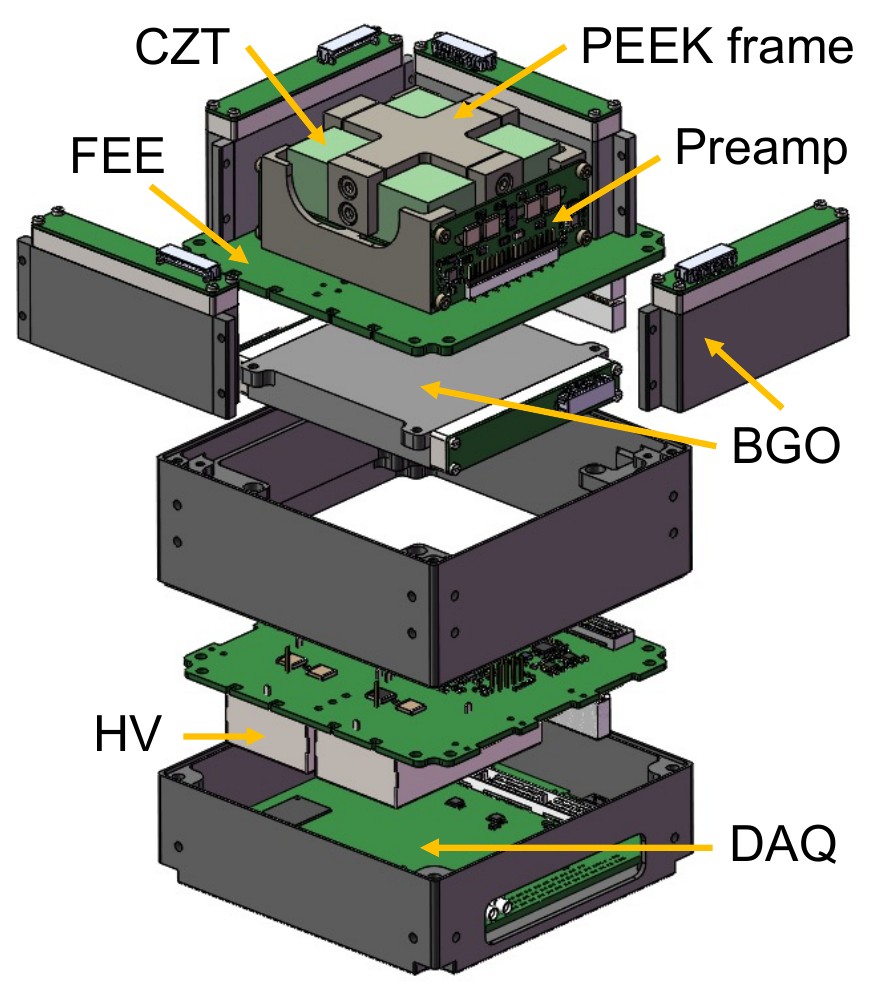}
\caption{Exploded view of the MASS-Cube payload. }
\label{fig:mass_cube}
\end{figure}

The purpose of MASS-Cube is for technical demonstration directly in the space. It is a 1U CubeSat payload consisting of 4 CZT detectors on the same plane, working as a miniature Compton telescope as well as a gamma-ray polarimeter. MASS-Cube is trying to make the first in-orbit test of 3D CZTs for space astronomy. BGO anti-coincidence detectors with a thickness of 0.8~cm are mounted on the five sides (except the top) of the CZTs, to test their efficiency. The CZT detectors with front-end electronics are mounted on a printed circuit board (PCB). The BGO detectors are mounted on the mechanical structure and connected to the PCB with flexible cables. The payload also contains a high voltage (HV) board and a data acquisition board.  A negative HV of $2 - 3$~kV provided by the CAEN\footnote{\url{https://www.caen.it/}} module A7504CN is supplied on the cathode to deplete the CZT. Two HV suppliers are used, each to power two CZT detectors. An FPGA is used for data acquisition and system control for the whole module. Around the 4 CZT detectors, a PEEK frame is used for structural reinforcement. A schematic drawing is shown in Fig.~\ref{fig:mass_cube}. The envelope size is $9.4 \times 9.4 \times 11$~cm$^3$ and the total mass is around 1.7~kg. The power consumption is about 10~W. Thus, MASS-Cube occupies roughly a standard 1U space in a CubeSat, similar to the PolarLight \cite{Feng2019} and GRID \cite{Wen2019} projects that we have done before.

\section{Simulation results}

Based on the current design, we simulated the detection of gamma-rays using the Geant4 package\footnote{\url{https://geant4.web.cern.ch}}, to evaluate the performance, background, and sensitivity of MASS and MASS-Cube. Uncertainties in both energy measurement and depth reconstruction are considered; we adopted the $z$ uncertainties as a function of depth \cite{Yang2020}, using the C/A charge ratio method if there is a single energy deposit in the unit detector, or the drift time method for multiple hits.

\subsection{ARM distribution}
\label{sec:arm}

The angular resolution measure (ARM) is calculated as the smallest angular distance between the reconstructed Compton cone and source location. The displacement is due to uncertainties on energy and position measurements, Doppler broadening, and misidentification of the hit sequence. In the case of our CZT detectors, the position uncertainty has the main contribution to the ARM. 

\begin{figure}
\centering
\includegraphics[width=0.32\columnwidth]{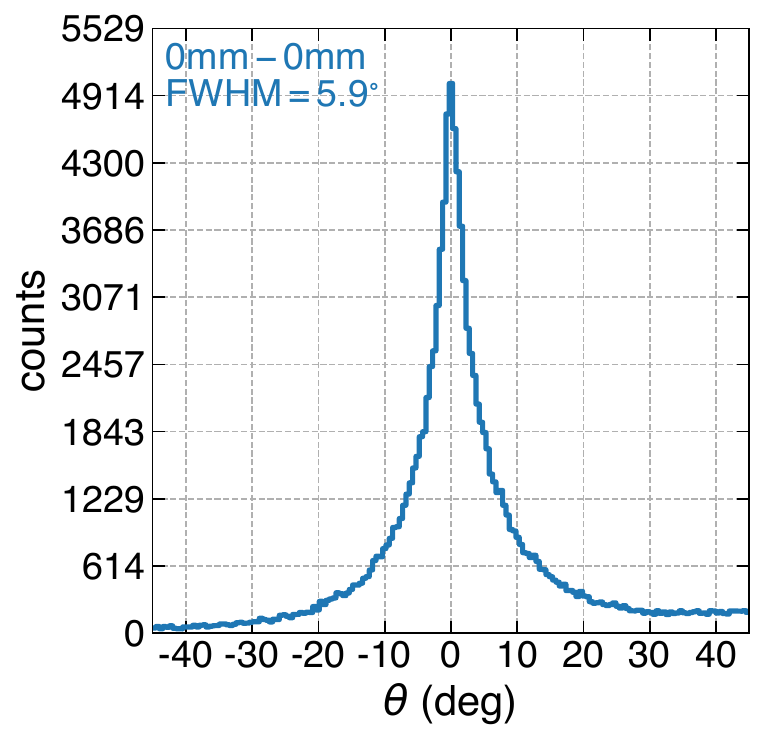}
\includegraphics[width=0.32\columnwidth]{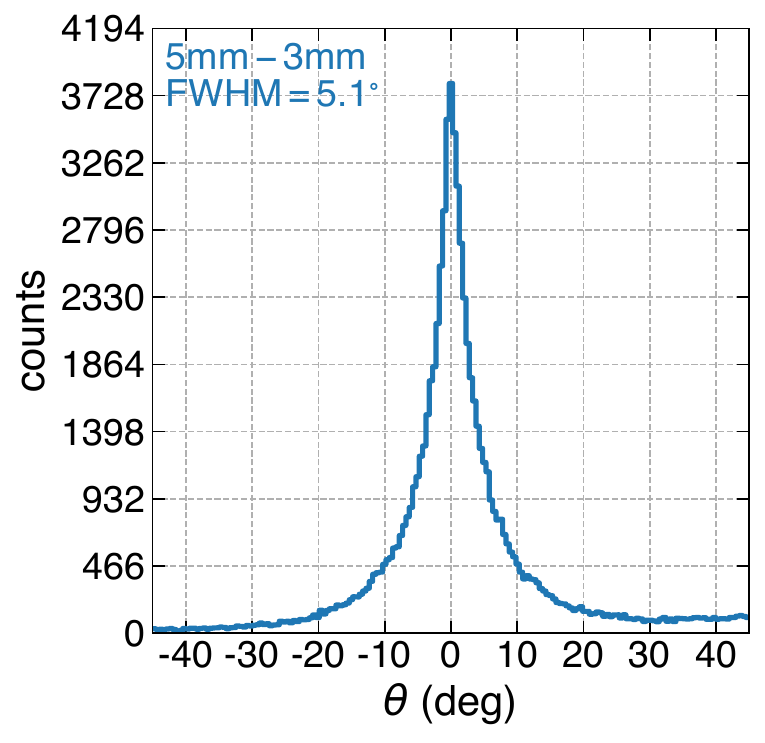}
\includegraphics[width=0.32\columnwidth]{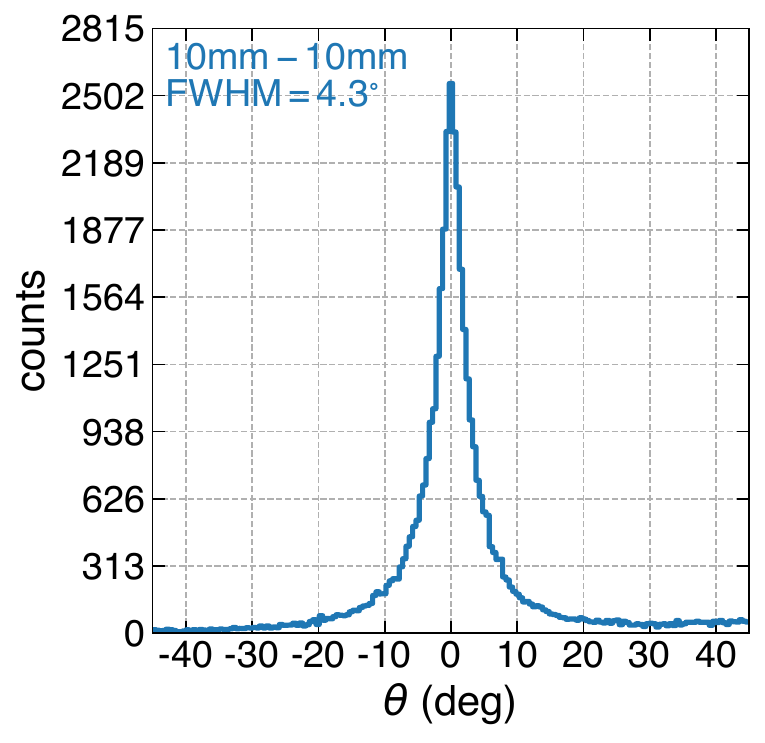}
\caption{Distribution of ARM ($\theta$) with different event selection criteria for 1~MeV on-axis gamma-rays. The event selection criterion is shown in the legend, with the first value indicating the minimum distance between the first and second hits, and the second one indicating the minimum distance between the following successive hits. ``0mm-0mm'' refers to the case with no requirement on the hit distance.}
\label{fig:arm}
\end{figure}

The ARM distribution depends on how we select the events. Here we simulated the ARM distribution for 1~MeV gamma-rays at on-axis with three different selection criteria, see Fig.~\ref{fig:arm}. First of all, we select full-energy events with at least 2 hits in the detectors without any other requirements (denoted as ``0mm-0mm''). Then, we further require a minimum distance of 5~mm between the first two hits, and 3~mm between the following successive hits (denoted as ``5mm-3mm''). In the third case, the distance between all successive hits is required to be at least 10~mm (denoted as ``10mm-10mm''). This is similar to the approach used in Ref.~\cite{Beechert2022}. Compared with 0mm-0mm events, 37.5\% events will be lost given a selection of 5mm-3mm, or 67.4\% are lost given 10mm-10mm, for 1~MeV full-energy events. The FWHM of the ARM distribution is found to be about 5.9$^\circ$, 5.1$^\circ$, and 4.3$^\circ$, respectively, for the three cases in the same order. In the case of 0mm-0mm, the ARM FWHM becomes 5.4$^\circ$, 5.8$^\circ$ and 5.1$^\circ$, respectively, at an off-axis angle of 30, 45 and 60 degrees. Reconstruction for the hit sequence is based on a probabilistic method~\cite{Yoneda2023}. We note that some advanced techniques such as the machine learning~\cite{Takashima2022} may improve the accuracy of reconstruction and narrow down the ARM distribution. This will be investigated in the future.  

\subsection{Effective area}
\label{sec:area}

\begin{figure}
\centering
\includegraphics[width=0.48\columnwidth]{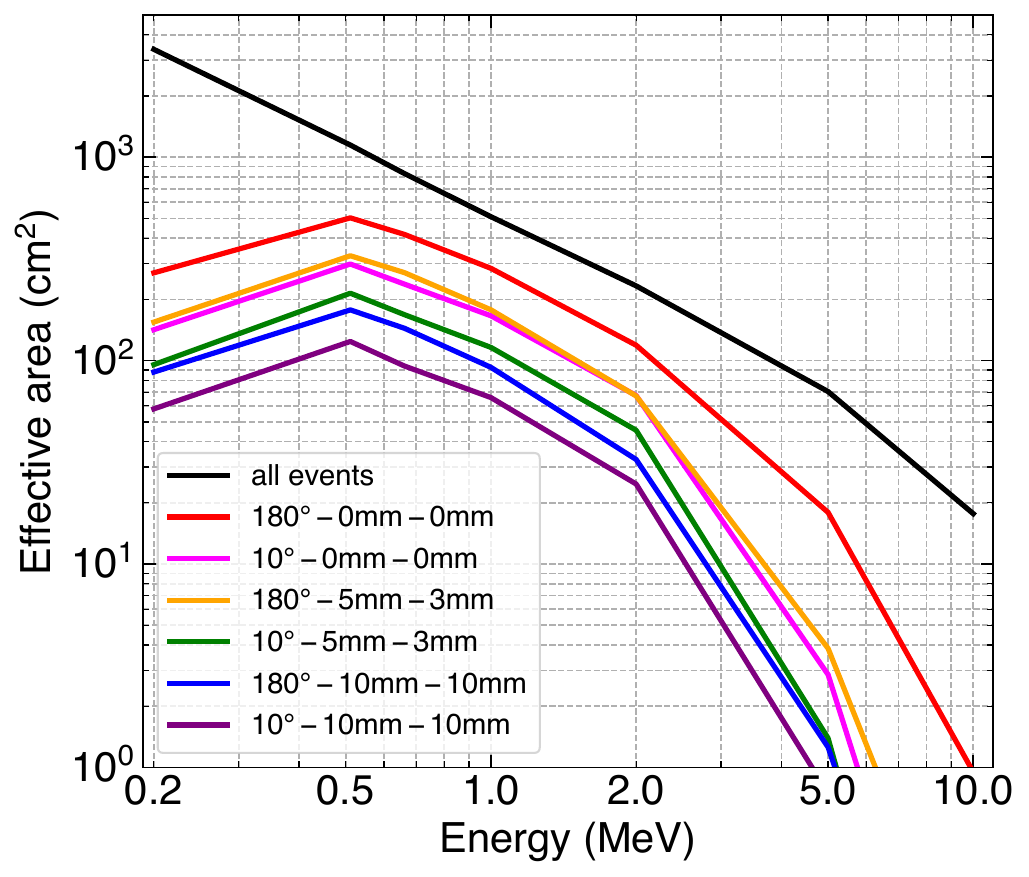}
\includegraphics[width=0.48\columnwidth]{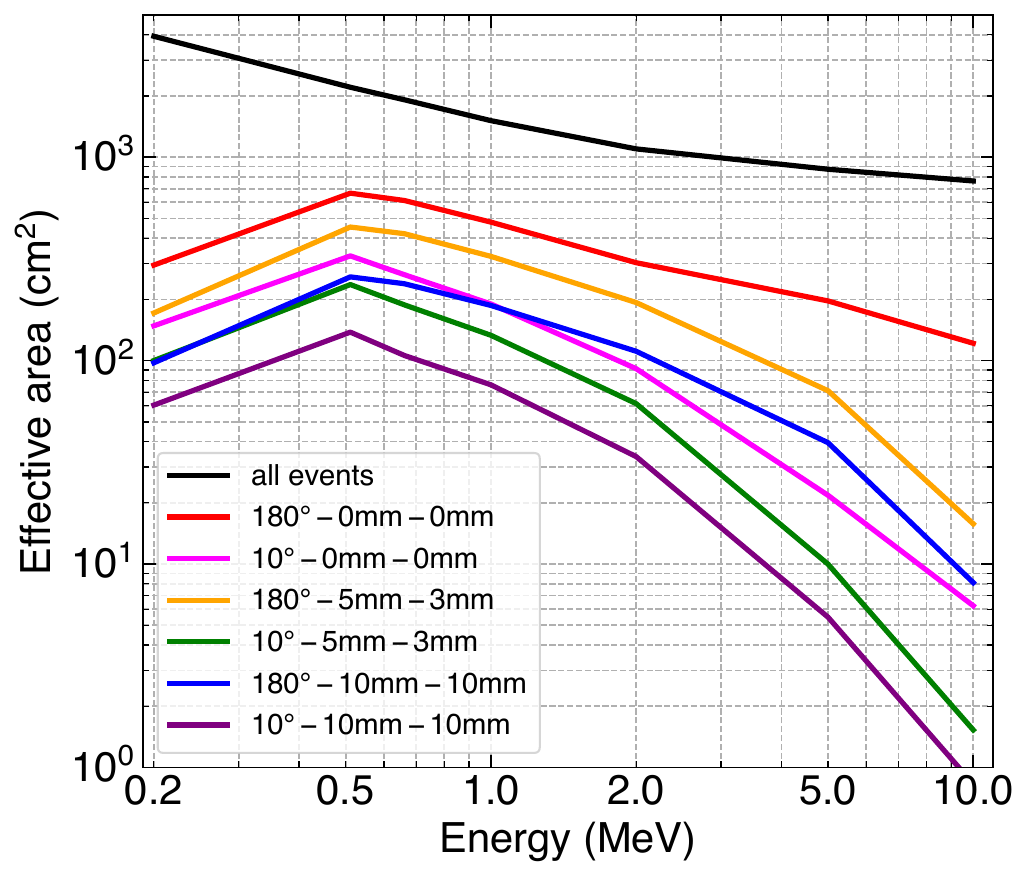}
\caption{Simulated effective area of MASS for full energy deposit ({\bf left}) and any energy deposit ({\bf right}). The event selection criterion is shown in the legend, with an additional requirement on the maximum $\lvert {\rm ARM} \lvert$ (180$^\circ$ means no requirement on the ARM) with respect to those explained in Fig.~\ref{fig:arm}. }
\label{fig:area}
\end{figure}

\begin{figure}
\centering
\includegraphics[width=0.7\columnwidth]{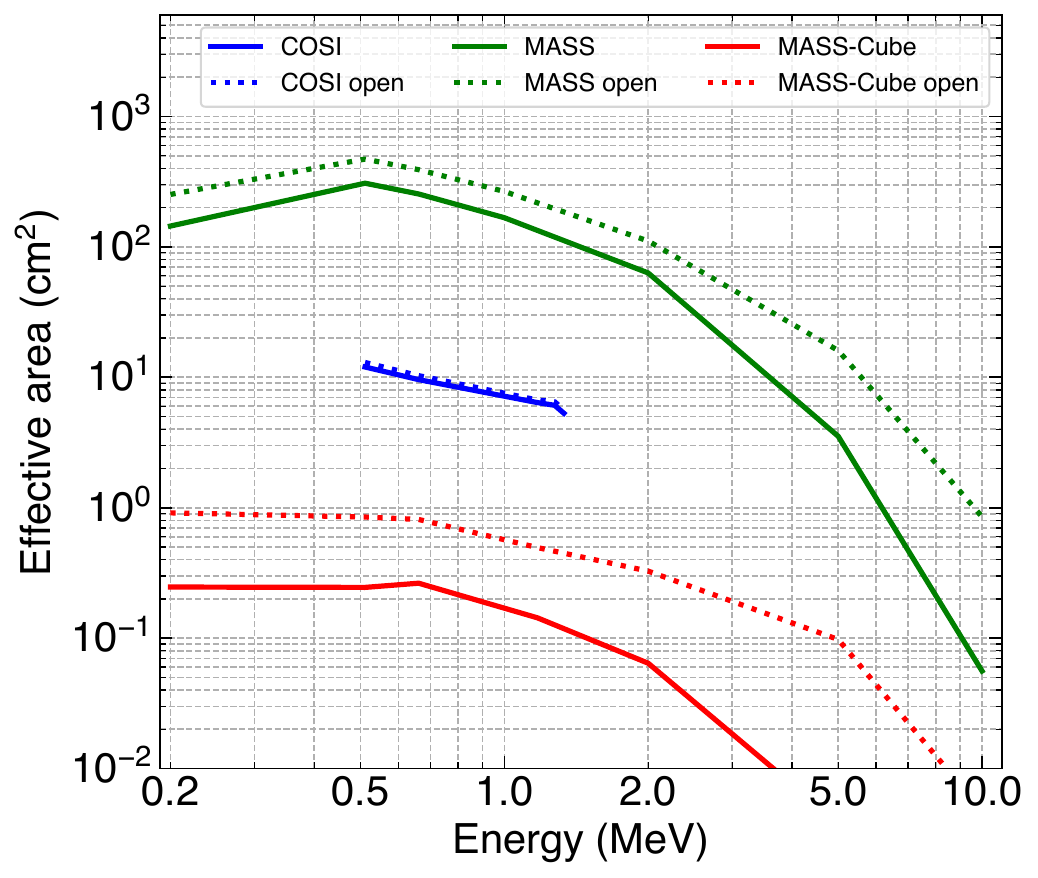}
\caption{Effective area for line detection of MASS (green), MASS-Cube (red), and COSI (blue; adopted from Ref.~\cite{Beechert2022}). The dashed curves (denoted as ``open'' following Ref.~\cite{Beechert2022}) represent the effective area for full energy ($\pm 2 \sigma$) Compton events (at least two hits), and the solid curves have a further requirement on the hit distance, at least 5~mm between the first and second hits, and 3~mm for the following ones.}
\label{fig:area_comp}
\end{figure}

The effective area of a Compton telescope also depends on the scheme of event selection. Here, we plot the effective area of MASS in Fig.~\ref{fig:area} with several selection criteria, including selection on the ARM and hit distance. In principle, more restricted selection criterion leads to a better angular resolution but a smaller effective area. Adopting the three selection criteria used for demonstrating the ARM distribution, we further require the $\lvert {\rm ARM} \lvert$ to be less than 10$^\circ$ in each case. In addition, we also plot the effective area for events with at least a single hit, which are not useful for Compton imaging but can be used for transient detection. The effective area for full energy deposit and any energy deposit are displayed separately in two panels; the former is useful for emission line measurement and the latter is for the continuum. In Fig.~\ref{fig:area_comp}, the effective areas of MASS and MASS-Cube are compared with those of COSI under the same selection criteria. As one can see, MASS is at least one order of magnitude larger than COSI in the energy range of 0.5--1 MeV. We note that the quoted effective area of COSI is based on a $2 \times 2 \times 3$ detector structure (the balloon design), while the satellite design will be $2 \times 2 \times 4$. An extra layer will certainly improve the effective area at the high energy end.

We also simulated the effective areas at an off-axis angle of 30$^\circ$, 45$^\circ$, and 60$^\circ$. Thanks to the thickness of the CZT detector and gaps between them, the effective area increases with increasing off-axis angle, because the intersection area of the unit detector becomes larger when viewed at these off-axis angles.

\begin{figure}
\centering
\includegraphics[width=0.7\columnwidth]{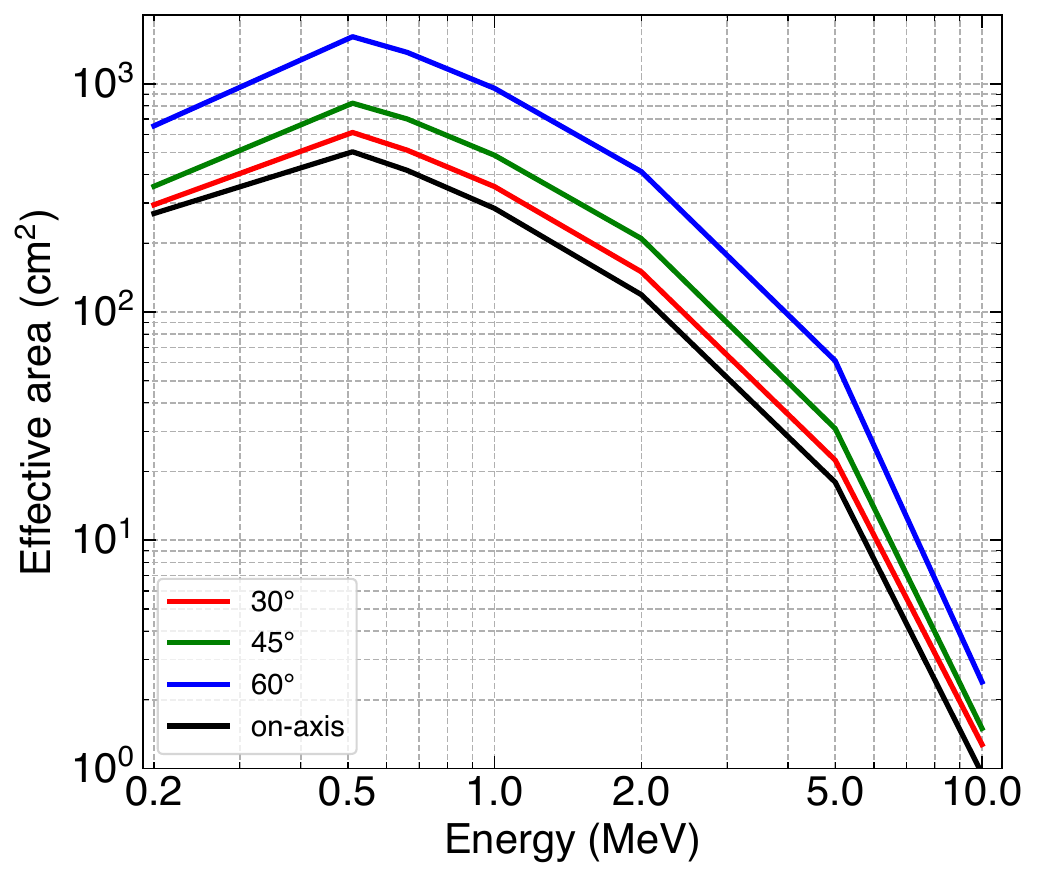}
\caption{Off-axis effective area of MASS for line detection with the selection criterion ``180$^\circ$-0mm-0mm'' (same as the red curve in Fig.~\ref{fig:area}). The on-axis effective area is also shown for comparison. }
\label{fig:area_off_axis}
\end{figure}

\section{Background}
\label{sec:bkg}

\begin{figure*}[th]
\centering
\includegraphics[width=0.3\textwidth]{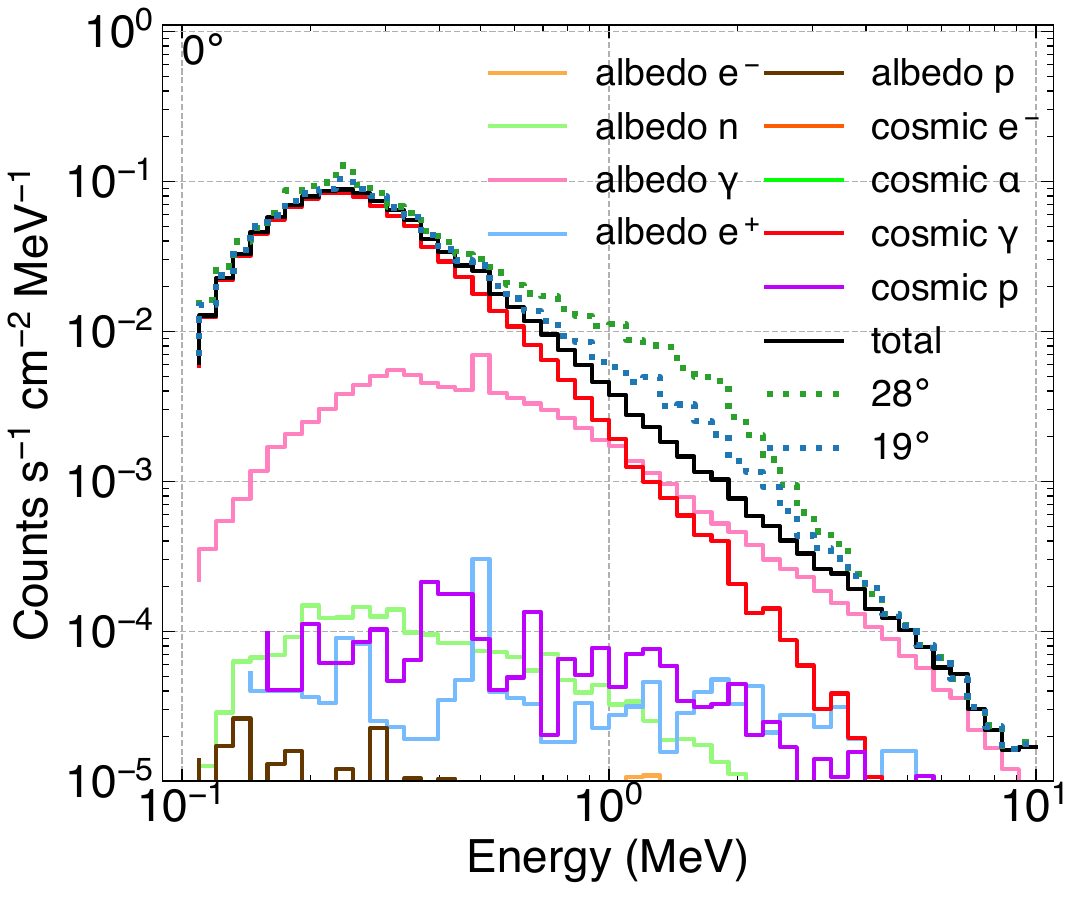}
\includegraphics[width=0.3\textwidth]{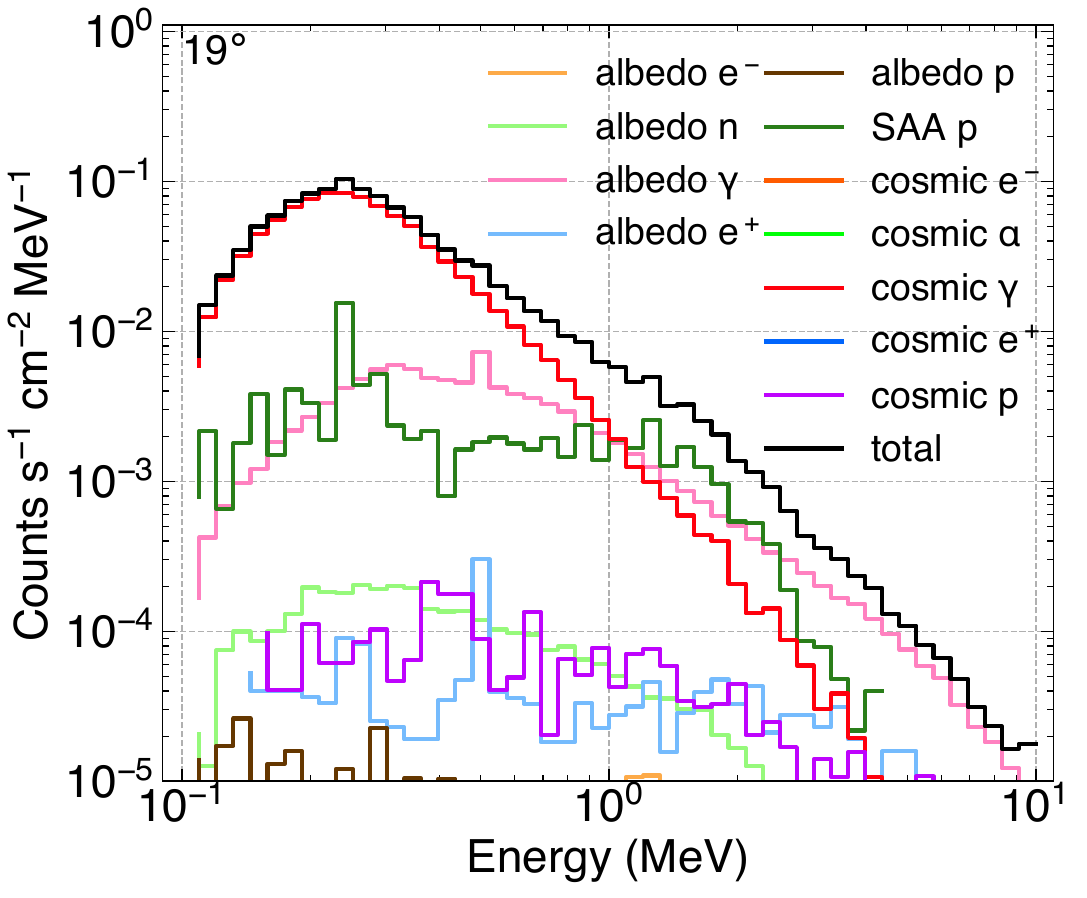}
\includegraphics[width=0.3\textwidth]{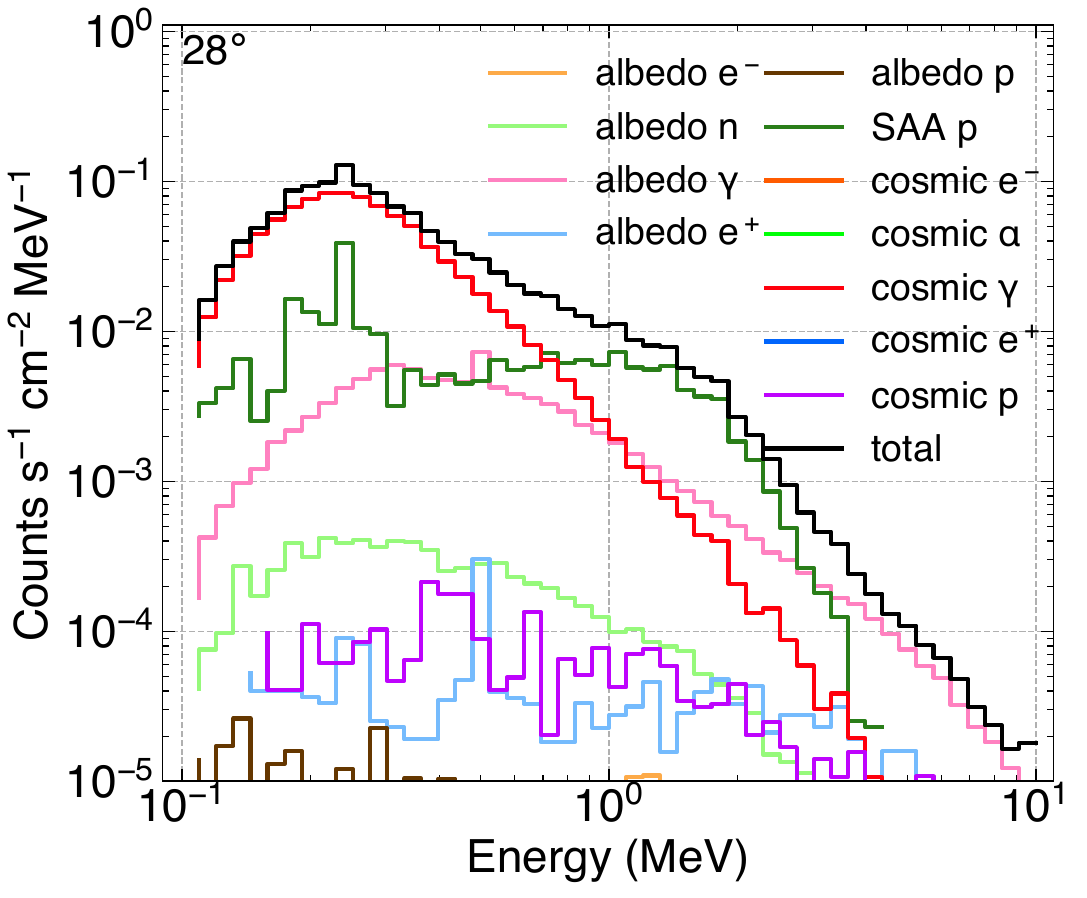}
\caption{Simulated background spectra of MASS in low-Earth orbits with different inclination angles, using events with $\lvert {\rm ARM} \lvert < 10^\circ$. The total background spectra for the 19$^\circ$ and 28$^\circ$ orbits are also plotted in the 0$^\circ$ orbit for comparison.}
\label{fig:bkg_mass}
\end{figure*}

Considering a 550~km circular low-Earth orbit with an inclination of 0$^\circ$, 19$^\circ$ (launched from Wenchang), or 28$^\circ$ (launched from Xichang), we simulated the MASS background spectra as a result of different incident components, including the
cosmic photons~\cite{Gruber1999}, 
cosmic electrons and positrons~\cite{Mizuno2004},
cosmic protons\footnote{\label{spenvis}Obtained from SPENVIS (\url{https://www.spenvis.oma.be}).}, 
cosmic alpha particles\textsuperscript{\ref{spenvis}}, 
albedo photons~\cite{Mizuno2004,Tuerler2010,Abdo2010}, 
albedo electrons and positrons~\cite{Alcaraz2000,Mizuno2004}, 
albedo protons~\cite{Mizuno2004}, and
albedo neutrons~\cite{Kole2015}. 
In the south Atlantic anomaly (SAA) region, we considered radioactivation of materials caused by high energy protons\textsuperscript{\ref{spenvis}}, and subsequent delayed emission after the satellite leaves the SAA. This component follows an exponential decay with time and its strength strongly depends on the orbital inclination or frequency/duration of SAA transits. Here, we considered background accumulation because of limited time between SAA transits, and calculated an orbit-averaged result at time intervals 10~min after the satellite leaving the SAA. The instrument always points toward the anti-Earth direction in the simulation. We note that, as a rough estimate, the rest of the payload and satellite components are not modeled at this stage. With additional surrounding materials, the particle induced background will increase, and the current results can be regarded as a lower limit. The background spectra are shown in Fig.~\ref{fig:bkg_mass} for the three cases, constructed using events with $\lvert {\rm ARM} \lvert < 10^\circ$. The selection on hit distance is not applied because it does not help improve the sensitivity based on the simulation data. 

The cosmic diffuse gamma-rays and albedo gamma-rays are the two major components that dominate at the low and high energies in the MASS band, respectively. The delayed emission due to SAA transit is highly dependent on the orbit inclination, and is important only in the 28$^\circ$ orbit in the three cases. In the first panel of Fig.~\ref{fig:bkg_mass}, the three background spectra are plotted together for comparison. Although the delayed background emerges in non-equatorial orbits and increases with increasing inclinations, the overall background spectra for the three orbits are comparable. To conclude, a lower inclination orbit is preferred but the three orbits are all acceptable. 

For MASS-Cube, the whole payload including detectors and surrounding materials shown in Fig.~\ref{fig:mass_cube} is modeled in Geant4 for simulation. The instrument is assumed to fly in a polar orbit at an altitude of 550~km, as this is the most likely type of orbits that a CubeSat will go. 

\begin{figure}[h!]
\centering
\includegraphics[width=0.3\textwidth]{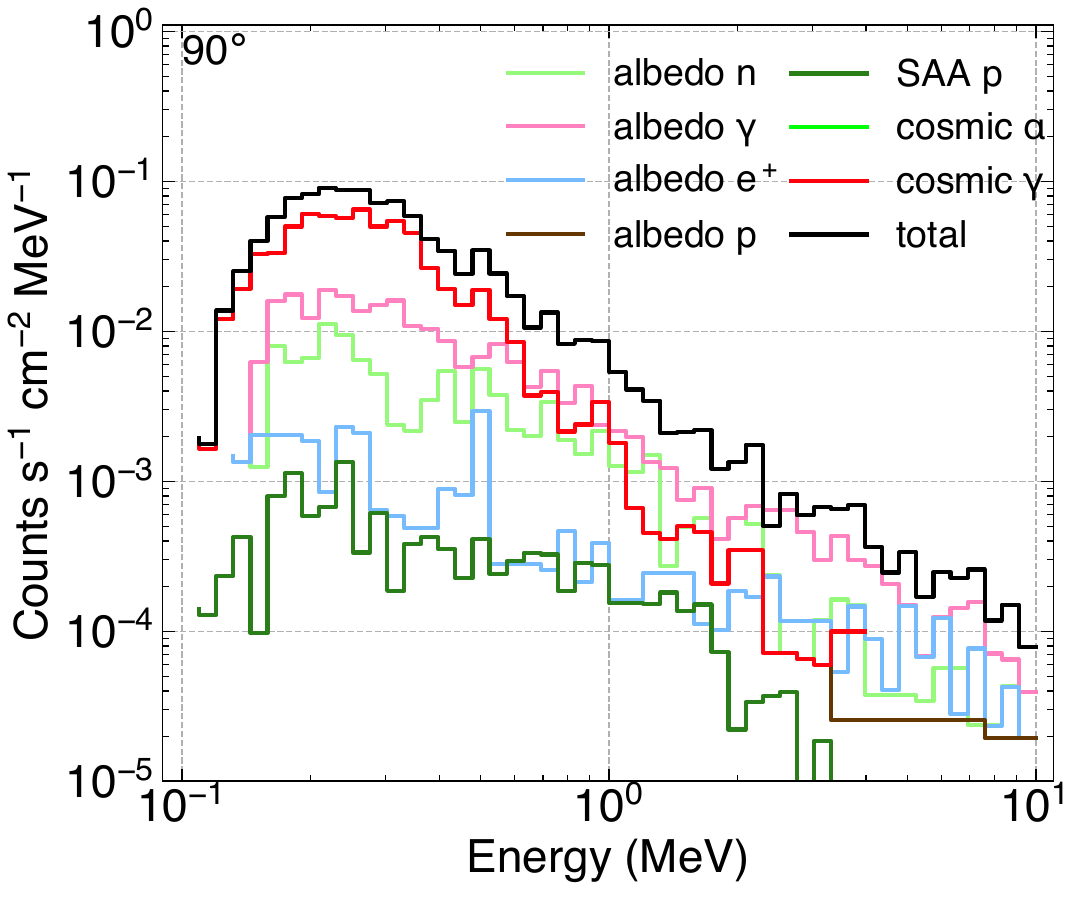}
\caption{Simulated background spectra of MASS-Cube in a polar orbit, using events with $\lvert {\rm ARM} \lvert < 10^\circ$.}
\label{fig:bkg_mass_cube}
\end{figure}

\section{Sensitivity}
\label{sec:sensitivity}

Given the effective area and background spectra, we calculated the sensitivities for MASS and MASS-Cube. The line sensitivity ($S_{\rm line}$) is calculated with the following equation \cite{Schoenfelder1993},
\begin{equation}
S_{\rm line}(E) = \frac{1}{A_{\rm eff}t}(\frac{n^2}{2}+\sqrt{\frac{n^4}{4} + n^2BA_{\rm eff}t\Delta E}) \, ,
\label{eq:line}
\end{equation}
where $A_{\rm eff}$ is the effective area, $B$ is the background flux, $n$ represents the significance in the unit of $\sigma$, $t$ is the observing time, and $\Delta E$ is the line width. We adopt the $\pm 3 \sigma$ line width for $\Delta E$. 

\begin{figure}
\centering
\includegraphics[width=0.7\columnwidth]{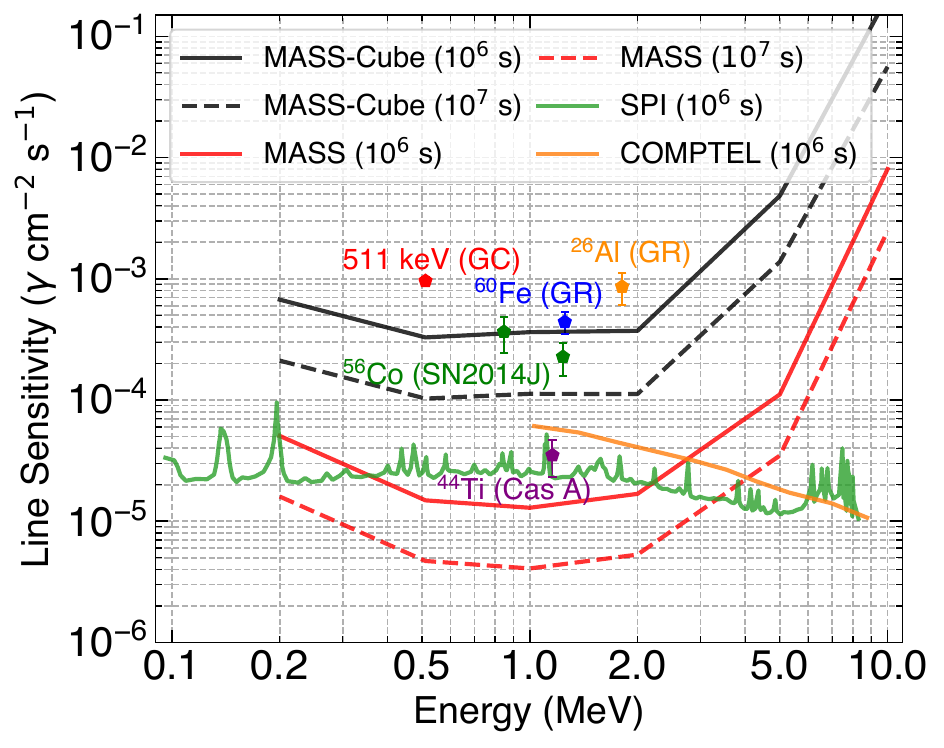}
\caption{$3\sigma$ line sensitivities for MASS and MASS-Cube. The background spectrum in the 19$^\circ$ orbit is adopted for MASS, and that in the polar orbit is used for MASS-Cube. The sensitivity of INTEGRAL/SPI~\cite{Winkler2003} and COMPTEL~\cite{Schoenfelder1993} are also plotted for comparison. The line fluxes from known sources are shown: the 0.511~MeV emission from the Galactic center~\cite{Siegert2016},  $^{60}$Fe~\cite{Wang2007} and $^{26}$Al~\cite{Wang2009} emission from the Galactic ridge, $^{56}$Co from SN~2014J~\cite{Diehl2015}, and $^{44}$Ti from Cas~A~\cite{Siegert2015}.} 
\label{fig:line}
\end{figure}

Assuming $n = 3$ and $t = 10^6$ or $10^7$~s, we plot the on-axis line sensitivities for MASS and MASS-Cube in Fig.~\ref{fig:line}, together with detected line fluxes from some known sources, including the 0.511~MeV from the Galactic center~\cite{Siegert2016}, $^{60}$Fe~\cite{Wang2007} and $^{26}$Al~\cite{Wang2009} emission from the Galactic ridge, $^{56}$Co from SN~2014J~\cite{Diehl2015}, and $^{44}$Ti from Cas~A~\cite{Siegert2015}.  The line sensitivities for INTEGRAL/SPI~\cite{Winkler2003} and COMPTEL~\cite{Schoenfelder1993} are displayed for comparison. We note that, compared with coded aperture mask instruments, the MASS sensitivity will not degrade for extended sources. As mentioned above, when surrounding materials are considered, the true background may be higher and lead to a slightly worse sensitivity.

\begin{figure}
\centering
\includegraphics[width=0.7\columnwidth]{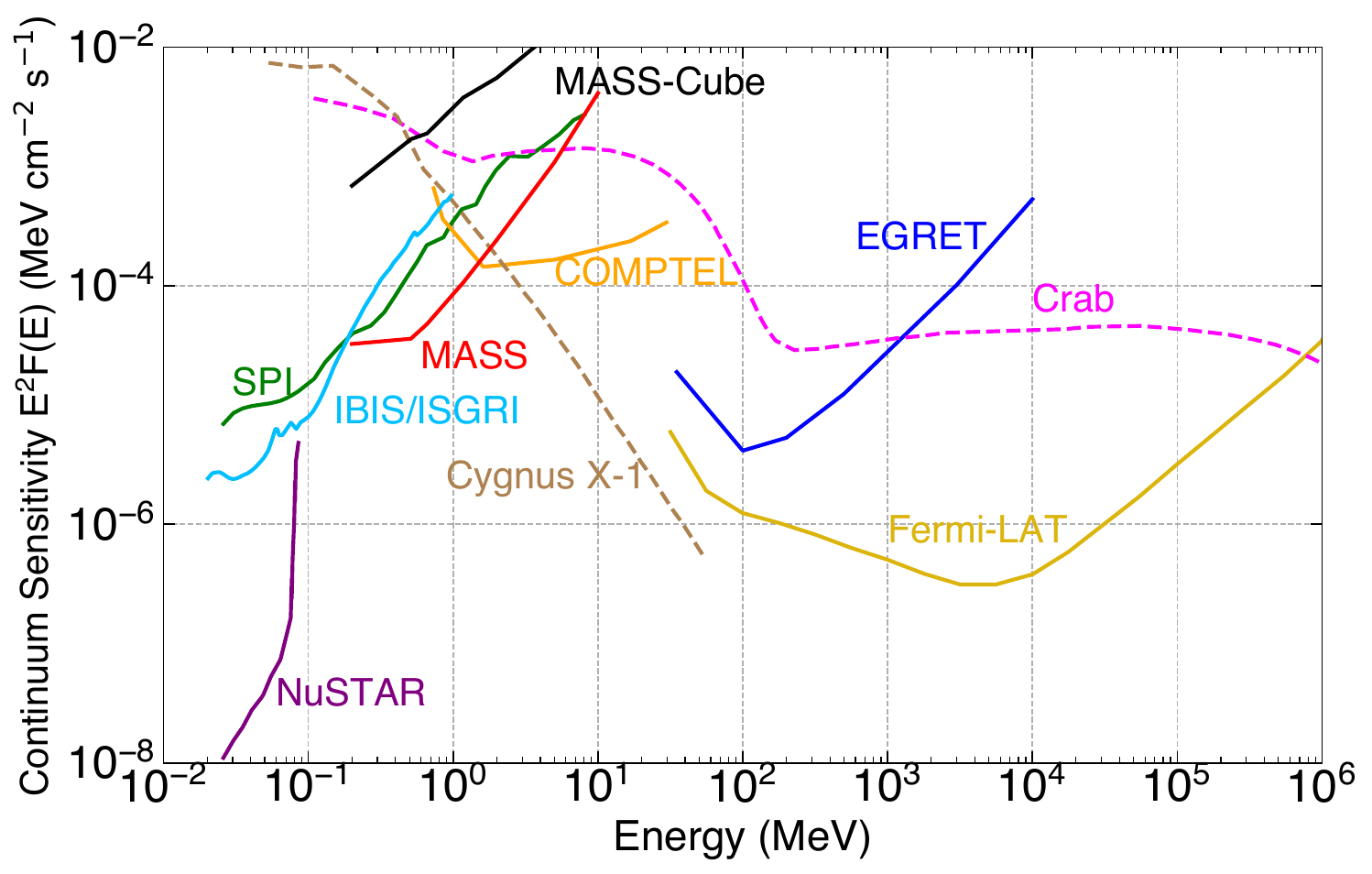}
\caption{$3\sigma$ continuum sensitivity with an exposure of $10^6$~s for MASS and MASS-Cube. The sensitivity curves for other instruments are plotted for comparison: NuSTAR~\cite{Harrison2013}, SPI~\cite{Winkler1995}, IBIS/ISGRI~\cite{Lebrun2003}, COMPTEL~\cite{Schoenfelder1993}, Fermi-LAT ($20\%$ efficiency over 5 years)~\cite{Atwood2009}, and EGRET($40\%$ efficiency over two weeks~\cite{Thompson1993}). The energy spectra of the Crab pulsar~\cite{deJager1996} and Cygnus X-1~\cite{McConnell2000} are also plotted.}
\label{fig:conti}
\end{figure}

The on-axis continuum sensitivity ($S_{\rm conti}$) is calculated as \cite{Schoenfelder1993}
\begin{equation}
S_{\rm conti}(E) = \frac{E}{A_{\rm eff}t}(\frac{n^2}{2}+\sqrt{\frac{n^4}{4} + \frac{n^2BA_{\rm eff}t}{E}}) \, ,
\label{eq:conti}
\end{equation}
where the parameters are the same as in Eq.~\ref{eq:line}. We plot the continuum sensitivity in Fig.~\ref{fig:conti}, in comparison with those of other instruments. 

\section{Discussion}
\label{sec:discuss}

Thanks to the recent progresses in the development of CZT detectors~\cite{Abraham2023,Yang2020}, we propose a large format Compton telescope MASS. The detector is 3D position sensitive and can be operated at room-temperature, allowing us to construct a large area instrument adapted to a micro- or mini-satellite. Compared with COSI, MASS will have a similar energy resolution \cite{Beechert2022} and a much larger effective area. The sensitivity of MASS degrades rapidly with increasing energy, as a result of two detector layers and consequently lower efficiency at higher energies. Thus, MASS could be a future mission successive to COSI to follow up MeV spectral features for nuclear astrophysics. 

We plan to test the technique in space with the pathfinder MASS-Cube in the near future. Currently, we have finished the design and assembly of the payload, and will perform tests and calibrations subsequently. The launch is scheduled around 2024, probably into a sun-synchronous orbit onboard a CubeSat. Besides the test of the CZT detector, MASS-Cube will also test the capability and efficiency of the BGO anti-coincidence. According to our simulations, the presence of the active shielding in MASS-Cube can reduce the Compton-events background by a factor of about 1.4. This is mainly because a Compton telescope itself can distinguish a considerable fraction of background based on Compton reconstruction. These results will be useful for the design of the veto system in MASS, or allow us to weigh whether or not we should have active shielding based on the cost and benefit. 
 
For semiconductor detectors, radiation damage in the orbit is always an issue of concern \cite{Shy2023}. For CZT detectors, non-ionization energy loss caused by high energy protons is argued to be the main cause of radiation damage~\cite{Kuvvetli2003}. In that case, additional trapping centers are generated, and consequently, the probability of electron capture during drift increases. This effect can be recovered with annealing~\cite{Fraboni2004}. With Geant4 simulations, we estimated the radiation damage after a 3-year operation in the orbit at a working temperature of 300~K, which can be regarded as a long-term annealing procedure. With an initial energy resolution of 0.6\% at 0.662~MeV, we found that the energy resolution after 3 years would become 0.6\%, 0.8\%, 1.3\%, and 1.4\%, respectively, with an orbital inclination of 0$^\circ$, 19$^\circ$, 28$^\circ$, and 90$^\circ$. These results argue in favor of a relatively low inclination orbit for MASS.

MASS will have the advantage of detecting gamma-ray lines from radioactive isotopes in the cosmos compared with the present and future gamma-ray detectors. Thus, the main science objectives will be strongly connected to breakthroughs in nuclear astrophysics with the study of emission lines due to positron-electron annihilation, $^{56}$Co, $^{44}$Ti, $^{26}$Al, $^{60}$Fe, etc. Searching for the 0.511~MeV emission from individual Galactic sources and nearby dwarf galaxies will probe the origin of positrons and annihilation signal of dark matter. Detection of $^{56}$Ni decay chain lines from nearby Ia supernovae within $\sim$10~Mpc will help resolve the supernova progenitors, and provide a direct test for the standard candle combining with the search of $^{44}$Ti lines from young Galactic supernova remnants~\cite{Wang2014,Wang2016}. $^{26}$Al and $^{60}$Fe may have the similar physical origin, but the productions are dominated by the different stages of massive stars. Thus, simultaneously detecting these two isotopes in the star forming regions will strongly constrain the physical environment of star formation process and explosion mechanism of core-collapse supernovae.

MASS is optimized for line detection, but observations of the continuum are also of great interest. For pulsars, which are poorly understood in the MeV band~\cite{Ruderman1975,Arons1981,Cheng1986,Harding2005,Harding2006,Kuiper2018}, the MeV spectra and pulse profiles may help distinguish high-energy emission models~\cite{Petri2015,Takata2017,Barnard2022,Iniguez-Pascual2022}, and constrain the maximum electron energy in the gaps~\cite{Torres2018,Torres2019,Acciari2021}. It is interesting to confirm the presence of a new population of MeV-peaked pulsars~\cite{Kuiper2015,Kuiper2018,CotiZelati2020,Hare2021} that challenge the current models. The MeV spectroscopy and polarimetry are powerful in differentiating the leptonic and hadronic models of blazars~\cite{Zhang2013,Petropoulou2015}, and may shed light on the nature of FR~0 radio galaxies~\cite{Baldi2019}. A Compton telescope is a natural gamma-ray polarimeter. Detailed simulation work and beam tests will be conducted in the near future and reported separately.

\backmatter
\bmhead{Acknowledgments}

HF acknowledges funding support from the National Natural Science Foundation of China under grants Nos.\ 12025301, 12103027, \& 11821303, the National Key R\&D Project under grant 2018YFA0404502, and the Tsinghua University Initiative Scientific Research Program.

\section*{Declarations}
\begin{itemize}
\item Funding

This work was supported by the National Natural Science Foundation of China under grants Nos.\ 12025301, 12103027, \& 11821303, the National Key R\&D Project under grant 2018YFA0404502, and the Tsinghua University Initiative Scientific Research Program.
\item Conflict of interest

The authors have no competing interests to declare that are relevant to the content of this article.
\item Availability of data and materials

Data sets generated during the current study are available from the corresponding author on reasonable request.
\item Authors' contributions

HF and MZ led the project, with a focus on science and instrumentation, respectively. J Zhu, CYH, JYH, HKC, and HL performed the simulations. XZ, HC, XP, GM, QW, and YL contributed to the development of the payload instrument. XB, MG, LJ, JL, YS, WW, XW, BZ, and J Zhang participated in the discussions of the science objectives. 
\end{itemize}

 

\begin{thebibliography}{124}
\ifx \bisbn   \undefined \def \bisbn  #1{ISBN #1}\fi
\ifx \binits  \undefined \def \binits#1{#1}\fi
\ifx \bauthor  \undefined \def \bauthor#1{#1}\fi
\ifx \batitle  \undefined \def \batitle#1{#1}\fi
\ifx \bjtitle  \undefined \def \bjtitle#1{#1}\fi
\ifx \bvolume  \undefined \def \bvolume#1{\textbf{#1}}\fi
\ifx \byear  \undefined \def \byear#1{#1}\fi
\ifx \bissue  \undefined \def \bissue#1{#1}\fi
\ifx \bfpage  \undefined \def \bfpage#1{#1}\fi
\ifx \blpage  \undefined \def \blpage #1{#1}\fi
\ifx \burl  \undefined \def \burl#1{\textsf{#1}}\fi
\ifx \doiurl  \undefined \def \doiurl#1{\url{https://doi.org/#1}}\fi
\ifx \betal  \undefined \def \betal{\textit{et al.}}\fi
\ifx \binstitute  \undefined \def \binstitute#1{#1}\fi
\ifx \binstitutionaled  \undefined \def \binstitutionaled#1{#1}\fi
\ifx \bctitle  \undefined \def \bctitle#1{#1}\fi
\ifx \beditor  \undefined \def \beditor#1{#1}\fi
\ifx \bpublisher  \undefined \def \bpublisher#1{#1}\fi
\ifx \bbtitle  \undefined \def \bbtitle#1{#1}\fi
\ifx \bedition  \undefined \def \bedition#1{#1}\fi
\ifx \bseriesno  \undefined \def \bseriesno#1{#1}\fi
\ifx \blocation  \undefined \def \blocation#1{#1}\fi
\ifx \bsertitle  \undefined \def \bsertitle#1{#1}\fi
\ifx \bsnm \undefined \def \bsnm#1{#1}\fi
\ifx \bsuffix \undefined \def \bsuffix#1{#1}\fi
\ifx \bparticle \undefined \def \bparticle#1{#1}\fi
\ifx \barticle \undefined \def \barticle#1{#1}\fi
\bibcommenthead
\ifx \bconfdate \undefined \def \bconfdate #1{#1}\fi
\ifx \botherref \undefined \def \botherref #1{#1}\fi
\ifx \url \undefined \def \url#1{\textsf{#1}}\fi
\ifx \bchapter \undefined \def \bchapter#1{#1}\fi
\ifx \bbook \undefined \def \bbook#1{#1}\fi
\ifx \bcomment \undefined \def \bcomment#1{#1}\fi
\ifx \oauthor \undefined \def \oauthor#1{#1}\fi
\ifx \citeauthoryear \undefined \def \citeauthoryear#1{#1}\fi
\ifx \endbibitem  \undefined \def \endbibitem {}\fi
\ifx \bconflocation  \undefined \def \bconflocation#1{#1}\fi
\ifx \arxivurl  \undefined \def \arxivurl#1{\textsf{#1}}\fi
\csname PreBibitemsHook\endcsname

\bibitem{Kierans2022}
\begin{bchapter}
\bauthor{\bsnm{{Kierans}}, \binits{C.}},
\bauthor{\bsnm{{Takahashi}}, \binits{T.}},
\bauthor{\bsnm{{Kanbach}}, \binits{G.}}:
\bctitle{{Compton Telescopes for Gamma-Ray Astrophysics}}.
In: \bbtitle{Handbook of X-ray and Gamma-ray Astrophysics},
p. \bfpage{18}
(\byear{2022}).
\doiurl{10.1007/978-981-16-4544-0_46-1}
\end{bchapter}
\endbibitem

\bibitem{Maoz2014}
\begin{barticle}
\bauthor{\bsnm{{Maoz}}, \binits{D.}},
\bauthor{\bsnm{{Mannucci}}, \binits{F.}},
\bauthor{\bsnm{{Nelemans}}, \binits{G.}}:
\batitle{{Observational Clues to the Progenitors of Type Ia Supernovae}}.
\bjtitle{\araa}
\bvolume{52},
\bfpage{107}--\blpage{170}
(\byear{2014})
{\href{https://arxiv.org/abs/1312.0628}{{arXiv:1312.0628}}}
{[astro-ph.CO]}.
\doiurl{10.1146/annurev-astro-082812-141031}
\end{barticle}
\endbibitem

\bibitem{Timmes2019}
\begin{barticle}
\bauthor{\bsnm{{Timmes}}, \binits{F.}},
\bauthor{\bsnm{{Fryer}}, \binits{C.}},
\bauthor{\bsnm{{Timmes}}, \binits{F.}},
\bauthor{\bsnm{{Hungerford}}, \binits{A.L.}},
\bauthor{\bsnm{{Couture}}, \binits{A.}},
\bauthor{\bsnm{{Adams}}, \binits{F.}},
\bauthor{\bsnm{{Aoki}}, \binits{W.}},
\bauthor{\bsnm{{Arcones}}, \binits{A.}},
\bauthor{\bsnm{{Arnett}}, \binits{D.}},
\bauthor{\bsnm{{Auchettl}}, \binits{K.}},
\bauthor{\bsnm{{Avila}}, \binits{M.}},
\bauthor{\bsnm{{Badenes}}, \binits{C.}},
\bauthor{\bsnm{{Baron}}, \binits{E.}},
\bauthor{\bsnm{{Bauswein}}, \binits{A.}},
\bauthor{\bsnm{{Beacom}}, \binits{J.}},
\bauthor{\bsnm{{Blackmon}}, \binits{J.}},
\bauthor{\bsnm{{Blondin}}, \binits{S.}},
\bauthor{\bsnm{{Bloser}}, \binits{P.}},
\bauthor{\bsnm{{Boggs}}, \binits{S.}},
\bauthor{\bsnm{{Boss}}, \binits{A.}},
\bauthor{\bsnm{{Brandt}}, \binits{T.}},
\bauthor{\bsnm{{Bravo}}, \binits{E.}},
\bauthor{\bsnm{{Brown}}, \binits{E.}},
\bauthor{\bsnm{{Brown}}, \binits{P.}},
\bauthor{\bsnm{{Bruenn}}, \binits{S.}},
\bauthor{\bsnm{{Budtz-J{\o}rgensen}}, \binits{C.}},
\bauthor{\bsnm{{Burns}}, \binits{E.}},
\bauthor{\bsnm{{Calder}}, \binits{A.}},
\bauthor{\bsnm{{Caputo}}, \binits{R.}},
\bauthor{\bsnm{{Champagne}}, \binits{A.}},
\bauthor{\bsnm{{Chevalier}}, \binits{R.}},
\bauthor{\bsnm{{Chieffi}}, \binits{A.}},
\bauthor{\bsnm{{Chipps}}, \binits{K.}},
\bauthor{\bsnm{{Cinabro}}, \binits{D.}},
\bauthor{\bsnm{{Clarkson}}, \binits{O.}},
\bauthor{\bsnm{{Clayton}}, \binits{D.}},
\bauthor{\bsnm{{Coc}}, \binits{A.}},
\bauthor{\bsnm{{Connolly}}, \binits{D.}},
\bauthor{\bsnm{{Conroy}}, \binits{C.}},
\bauthor{\bsnm{{C{\^o}t{\'e}}}, \binits{B.}},
\bauthor{\bsnm{{Couch}}, \binits{S.}},
\bauthor{\bsnm{{Dauphas}}, \binits{N.}},
\bauthor{\bsnm{{deBoer}}, \binits{R.J.}},
\bauthor{\bsnm{{Deibel}}, \binits{C.}},
\bauthor{\bsnm{{Denisenkov}}, \binits{P.}},
\bauthor{\bsnm{{Desch}}, \binits{S.}},
\bauthor{\bsnm{{Dessart}}, \binits{L.}},
\bauthor{\bsnm{{Diehl}}, \binits{R.}},
\bauthor{\bsnm{{Doherty}}, \binits{C.}},
\bauthor{\bsnm{{Dom{\'\i}nguez}}, \binits{I.}},
\bauthor{\bsnm{{Dong}}, \binits{S.}},
\bauthor{\bsnm{{Dwarkadas}}, \binits{V.}},
\bauthor{\bsnm{{Fan}}, \binits{D.}},
\bauthor{\bsnm{{Fields}}, \binits{B.}},
\bauthor{\bsnm{{Fields}}, \binits{C.}},
\bauthor{\bsnm{{Filippenko}}, \binits{A.}},
\bauthor{\bsnm{{Fisher}}, \binits{R.}},
\bauthor{\bsnm{{Foucart}}, \binits{F.}},
\bauthor{\bsnm{{Fransson}}, \binits{C.}},
\bauthor{\bsnm{{Fr{\"o}hlich}}, \binits{C.}},
\bauthor{\bsnm{{Fuller}}, \binits{G.}},
\bauthor{\bsnm{{Gibson}}, \binits{B.}},
\bauthor{\bsnm{{Giryanskaya}}, \binits{V.}},
\bauthor{\bsnm{{G{\"o}rres}}, \binits{J.}},
\bauthor{\bsnm{{Goriely}}, \binits{S.}},
\bauthor{\bsnm{{Grebenev}}, \binits{S.}},
\bauthor{\bsnm{{Grefenstette}}, \binits{B.}},
\bauthor{\bsnm{{Grohs}}, \binits{E.}},
\bauthor{\bsnm{{Guillochon}}, \binits{J.}},
\bauthor{\bsnm{{Harpole}}, \binits{A.}},
\bauthor{\bsnm{{Harris}}, \binits{C.}},
\bauthor{\bsnm{{Harris}}, \binits{J.A.}},
\bauthor{\bsnm{{Harrison}}, \binits{F.}},
\bauthor{\bsnm{{Hartmann}}, \binits{D.}},
\bauthor{\bsnm{{Hashimoto}}, \binits{M.-a.}},
\bauthor{\bsnm{{Heger}}, \binits{A.}},
\bauthor{\bsnm{{Hernanz}}, \binits{M.}},
\bauthor{\bsnm{{Herwig}}, \binits{F.}},
\bauthor{\bsnm{{Hirschi}}, \binits{R.}},
\bauthor{\bsnm{{Hix}}, \binits{W.R.}},
\bauthor{\bsnm{{H{\"o}flich}}, \binits{P.}},
\bauthor{\bsnm{{Hoffman}}, \binits{R.}},
\bauthor{\bsnm{{Holcomb}}, \binits{C.}},
\bauthor{\bsnm{{Hsiao}}, \binits{E.}},
\bauthor{\bsnm{{Iliadis}}, \binits{C.}},
\bauthor{\bsnm{{Janiuk}}, \binits{A.}},
\bauthor{\bsnm{{Janka}}, \binits{T.}},
\bauthor{\bsnm{{Jerkstrand}}, \binits{A.}},
\bauthor{\bsnm{{Johns}}, \binits{L.}},
\bauthor{\bsnm{{Jones}}, \binits{S.}},
\bauthor{\bsnm{{Jos{\'e}}}, \binits{J.}},
\bauthor{\bsnm{{Kajino}}, \binits{T.}},
\bauthor{\bsnm{{Karakas}}, \binits{A.}},
\bauthor{\bsnm{{Karpov}}, \binits{P.}},
\bauthor{\bsnm{{Kasen}}, \binits{D.}},
\bauthor{\bsnm{{Kierans}}, \binits{C.}},
\bauthor{\bsnm{{Kippen}}, \binits{M.}},
\bauthor{\bsnm{{Korobkin}}, \binits{O.}},
\bauthor{\bsnm{{Kobayashi}}, \binits{C.}},
\bauthor{\bsnm{{Kozma}}, \binits{C.}},
\bauthor{\bsnm{{Krot}}, \binits{S.}},
\bauthor{\bsnm{{Kumar}}, \binits{P.}},
\bauthor{\bsnm{{Kuvvetli}}, \binits{I.}},
\bauthor{\bsnm{{Laird}}, \binits{A.}},
\bauthor{\bsnm{{Laming}}, \binits{J.M.}},
\bauthor{\bsnm{{Larsson}}, \binits{J.}},
\bauthor{\bsnm{{Lattanzio}}, \binits{J.}},
\bauthor{\bsnm{{Lattimer}}, \binits{J.}},
\bauthor{\bsnm{{Leising}}, \binits{M.}},
\bauthor{\bsnm{{Lennarz}}, \binits{A.}},
\bauthor{\bsnm{{Lentz}}, \binits{E.}},
\bauthor{\bsnm{{Limongi}}, \binits{M.}},
\bauthor{\bsnm{{Lippuner}}, \binits{J.}},
\bauthor{\bsnm{{Livne}}, \binits{E.}},
\bauthor{\bsnm{{Lloyd-Ronning}}, \binits{N.}},
\bauthor{\bsnm{{Longland}}, \binits{R.}},
\bauthor{\bsnm{{Lopez}}, \binits{L.A.}},
\bauthor{\bsnm{{Lugaro}}, \binits{M.}},
\bauthor{\bsnm{{Lutovinov}}, \binits{A.}},
\bauthor{\bsnm{{Madsen}}, \binits{K.}},
\bauthor{\bsnm{{Malone}}, \binits{C.}},
\bauthor{\bsnm{{Matteucci}}, \binits{F.}},
\bauthor{\bsnm{{McEnery}}, \binits{J.}},
\bauthor{\bsnm{{Meisel}}, \binits{Z.}},
\bauthor{\bsnm{{Messer}}, \binits{B.}},
\bauthor{\bsnm{{Metzger}}, \binits{B.}},
\bauthor{\bsnm{{Meyer}}, \binits{B.}},
\bauthor{\bsnm{{Meynet}}, \binits{G.}},
\bauthor{\bsnm{{Mezzacappa}}, \binits{A.}},
\bauthor{\bsnm{{Miller}}, \binits{J.}},
\bauthor{\bsnm{{Miller}}, \binits{R.}},
\bauthor{\bsnm{{Milne}}, \binits{P.}},
\bauthor{\bsnm{{Misch}}, \binits{W.}},
\bauthor{\bsnm{{Mitchell}}, \binits{L.}},
\bauthor{\bsnm{{M{\"o}sta}}, \binits{P.}},
\bauthor{\bsnm{{Motizuki}}, \binits{Y.}},
\bauthor{\bsnm{{M{\"u}ller}}, \binits{B.}},
\bauthor{\bsnm{{Mumpower}}, \binits{M.}},
\bauthor{\bsnm{{Murphy}}, \binits{J.}},
\bauthor{\bsnm{{Nagataki}}, \binits{S.}},
\bauthor{\bsnm{{Nakar}}, \binits{E.}},
\bauthor{\bsnm{{Nomoto}}, \binits{K.}},
\bauthor{\bsnm{{Nugent}}, \binits{P.}},
\bauthor{\bsnm{{Nunes}}, \binits{F.}},
\bauthor{\bsnm{{O'Shea}}, \binits{B.}},
\bauthor{\bsnm{{Oberlack}}, \binits{U.}},
\bauthor{\bsnm{{Pain}}, \binits{S.}},
\bauthor{\bsnm{{Parker}}, \binits{L.}},
\bauthor{\bsnm{{Perego}}, \binits{A.}},
\bauthor{\bsnm{{Pignatari}}, \binits{M.}},
\bauthor{\bsnm{{Pinedo}}, \binits{G.M.}},
\bauthor{\bsnm{{Plewa}}, \binits{T.}},
\bauthor{\bsnm{{Poznanski}}, \binits{D.}},
\bauthor{\bsnm{{Priedhorsky}}, \binits{W.}},
\bauthor{\bsnm{{Pritychenko}}, \binits{B.}},
\bauthor{\bsnm{{Radice}}, \binits{D.}},
\bauthor{\bsnm{{Ramirez-Ruiz}}, \binits{E.}},
\bauthor{\bsnm{{Rauscher}}, \binits{T.}},
\bauthor{\bsnm{{Reddy}}, \binits{S.}},
\bauthor{\bsnm{{Rehm}}, \binits{E.}},
\bauthor{\bsnm{{Reifarth}}, \binits{R.}},
\bauthor{\bsnm{{Richman}}, \binits{D.}},
\bauthor{\bsnm{{Ricker}}, \binits{P.}},
\bauthor{\bsnm{{Rijal}}, \binits{N.}},
\bauthor{\bsnm{{Roberts}}, \binits{L.}},
\bauthor{\bsnm{{R{\"o}pke}}, \binits{F.}},
\bauthor{\bsnm{{Rosswog}}, \binits{S.}},
\bauthor{\bsnm{{Ruiter}}, \binits{A.J.}},
\bauthor{\bsnm{{Ruiz}}, \binits{C.}},
\bauthor{\bsnm{{Savin}}, \binits{D.W.}},
\bauthor{\bsnm{{Schatz}}, \binits{H.}},
\bauthor{\bsnm{{Schneider}}, \binits{D.}},
\bauthor{\bsnm{{Schwab}}, \binits{J.}},
\bauthor{\bsnm{{Seitenzahl}}, \binits{I.}},
\bauthor{\bsnm{{Shen}}, \binits{K.}},
\bauthor{\bsnm{{Siegert}}, \binits{T.}},
\bauthor{\bsnm{{Sim}}, \binits{S.}},
\bauthor{\bsnm{{Smith}}, \binits{D.}},
\bauthor{\bsnm{{Smith}}, \binits{K.}},
\bauthor{\bsnm{{Smith}}, \binits{M.}},
\bauthor{\bsnm{{Sollerman}}, \binits{J.}},
\bauthor{\bsnm{{Sprouse}}, \binits{T.}},
\bauthor{\bsnm{{Spyrou}}, \binits{A.}},
\bauthor{\bsnm{{Starrfield}}, \binits{S.}},
\bauthor{\bsnm{{Steiner}}, \binits{A.}},
\bauthor{\bsnm{{Strong}}, \binits{A.W.}},
\bauthor{\bsnm{{Sukhbold}}, \binits{T.}},
\bauthor{\bsnm{{Suntzeff}}, \binits{N.}},
\bauthor{\bsnm{{Surman}}, \binits{R.}},
\bauthor{\bsnm{{Tanimori}}, \binits{T.}},
\bauthor{\bsnm{{The}}, \binits{L.-S.}},
\bauthor{\bsnm{{Thielemann}}, \binits{F.-K.}},
\bauthor{\bsnm{{Tolstov}}, \binits{A.}},
\bauthor{\bsnm{{Tominaga}}, \binits{N.}},
\bauthor{\bsnm{{Tomsick}}, \binits{J.}},
\bauthor{\bsnm{{Townsley}}, \binits{D.}},
\bauthor{\bsnm{{Tsintari}}, \binits{P.}},
\bauthor{\bsnm{{Tsygankov}}, \binits{S.}},
\bauthor{\bsnm{{Vartanyan}}, \binits{D.}},
\bauthor{\bsnm{{Venters}}, \binits{T.}},
\bauthor{\bsnm{{Vestrand}}, \binits{T.}},
\bauthor{\bsnm{{Vink}}, \binits{J.}},
\bauthor{\bsnm{{Waldman}}, \binits{R.}},
\bauthor{\bsnm{{Wang}}, \binits{L.}},
\bauthor{\bsnm{{Wang}}, \binits{X.}},
\bauthor{\bsnm{{Warren}}, \binits{M.}},
\bauthor{\bsnm{{West}}, \binits{C.}},
\bauthor{\bsnm{{Wheeler}}, \binits{J.C.}},
\bauthor{\bsnm{{Wiescher}}, \binits{M.}},
\bauthor{\bsnm{{Winkler}}, \binits{C.}},
\bauthor{\bsnm{{Winter}}, \binits{L.}},
\bauthor{\bsnm{{Wolf}}, \binits{B.}},
\bauthor{\bsnm{{Woolf}}, \binits{R.}},
\bauthor{\bsnm{{Woosley}}, \binits{S.}},
\bauthor{\bsnm{{Wu}}, \binits{J.}},
\bauthor{\bsnm{{Wrede}}, \binits{C.}},
\bauthor{\bsnm{{Yamada}}, \binits{S.}},
\bauthor{\bsnm{{Young}}, \binits{P.}},
\bauthor{\bsnm{{Zegers}}, \binits{R.}},
\bauthor{\bsnm{{Zingale}}, \binits{M.}},
\bauthor{\bsnm{{Portegies Zwart}}, \binits{S.}}:
\batitle{{Catching Element Formation In The Act ; The Case for a New MeV
  Gamma-Ray Mission: Radionuclide Astronomy in the 2020s}}.
\bjtitle{\baas}
\bvolume{51}(\bissue{3}),
\bfpage{2}
(\byear{2019})
{\href{https://arxiv.org/abs/1902.02915}{{arXiv:1902.02915}}}
{[astro-ph.HE]}.
\doiurl{10.48550/arXiv.1902.02915}
\end{barticle}
\endbibitem

\bibitem{Phillips1993}
\begin{barticle}
\bauthor{\bsnm{{Phillips}}, \binits{M.M.}}:
\batitle{{The Absolute Magnitudes of Type IA Supernovae}}.
\bjtitle{\apjl}
\bvolume{413},
\bfpage{105}
(\byear{1993}).
\doiurl{10.1086/186970}
\end{barticle}
\endbibitem

\bibitem{Phillips1999}
\begin{barticle}
\bauthor{\bsnm{{Phillips}}, \binits{M.M.}},
\bauthor{\bsnm{{Lira}}, \binits{P.}},
\bauthor{\bsnm{{Suntzeff}}, \binits{N.B.}},
\bauthor{\bsnm{{Schommer}}, \binits{R.A.}},
\bauthor{\bsnm{{Hamuy}}, \binits{M.}},
\bauthor{\bsnm{{Maza}}, \binits{J.}}:
\batitle{{The Reddening-Free Decline Rate Versus Luminosity Relationship for
  Type IA Supernovae}}.
\bjtitle{\aj}
\bvolume{118}(\bissue{4}),
\bfpage{1766}--\blpage{1776}
(\byear{1999})
{\href{https://arxiv.org/abs/astro-ph/9907052}{{arXiv:astro-ph/9907052}}}
{[astro-ph]}.
\doiurl{10.1086/301032}
\end{barticle}
\endbibitem

\bibitem{Wang2019}
\begin{barticle}
\bauthor{\bsnm{{Wang}}, \binits{X.}},
\bauthor{\bsnm{{Fields}}, \binits{B.D.}},
\bauthor{\bsnm{{Lien}}, \binits{A.Y.}}:
\batitle{{Using gamma ray monitoring to avoid missing the next Milky Way Type
  Ia supernova}}.
\bjtitle{\mnras}
\bvolume{486}(\bissue{2}),
\bfpage{2910}--\blpage{2918}
(\byear{2019})
{\href{https://arxiv.org/abs/1904.04310}{{arXiv:1904.04310}}}
{[astro-ph.HE]}.
\doiurl{10.1093/mnras/stz993}
\end{barticle}
\endbibitem

\bibitem{The1998}
\begin{barticle}
\bauthor{\bsnm{{The}}, \binits{L.-S.}},
\bauthor{\bsnm{{Clayton}}, \binits{D.D.}},
\bauthor{\bsnm{{Jin}}, \binits{L.}},
\bauthor{\bsnm{{Meyer}}, \binits{B.S.}}:
\batitle{{Nuclear Reactions Governing the Nucleosynthesis of $^{44}$Ti}}.
\bjtitle{\apj}
\bvolume{504}(\bissue{1}),
\bfpage{500}--\blpage{515}
(\byear{1998})
{\href{https://arxiv.org/abs/astro-ph/9806211}{{arXiv:astro-ph/9806211}}}
{[astro-ph]}.
\doiurl{10.1086/306057}
\end{barticle}
\endbibitem

\bibitem{Wongwathanarat2017}
\begin{barticle}
\bauthor{\bsnm{{Wongwathanarat}}, \binits{A.}},
\bauthor{\bsnm{{Janka}}, \binits{H.-T.}},
\bauthor{\bsnm{{M{\"u}ller}}, \binits{E.}},
\bauthor{\bsnm{{Pllumbi}}, \binits{E.}},
\bauthor{\bsnm{{Wanajo}}, \binits{S.}}:
\batitle{{Production and Distribution of $^{44}$Ti and $^{56}$Ni in a
  Three-dimensional Supernova Model Resembling Cassiopeia A}}.
\bjtitle{\apj}
\bvolume{842}(\bissue{1}),
\bfpage{13}
(\byear{2017})
{\href{https://arxiv.org/abs/1610.05643}{{arXiv:1610.05643}}}
{[astro-ph.HE]}.
\doiurl{10.3847/1538-4357/aa72de}
\end{barticle}
\endbibitem

\bibitem{Matz1988}
\begin{barticle}
\bauthor{\bsnm{{Matz}}, \binits{S.M.}},
\bauthor{\bsnm{{Share}}, \binits{G.H.}},
\bauthor{\bsnm{{Leising}}, \binits{M.D.}},
\bauthor{\bsnm{{Chupp}}, \binits{E.L.}},
\bauthor{\bsnm{{Vestrand}}, \binits{W.T.}},
\bauthor{\bsnm{{Purcell}}, \binits{W.R.}},
\bauthor{\bsnm{{Strickman}}, \binits{M.S.}},
\bauthor{\bsnm{{Reppin}}, \binits{C.}}:
\batitle{{Gamma-ray line emission from SN1987A}}.
\bjtitle{\nat}
\bvolume{331}(\bissue{6155}),
\bfpage{416}--\blpage{418}
(\byear{1988}).
\doiurl{10.1038/331416a0}
\end{barticle}
\endbibitem

\bibitem{Churazov2015}
\begin{barticle}
\bauthor{\bsnm{{Churazov}}, \binits{E.}},
\bauthor{\bsnm{{Sunyaev}}, \binits{R.}},
\bauthor{\bsnm{{Isern}}, \binits{J.}},
\bauthor{\bsnm{{Bikmaev}}, \binits{I.}},
\bauthor{\bsnm{{Bravo}}, \binits{E.}},
\bauthor{\bsnm{{Chugai}}, \binits{N.}},
\bauthor{\bsnm{{Grebenev}}, \binits{S.}},
\bauthor{\bsnm{{Jean}}, \binits{P.}},
\bauthor{\bsnm{{Kn{\"o}dlseder}}, \binits{J.}},
\bauthor{\bsnm{{Lebrun}}, \binits{F.}},
\bauthor{\bsnm{{Kuulkers}}, \binits{E.}}:
\batitle{{Gamma-rays from Type Ia Supernova SN2014J}}.
\bjtitle{\apj}
\bvolume{812}(\bissue{1}),
\bfpage{62}
(\byear{2015})
{\href{https://arxiv.org/abs/1502.00255}{{arXiv:1502.00255}}}
{[astro-ph.HE]}.
\doiurl{10.1088/0004-637X/812/1/62}
\end{barticle}
\endbibitem

\bibitem{Renaud2006}
\begin{barticle}
\bauthor{\bsnm{{Renaud}}, \binits{M.}},
\bauthor{\bsnm{{Vink}}, \binits{J.}},
\bauthor{\bsnm{{Decourchelle}}, \binits{A.}},
\bauthor{\bsnm{{Lebrun}}, \binits{F.}},
\bauthor{\bsnm{{den Hartog}}, \binits{P.R.}},
\bauthor{\bsnm{{Terrier}}, \binits{R.}},
\bauthor{\bsnm{{Couvreur}}, \binits{C.}},
\bauthor{\bsnm{{Kn{\"o}dlseder}}, \binits{J.}},
\bauthor{\bsnm{{Martin}}, \binits{P.}},
\bauthor{\bsnm{{Prantzos}}, \binits{N.}},
\bauthor{\bsnm{{Bykov}}, \binits{A.M.}},
\bauthor{\bsnm{{Bloemen}}, \binits{H.}}:
\batitle{{The Signature of $^{44}$Ti in Cassiopeia A Revealed by IBIS/ISGRI on
  INTEGRAL}}.
\bjtitle{\apjl}
\bvolume{647}(\bissue{1}),
\bfpage{41}--\blpage{44}
(\byear{2006})
{\href{https://arxiv.org/abs/astro-ph/0606736}{{arXiv:astro-ph/0606736}}}
{[astro-ph]}.
\doiurl{10.1086/507300}
\end{barticle}
\endbibitem

\bibitem{The1990}
\begin{barticle}
\bauthor{\bsnm{{The}}, \binits{L.-S.}},
\bauthor{\bsnm{{Burrows}}, \binits{A.}},
\bauthor{\bsnm{{Bussard}}, \binits{R.}}:
\batitle{{X-Ray and Gamma-Ray Fluxes from SN 1987A}}.
\bjtitle{\apj}
\bvolume{352},
\bfpage{731}
(\byear{1990}).
\doiurl{10.1086/168575}
\end{barticle}
\endbibitem

\bibitem{Tsygankov2016}
\begin{barticle}
\bauthor{\bsnm{{Tsygankov}}, \binits{S.S.}},
\bauthor{\bsnm{{Krivonos}}, \binits{R.A.}},
\bauthor{\bsnm{{Lutovinov}}, \binits{A.A.}},
\bauthor{\bsnm{{Revnivtsev}}, \binits{M.G.}},
\bauthor{\bsnm{{Churazov}}, \binits{E.M.}},
\bauthor{\bsnm{{Sunyaev}}, \binits{R.A.}},
\bauthor{\bsnm{{Grebenev}}, \binits{S.A.}}:
\batitle{{Galactic survey of $^{44}$Ti sources with the IBIS telescope onboard
  INTEGRAL}}.
\bjtitle{\mnras}
\bvolume{458}(\bissue{4}),
\bfpage{3411}--\blpage{3419}
(\byear{2016})
{\href{https://arxiv.org/abs/1603.01264}{{arXiv:1603.01264}}}
{[astro-ph.HE]}.
\doiurl{10.1093/mnras/stw549}
\end{barticle}
\endbibitem

\bibitem{Burbidge1957}
\begin{barticle}
\bauthor{\bsnm{{Burbidge}}, \binits{E.M.}},
\bauthor{\bsnm{{Burbidge}}, \binits{G.R.}},
\bauthor{\bsnm{{Fowler}}, \binits{W.A.}},
\bauthor{\bsnm{{Hoyle}}, \binits{F.}}:
\batitle{{Synthesis of the Elements in Stars}}.
\bjtitle{Reviews of Modern Physics}
\bvolume{29}(\bissue{4}),
\bfpage{547}--\blpage{650}
(\byear{1957}).
\doiurl{10.1103/RevModPhys.29.547}
\end{barticle}
\endbibitem

\bibitem{Wang2020}
\begin{barticle}
\bauthor{\bsnm{{Wang}}, \binits{W.}},
\bauthor{\bsnm{{Siegert}}, \binits{T.}},
\bauthor{\bsnm{{Dai}}, \binits{Z.G.}},
\bauthor{\bsnm{{Diehl}}, \binits{R.}},
\bauthor{\bsnm{{Greiner}}, \binits{J.}},
\bauthor{\bsnm{{Heger}}, \binits{A.}},
\bauthor{\bsnm{{Krause}}, \binits{M.}},
\bauthor{\bsnm{{Lang}}, \binits{M.}},
\bauthor{\bsnm{{Pleintinger}}, \binits{M.M.M.}},
\bauthor{\bsnm{{Zhang}}, \binits{X.L.}}:
\batitle{{Gamma-Ray Emission of $^{60}$Fe and $^{26}$Al Radioactivity in Our
  Galaxy}}.
\bjtitle{\apj}
\bvolume{889}(\bissue{2}),
\bfpage{169}
(\byear{2020})
{\href{https://arxiv.org/abs/1912.07874}{{arXiv:1912.07874}}}
{[astro-ph.HE]}.
\doiurl{10.3847/1538-4357/ab6336}
\end{barticle}
\endbibitem

\bibitem{Diehl2006}
\begin{barticle}
\bauthor{\bsnm{{Diehl}}, \binits{R.}},
\bauthor{\bsnm{{Halloin}}, \binits{H.}},
\bauthor{\bsnm{{Kretschmer}}, \binits{K.}},
\bauthor{\bsnm{{Strong}}, \binits{A.W.}},
\bauthor{\bsnm{{Wang}}, \binits{W.}},
\bauthor{\bsnm{{Jean}}, \binits{P.}},
\bauthor{\bsnm{{Lichti}}, \binits{G.G.}},
\bauthor{\bsnm{{Kn{\"o}dlseder}}, \binits{J.}},
\bauthor{\bsnm{{Roques}}, \binits{J.-P.}},
\bauthor{\bsnm{{Schanne}}, \binits{S.}},
\bauthor{\bsnm{{Sch{\"o}nfelder}}, \binits{V.}},
\bauthor{\bsnm{{von Kienlin}}, \binits{A.}},
\bauthor{\bsnm{{Weidenspointner}}, \binits{G.}},
\bauthor{\bsnm{{Winkler}}, \binits{C.}},
\bauthor{\bsnm{{Wunderer}}, \binits{C.}}:
\batitle{{$^{26}$Al in the inner Galaxy. Large-scale spectral characteristics
  derived with SPI/INTEGRAL}}.
\bjtitle{\aap}
\bvolume{449}(\bissue{3}),
\bfpage{1025}--\blpage{1031}
(\byear{2006})
{\href{https://arxiv.org/abs/astro-ph/0512334}{{arXiv:astro-ph/0512334}}}
{[astro-ph]}.
\doiurl{10.1051/0004-6361:20054301}
\end{barticle}
\endbibitem

\bibitem{Wang2007}
\begin{barticle}
\bauthor{\bsnm{{Wang}}, \binits{W.}},
\bauthor{\bsnm{{Harris}}, \binits{M.J.}},
\bauthor{\bsnm{{Diehl}}, \binits{R.}},
\bauthor{\bsnm{{Halloin}}, \binits{H.}},
\bauthor{\bsnm{{Cordier}}, \binits{B.}},
\bauthor{\bsnm{{Strong}}, \binits{A.W.}},
\bauthor{\bsnm{{Kretschmer}}, \binits{K.}},
\bauthor{\bsnm{{Kn{\"o}dlseder}}, \binits{J.}},
\bauthor{\bsnm{{Jean}}, \binits{P.}},
\bauthor{\bsnm{{Lichti}}, \binits{G.G.}},
\bauthor{\bsnm{{Roques}}, \binits{J.P.}},
\bauthor{\bsnm{{Schanne}}, \binits{S.}},
\bauthor{\bsnm{{von Kienlin}}, \binits{A.}},
\bauthor{\bsnm{{Weidenspointner}}, \binits{G.}},
\bauthor{\bsnm{{Wunderer}}, \binits{C.}}:
\batitle{{SPI observations of the diffuse $^{60}$Fe emission in the Galaxy}}.
\bjtitle{\aap}
\bvolume{469}(\bissue{3}),
\bfpage{1005}--\blpage{1012}
(\byear{2007})
{\href{https://arxiv.org/abs/0704.3895}{{arXiv:0704.3895}}}
{[astro-ph]}.
\doiurl{10.1051/0004-6361:20066982}
\end{barticle}
\endbibitem

\bibitem{Wu2019}
\begin{barticle}
\bauthor{\bsnm{{Wu}}, \binits{M.-R.}},
\bauthor{\bsnm{{Banerjee}}, \binits{P.}},
\bauthor{\bsnm{{Metzger}}, \binits{B.D.}},
\bauthor{\bsnm{{Mart{\'\i}nez-Pinedo}}, \binits{G.}},
\bauthor{\bsnm{{Aramaki}}, \binits{T.}},
\bauthor{\bsnm{{Burns}}, \binits{E.}},
\bauthor{\bsnm{{Hailey}}, \binits{C.J.}},
\bauthor{\bsnm{{Barnes}}, \binits{J.}},
\bauthor{\bsnm{{Karagiorgi}}, \binits{G.}}:
\batitle{{Finding the Remnants of the Milky Way's Last Neutron Star Mergers}}.
\bjtitle{\apj}
\bvolume{880}(\bissue{1}),
\bfpage{23}
(\byear{2019})
{\href{https://arxiv.org/abs/1905.03793}{{arXiv:1905.03793}}}
{[astro-ph.HE]}.
\doiurl{10.3847/1538-4357/ab2593}
\end{barticle}
\endbibitem

\bibitem{Knoedlseder2005}
\begin{barticle}
\bauthor{\bsnm{{Kn{\"o}dlseder}}, \binits{J.}},
\bauthor{\bsnm{{Jean}}, \binits{P.}},
\bauthor{\bsnm{{Lonjou}}, \binits{V.}},
\bauthor{\bsnm{{Weidenspointner}}, \binits{G.}},
\bauthor{\bsnm{{Guessoum}}, \binits{N.}},
\bauthor{\bsnm{{Gillard}}, \binits{W.}},
\bauthor{\bsnm{{Skinner}}, \binits{G.}},
\bauthor{\bsnm{{von Ballmoos}}, \binits{P.}},
\bauthor{\bsnm{{Vedrenne}}, \binits{G.}},
\bauthor{\bsnm{{Roques}}, \binits{J.-P.}},
\bauthor{\bsnm{{Schanne}}, \binits{S.}},
\bauthor{\bsnm{{Teegarden}}, \binits{B.}},
\bauthor{\bsnm{{Sch{\"o}nfelder}}, \binits{V.}},
\bauthor{\bsnm{{Winkler}}, \binits{C.}}:
\batitle{{The all-sky distribution of 511 keV electron-positron annihilation
  emission}}.
\bjtitle{\aap}
\bvolume{441}(\bissue{2}),
\bfpage{513}--\blpage{532}
(\byear{2005})
{\href{https://arxiv.org/abs/astro-ph/0506026}{{arXiv:astro-ph/0506026}}}
{[astro-ph]}.
\doiurl{10.1051/0004-6361:20042063}
\end{barticle}
\endbibitem

\bibitem{Siegert2016a}
\begin{barticle}
\bauthor{\bsnm{{Siegert}}, \binits{T.}},
\bauthor{\bsnm{{Diehl}}, \binits{R.}},
\bauthor{\bsnm{{Khachatryan}}, \binits{G.}},
\bauthor{\bsnm{{Krause}}, \binits{M.G.H.}},
\bauthor{\bsnm{{Guglielmetti}}, \binits{F.}},
\bauthor{\bsnm{{Greiner}}, \binits{J.}},
\bauthor{\bsnm{{Strong}}, \binits{A.W.}},
\bauthor{\bsnm{{Zhang}}, \binits{X.}}:
\batitle{{Gamma-ray spectroscopy of positron annihilation in the Milky Way}}.
\bjtitle{\aap}
\bvolume{586},
\bfpage{84}
(\byear{2016})
{\href{https://arxiv.org/abs/1512.00325}{{arXiv:1512.00325}}}
{[astro-ph.HE]}.
\doiurl{10.1051/0004-6361/201527510}
\end{barticle}
\endbibitem

\bibitem{Fuller2019}
\begin{barticle}
\bauthor{\bsnm{{Fuller}}, \binits{G.M.}},
\bauthor{\bsnm{{Kusenko}}, \binits{A.}},
\bauthor{\bsnm{{Radice}}, \binits{D.}},
\bauthor{\bsnm{{Takhistov}}, \binits{V.}}:
\batitle{{Positrons and 511 keV Radiation as Tracers of Recent Binary Neutron
  Star Mergers}}.
\bjtitle{\prl}
\bvolume{122}(\bissue{12}),
\bfpage{121101}
(\byear{2019})
{\href{https://arxiv.org/abs/1811.00133}{{arXiv:1811.00133}}}
{[astro-ph.HE]}.
\doiurl{10.1103/PhysRevLett.122.121101}
\end{barticle}
\endbibitem

\bibitem{Prantzos2011}
\begin{barticle}
\bauthor{\bsnm{{Prantzos}}, \binits{N.}},
\bauthor{\bsnm{{Boehm}}, \binits{C.}},
\bauthor{\bsnm{{Bykov}}, \binits{A.M.}},
\bauthor{\bsnm{{Diehl}}, \binits{R.}},
\bauthor{\bsnm{{Ferri{\`e}re}}, \binits{K.}},
\bauthor{\bsnm{{Guessoum}}, \binits{N.}},
\bauthor{\bsnm{{Jean}}, \binits{P.}},
\bauthor{\bsnm{{Knoedlseder}}, \binits{J.}},
\bauthor{\bsnm{{Marcowith}}, \binits{A.}},
\bauthor{\bsnm{{Moskalenko}}, \binits{I.V.}},
\bauthor{\bsnm{{Strong}}, \binits{A.}},
\bauthor{\bsnm{{Weidenspointner}}, \binits{G.}}:
\batitle{{The 511 keV emission from positron annihilation in the Galaxy}}.
\bjtitle{Reviews of Modern Physics}
\bvolume{83}(\bissue{3}),
\bfpage{1001}--\blpage{1056}
(\byear{2011})
{\href{https://arxiv.org/abs/1009.4620}{{arXiv:1009.4620}}}
{[astro-ph.HE]}.
\doiurl{10.1103/RevModPhys.83.1001}
\end{barticle}
\endbibitem

\bibitem{Totani2006}
\begin{barticle}
\bauthor{\bsnm{{Totani}}, \binits{T.}}:
\batitle{{A RIAF Interpretation for the Past Higher Activity of the Galactic
  Center Black Hole and the 511 keV Annihilation Emission}}.
\bjtitle{\pasj}
\bvolume{58},
\bfpage{965}--\blpage{977}
(\byear{2006})
{\href{https://arxiv.org/abs/astro-ph/0607414}{{arXiv:astro-ph/0607414}}}
{[astro-ph]}.
\doiurl{10.1093/pasj/58.6.965}
\end{barticle}
\endbibitem

\bibitem{Alexis2014}
\begin{barticle}
\bauthor{\bsnm{{Alexis}}, \binits{A.}},
\bauthor{\bsnm{{Jean}}, \binits{P.}},
\bauthor{\bsnm{{Martin}}, \binits{P.}},
\bauthor{\bsnm{{Ferri{\`e}re}}, \binits{K.}}:
\batitle{{Monte Carlo modelling of the propagation and annihilation of
  nucleosynthesis positrons in the Galaxy}}.
\bjtitle{\aap}
\bvolume{564},
\bfpage{108}
(\byear{2014})
{\href{https://arxiv.org/abs/1402.6110}{{arXiv:1402.6110}}}
{[astro-ph.HE]}.
\doiurl{10.1051/0004-6361/201322393}
\end{barticle}
\endbibitem

\bibitem{Bartels2018}
\begin{barticle}
\bauthor{\bsnm{{Bartels}}, \binits{R.}},
\bauthor{\bsnm{{Calore}}, \binits{F.}},
\bauthor{\bsnm{{Storm}}, \binits{E.}},
\bauthor{\bsnm{{Weniger}}, \binits{C.}}:
\batitle{{Galactic binaries can explain the Fermi Galactic centre excess and
  511 keV emission}}.
\bjtitle{\mnras}
\bvolume{480}(\bissue{3}),
\bfpage{3826}--\blpage{3841}
(\byear{2018})
{\href{https://arxiv.org/abs/1803.04370}{{arXiv:1803.04370}}}
{[astro-ph.HE]}.
\doiurl{10.1093/mnras/sty2135}
\end{barticle}
\endbibitem

\bibitem{Pospelov2007}
\begin{barticle}
\bauthor{\bsnm{{Pospelov}}, \binits{M.}},
\bauthor{\bsnm{{Ritz}}, \binits{A.}}:
\batitle{{The galactic 511 keV line from electroweak scale WIMPs}}.
\bjtitle{Physics Letters B}
\bvolume{651}(\bissue{2-3}),
\bfpage{208}--\blpage{215}
(\byear{2007})
{\href{https://arxiv.org/abs/hep-ph/0703128}{{arXiv:hep-ph/0703128}}}
{[hep-ph]}.
\doiurl{10.1016/j.physletb.2007.06.027}
\end{barticle}
\endbibitem

\bibitem{Hooper2008}
\begin{barticle}
\bauthor{\bsnm{{Hooper}}, \binits{D.}},
\bauthor{\bsnm{{Zurek}}, \binits{K.M.}}:
\batitle{{Natural supersymmetric model with MeV dark matter}}.
\bjtitle{\prd}
\bvolume{77}(\bissue{8}),
\bfpage{087302}
(\byear{2008})
{\href{https://arxiv.org/abs/0801.3686}{{arXiv:0801.3686}}}
{[hep-ph]}.
\doiurl{10.1103/PhysRevD.77.087302}
\end{barticle}
\endbibitem

\bibitem{Khalil2008}
\begin{barticle}
\bauthor{\bsnm{{Khalil}}, \binits{S.}},
\bauthor{\bsnm{{Seto}}, \binits{O.}}:
\batitle{{Sterile neutrino dark matter in B-L extension of the standard model
  and galactic 511 keV line}}.
\bjtitle{\jcap}
\bvolume{2008}(\bissue{10}),
\bfpage{024}
(\byear{2008})
{\href{https://arxiv.org/abs/0804.0336}{{arXiv:0804.0336}}}
{[hep-ph]}.
\doiurl{10.1088/1475-7516/2008/10/024}
\end{barticle}
\endbibitem

\bibitem{Keith2021}
\begin{barticle}
\bauthor{\bsnm{{Keith}}, \binits{C.}},
\bauthor{\bsnm{{Hooper}}, \binits{D.}}:
\batitle{{511 keV excess and primordial black holes}}.
\bjtitle{\prd}
\bvolume{104}(\bissue{6}),
\bfpage{063033}
(\byear{2021})
{\href{https://arxiv.org/abs/2103.08611}{{arXiv:2103.08611}}}
{[astro-ph.CO]}.
\doiurl{10.1103/PhysRevD.104.063033}
\end{barticle}
\endbibitem

\bibitem{Cai2021}
\begin{barticle}
\bauthor{\bsnm{{Cai}}, \binits{R.-G.}},
\bauthor{\bsnm{{Ding}}, \binits{Y.-C.}},
\bauthor{\bsnm{{Yang}}, \binits{X.-Y.}},
\bauthor{\bsnm{{Zhou}}, \binits{Y.-F.}}:
\batitle{{Constraints on a mixed model of dark matter particles and primordial
  black holes from the galactic 511 keV line}}.
\bjtitle{\jcap}
\bvolume{2021}(\bissue{3}),
\bfpage{057}
(\byear{2021})
{\href{https://arxiv.org/abs/2007.11804}{{arXiv:2007.11804}}}
{[astro-ph.CO]}.
\doiurl{10.1088/1475-7516/2021/03/057}
\end{barticle}
\endbibitem

\bibitem{Siegert2016b}
\begin{barticle}
\bauthor{\bsnm{{Siegert}}, \binits{T.}},
\bauthor{\bsnm{{Diehl}}, \binits{R.}},
\bauthor{\bsnm{{Vincent}}, \binits{A.C.}},
\bauthor{\bsnm{{Guglielmetti}}, \binits{F.}},
\bauthor{\bsnm{{Krause}}, \binits{M.G.H.}},
\bauthor{\bsnm{{Boehm}}, \binits{C.}}:
\batitle{{Search for 511 keV emission in satellite galaxies of the Milky Way
  with INTEGRAL/SPI}}.
\bjtitle{\aap}
\bvolume{595},
\bfpage{25}
(\byear{2016})
{\href{https://arxiv.org/abs/1608.00393}{{arXiv:1608.00393}}}
{[astro-ph.HE]}.
\doiurl{10.1051/0004-6361/201629136}
\end{barticle}
\endbibitem

\bibitem{Margon1979}
\begin{barticle}
\bauthor{\bsnm{{Margon}}, \binits{B.}},
\bauthor{\bsnm{{Ford}}, \binits{H.C.}},
\bauthor{\bsnm{{Katz}}, \binits{J.I.}},
\bauthor{\bsnm{{Kwitter}}, \binits{K.B.}},
\bauthor{\bsnm{{Ulrich}}, \binits{R.K.}},
\bauthor{\bsnm{{Stone}}, \binits{R.P.S.}},
\bauthor{\bsnm{{Klemola}}, \binits{A.}}:
\batitle{{The bizarre spectrum of SS 433.}}
\bjtitle{\apjl}
\bvolume{230},
\bfpage{41}--\blpage{45}
(\byear{1979}).
\doiurl{10.1086/182958}
\end{barticle}
\endbibitem

\bibitem{Liu2015}
\begin{barticle}
\bauthor{\bsnm{{Liu}}, \binits{J.-F.}},
\bauthor{\bsnm{{Bai}}, \binits{Y.}},
\bauthor{\bsnm{{Wang}}, \binits{S.}},
\bauthor{\bsnm{{Justham}}, \binits{S.}},
\bauthor{\bsnm{{Lu}}, \binits{Y.-J.}},
\bauthor{\bsnm{{Gu}}, \binits{W.-M.}},
\bauthor{\bsnm{{Liu}}, \binits{Q.-Z.}},
\bauthor{\bsnm{{di Stefano}}, \binits{R.}},
\bauthor{\bsnm{{Guo}}, \binits{J.-C.}},
\bauthor{\bsnm{{Cabrera-Lavers}}, \binits{A.}},
\bauthor{\bsnm{{{\'A}lvarez}}, \binits{P.}},
\bauthor{\bsnm{{Cao}}, \binits{Y.}},
\bauthor{\bsnm{{Kulkarni}}, \binits{S.}}:
\batitle{{Relativistic baryonic jets from an ultraluminous supersoft X-ray
  source}}.
\bjtitle{\nat}
\bvolume{528}(\bissue{7580}),
\bfpage{108}--\blpage{110}
(\byear{2015})
{\href{https://arxiv.org/abs/1511.09200}{{arXiv:1511.09200}}}
{[astro-ph.HE]}.
\doiurl{10.1038/nature15751}
\end{barticle}
\endbibitem

\bibitem{Begelman1984}
\begin{barticle}
\bauthor{\bsnm{{Begelman}}, \binits{M.C.}},
\bauthor{\bsnm{{Blandford}}, \binits{R.D.}},
\bauthor{\bsnm{{Rees}}, \binits{M.J.}}:
\batitle{{Theory of extragalactic radio sources}}.
\bjtitle{Reviews of Modern Physics}
\bvolume{56}(\bissue{2}),
\bfpage{255}--\blpage{351}
(\byear{1984}).
\doiurl{10.1103/RevModPhys.56.255}
\end{barticle}
\endbibitem

\bibitem{Reynolds1996}
\begin{barticle}
\bauthor{\bsnm{{Reynolds}}, \binits{C.S.}},
\bauthor{\bsnm{{Fabian}}, \binits{A.C.}},
\bauthor{\bsnm{{Celotti}}, \binits{A.}},
\bauthor{\bsnm{{Rees}}, \binits{M.J.}}:
\batitle{{The matter content of the jet in M87: evidence for an
  electron-positron jet}}.
\bjtitle{\mnras}
\bvolume{283}(\bissue{3}),
\bfpage{873}--\blpage{880}
(\byear{1996})
{\href{https://arxiv.org/abs/astro-ph/9603140}{{arXiv:astro-ph/9603140}}}
{[astro-ph]}.
\doiurl{10.1093/mnras/283.3.873}
\end{barticle}
\endbibitem

\bibitem{Marscher2007}
\begin{barticle}
\bauthor{\bsnm{{Marscher}}, \binits{A.P.}},
\bauthor{\bsnm{{Jorstad}}, \binits{S.G.}},
\bauthor{\bsnm{{G{\'o}mez}}, \binits{J.L.}},
\bauthor{\bsnm{{McHardy}}, \binits{I.M.}},
\bauthor{\bsnm{{Krichbaum}}, \binits{T.P.}},
\bauthor{\bsnm{{Agudo}}, \binits{I.}}:
\batitle{{Search for Electron-Positron Annihilation Radiation from the Jet in
  3C 120}}.
\bjtitle{\apj}
\bvolume{665}(\bissue{1}),
\bfpage{232}--\blpage{236}
(\byear{2007}).
\doiurl{10.1086/519481}
\end{barticle}
\endbibitem

\bibitem{Zhang2010}
\begin{barticle}
\bauthor{\bsnm{{Zhang}}, \binits{S.}},
\bauthor{\bsnm{{Collmar}}, \binits{W.}},
\bauthor{\bsnm{{Torres}}, \binits{D.F.}},
\bauthor{\bsnm{{Wang}}, \binits{J.-M.}},
\bauthor{\bsnm{{Lang}}, \binits{M.}},
\bauthor{\bsnm{{Zhang}}, \binits{S.-N.}}:
\batitle{{INTEGRAL and Swift/XRT observations of the source PKS 0208-512}}.
\bjtitle{\aap}
\bvolume{514},
\bfpage{69}
(\byear{2010})
{\href{https://arxiv.org/abs/1002.4030}{{arXiv:1002.4030}}}
{[astro-ph.CO]}.
\doiurl{10.1051/0004-6361/200913655}
\end{barticle}
\endbibitem

\bibitem{Siegert2016}
\begin{barticle}
\bauthor{\bsnm{{Siegert}}, \binits{T.}},
\bauthor{\bsnm{{Diehl}}, \binits{R.}},
\bauthor{\bsnm{{Greiner}}, \binits{J.}},
\bauthor{\bsnm{{Krause}}, \binits{M.G.H.}},
\bauthor{\bsnm{{Beloborodov}}, \binits{A.M.}},
\bauthor{\bsnm{{Bel}}, \binits{M.C.}},
\bauthor{\bsnm{{Guglielmetti}}, \binits{F.}},
\bauthor{\bsnm{{Rodriguez}}, \binits{J.}},
\bauthor{\bsnm{{Strong}}, \binits{A.W.}},
\bauthor{\bsnm{{Zhang}}, \binits{X.}}:
\batitle{{Positron annihilation signatures associated with the outburst of the
  microquasar V404 Cygni}}.
\bjtitle{\nat}
\bvolume{531}(\bissue{7594}),
\bfpage{341}--\blpage{343}
(\byear{2016})
{\href{https://arxiv.org/abs/1603.01169}{{arXiv:1603.01169}}}
{[astro-ph.HE]}.
\doiurl{10.1038/nature16978}
\end{barticle}
\endbibitem

\bibitem{Roques2016}
\begin{botherref}
\oauthor{\bsnm{{Roques}}, \binits{J.P.}},
\oauthor{\bsnm{{Jourdain}}, \binits{E.}}:
{High Energy Emission of V404 Cygni during 2015 outburst with INTEGRAL/SPI:
  Spectral analysis issues and solutions}.
arXiv e-prints,
1601--05289
(2016)
{\href{https://arxiv.org/abs/1601.05289}{{arXiv:1601.05289}}}
{[astro-ph.HE]}.
\doiurl{10.48550/arXiv.1601.05289}
\end{botherref}
\endbibitem

\bibitem{EdvigeRavasio2023}
\begin{botherref}
\oauthor{\bsnm{{Edvige Ravasio}}, \binits{M.}},
\oauthor{\bsnm{{Sharan Salafia}}, \binits{O.}},
\oauthor{\bsnm{{Oganesyan}}, \binits{G.}},
\oauthor{\bsnm{{Mei}}, \binits{A.}},
\oauthor{\bsnm{{Ghirlanda}}, \binits{G.}},
\oauthor{\bsnm{{Ascenzi}}, \binits{S.}},
\oauthor{\bsnm{{Banerjee}}, \binits{B.}},
\oauthor{\bsnm{{Macera}}, \binits{S.}},
\oauthor{\bsnm{{Branchesi}}, \binits{M.}},
\oauthor{\bsnm{{Jonker}}, \binits{P.G.}},
\oauthor{\bsnm{{Levan}}, \binits{A.J.}},
\oauthor{\bsnm{{Malesani}}, \binits{D.B.}},
\oauthor{\bsnm{{Mulrey}}, \binits{K.B.}},
\oauthor{\bsnm{{Giuliani}}, \binits{A.}},
\oauthor{\bsnm{{Celotti}}, \binits{A.}},
\oauthor{\bsnm{{Ghisellini}}, \binits{G.}}:
{A bright megaelectronvolt emission line in $\gamma$-ray burst GRB 221009A}.
arXiv e-prints,
2303--16223
(2023)
{\href{https://arxiv.org/abs/2303.16223}{{arXiv:2303.16223}}}
{[astro-ph.HE]}.
\doiurl{10.48550/arXiv.2303.16223}
\end{botherref}
\endbibitem

\bibitem{Reina1974}
\begin{barticle}
\bauthor{\bsnm{{Reina}}, \binits{C.}},
\bauthor{\bsnm{{Treves}}, \binits{A.}},
\bauthor{\bsnm{{Tarenghi}}, \binits{M.}}:
\batitle{{Gamma-ray Lines from Accreting Neutron Stars}}.
\bjtitle{\aap}
\bvolume{32},
\bfpage{317}
(\byear{1974})
\end{barticle}
\endbibitem

\bibitem{Brecher1980}
\begin{barticle}
\bauthor{\bsnm{{Brecher}}, \binits{K.}},
\bauthor{\bsnm{{Burrows}}, \binits{A.}}:
\batitle{{Gamma-ray lines from accreting neutron stars}}.
\bjtitle{\apj}
\bvolume{240},
\bfpage{642}--\blpage{647}
(\byear{1980}).
\doiurl{10.1086/158270}
\end{barticle}
\endbibitem

\bibitem{Agaronian1984}
\begin{barticle}
\bauthor{\bsnm{{Agaronian}}, \binits{F.A.}},
\bauthor{\bsnm{{Sunyaev}}, \binits{R.A.}}:
\batitle{{Gamma-ray line emission, nuclear destruction and neutron production
  in hot astrophysical plasmas.The deuterium boiler as a gamma-ray source.}}
\bjtitle{\mnras}
\bvolume{210},
\bfpage{257}--\blpage{277}
(\byear{1984}).
\doiurl{10.1093/mnras/210.2.257}
\end{barticle}
\endbibitem

\bibitem{Bildsten1993}
\begin{barticle}
\bauthor{\bsnm{{Bildsten}}, \binits{L.}},
\bauthor{\bsnm{{Salpeter}}, \binits{E.E.}},
\bauthor{\bsnm{{Wasserman}}, \binits{I.}}:
\batitle{{Helium Destruction and Gamma-Ray Line Emission in Accreting Neutron
  Stars}}.
\bjtitle{\apj}
\bvolume{408},
\bfpage{615}
(\byear{1993}).
\doiurl{10.1086/172621}
\end{barticle}
\endbibitem

\bibitem{Jean2001}
\begin{barticle}
\bauthor{\bsnm{{Jean}}, \binits{P.}},
\bauthor{\bsnm{{Guessoum}}, \binits{N.}}:
\batitle{{Neutron-capture and 2.22 MeV emission in the atmosphere of the
  secondary of an X-ray binary}}.
\bjtitle{\aap}
\bvolume{378},
\bfpage{509}--\blpage{521}
(\byear{2001})
{\href{https://arxiv.org/abs/astro-ph/0109185}{{arXiv:astro-ph/0109185}}}
{[astro-ph]}.
\doiurl{10.1051/0004-6361:20011201}
\end{barticle}
\endbibitem

\bibitem{Guessoum2002}
\begin{barticle}
\bauthor{\bsnm{{Guessoum}}, \binits{N.}},
\bauthor{\bsnm{{Jean}}, \binits{P.}}:
\batitle{{Detectability and characteristics of the 2.223 MeV line emission from
  nearby X-ray binaries}}.
\bjtitle{\aap}
\bvolume{396},
\bfpage{157}--\blpage{169}
(\byear{2002}).
\doiurl{10.1051/0004-6361:20021376}
\end{barticle}
\endbibitem

\bibitem{Boggs2006}
\begin{barticle}
\bauthor{\bsnm{{Boggs}}, \binits{S.E.}},
\bauthor{\bsnm{{Smith}}, \binits{D.M.}}:
\batitle{{Search for Neutron-Capture Gamma-Ray Lines from A0535+26 in
  Outburst}}.
\bjtitle{\apjl}
\bvolume{637}(\bissue{2}),
\bfpage{121}--\blpage{124}
(\byear{2006}).
\doiurl{10.1086/500690}
\end{barticle}
\endbibitem

\bibitem{Caliskan2009}
\begin{barticle}
\bauthor{\bsnm{{{\c{C}}ali{\c{s}}kan}}, \binits{{\c{S}}.}},
\bauthor{\bsnm{{Kalemci}}, \binits{E.}},
\bauthor{\bsnm{{Baring}}, \binits{M.G.}},
\bauthor{\bsnm{{Boggs}}, \binits{S.E.}},
\bauthor{\bsnm{{Kretschmar}}, \binits{P.}}:
\batitle{{Search for a Redshifted 2.2 MeV Neutron Capture Line from A0535+262
  in Outburst}}.
\bjtitle{\apj}
\bvolume{694}(\bissue{1}),
\bfpage{593}--\blpage{598}
(\byear{2009})
{\href{https://arxiv.org/abs/0812.2742}{{arXiv:0812.2742}}}
{[astro-ph]}.
\doiurl{10.1088/0004-637X/694/1/593}
\end{barticle}
\endbibitem

\bibitem{Jacobsen2008}
\begin{barticle}
\bauthor{\bsnm{{Jacobsen}}, \binits{B.}},
\bauthor{\bsnm{{Yin}}, \binits{Q.-z.}},
\bauthor{\bsnm{{Moynier}}, \binits{F.}},
\bauthor{\bsnm{{Amelin}}, \binits{Y.}},
\bauthor{\bsnm{{Krot}}, \binits{A.N.}},
\bauthor{\bsnm{{Nagashima}}, \binits{K.}},
\bauthor{\bsnm{{Hutcheon}}, \binits{I.D.}},
\bauthor{\bsnm{{Palme}}, \binits{H.}}:
\batitle{{$^{26}$Al- $^{26}$Mg and $^{207}$Pb- $^{206}$Pb systematics of
  Allende CAIs: Canonical solar initial $^{26}$Al/ $^{27}$Al ratio
  reinstated}}.
\bjtitle{Earth and Planetary Science Letters}
\bvolume{272}(\bissue{1-2}),
\bfpage{353}--\blpage{364}
(\byear{2008}).
\doiurl{10.1016/j.epsl.2008.05.003}
\end{barticle}
\endbibitem

\bibitem{Turner2009}
\begin{barticle}
\bauthor{\bsnm{{Turner}}, \binits{N.J.}},
\bauthor{\bsnm{{Drake}}, \binits{J.F.}}:
\batitle{{Energetic Protons, Radionuclides, and Magnetic Activity in
  Protostellar Disks}}.
\bjtitle{\apj}
\bvolume{703}(\bissue{2}),
\bfpage{2152}--\blpage{2159}
(\byear{2009})
{\href{https://arxiv.org/abs/0908.3874}{{arXiv:0908.3874}}}
{[astro-ph.SR]}.
\doiurl{10.1088/0004-637X/703/2/2152}
\end{barticle}
\endbibitem

\bibitem{Lee1976}
\begin{bchapter}
\bauthor{\bsnm{{Lee}}, \binits{T.}},
\bauthor{\bsnm{{Papanastassiou}}, \binits{D.A.}},
\bauthor{\bsnm{{Wasserburg}}, \binits{G.J.}}:
\bctitle{{The Presence of $^{26}$Al in the Early Solar Nebula}}.
In: \bbtitle{Bulletin of the American Astronomical Society},
vol. \bseriesno{8},
p. \bfpage{457}
(\byear{1976})
\end{bchapter}
\endbibitem

\bibitem{Gounelle2008}
\begin{barticle}
\bauthor{\bsnm{{Gounelle}}, \binits{M.}},
\bauthor{\bsnm{{Meibom}}, \binits{A.}}:
\batitle{{The Origin of Short-lived Radionuclides and the Astrophysical
  Environment of Solar System Formation}}.
\bjtitle{\apj}
\bvolume{680}(\bissue{1}),
\bfpage{781}--\blpage{792}
(\byear{2008})
{\href{https://arxiv.org/abs/0805.0569}{{arXiv:0805.0569}}}
{[astro-ph]}.
\doiurl{10.1086/587613}
\end{barticle}
\endbibitem

\bibitem{Gaidos2009}
\begin{barticle}
\bauthor{\bsnm{{Gaidos}}, \binits{E.}},
\bauthor{\bsnm{{Krot}}, \binits{A.N.}},
\bauthor{\bsnm{{Williams}}, \binits{J.P.}},
\bauthor{\bsnm{{Raymond}}, \binits{S.N.}}:
\batitle{{$^{26}$Al and the Formation of the Solar System from a Molecular
  Cloud Contaminated by Wolf-Rayet Winds}}.
\bjtitle{\apj}
\bvolume{696}(\bissue{2}),
\bfpage{1854}--\blpage{1863}
(\byear{2009})
{\href{https://arxiv.org/abs/0901.3364}{{arXiv:0901.3364}}}
{[astro-ph.EP]}.
\doiurl{10.1088/0004-637X/696/2/1854}
\end{barticle}
\endbibitem

\bibitem{Gritschneder2012}
\begin{barticle}
\bauthor{\bsnm{{Gritschneder}}, \binits{M.}},
\bauthor{\bsnm{{Lin}}, \binits{D.N.C.}},
\bauthor{\bsnm{{Murray}}, \binits{S.D.}},
\bauthor{\bsnm{{Yin}}, \binits{Q.-Z.}},
\bauthor{\bsnm{{Gong}}, \binits{M.-N.}}:
\batitle{{The Supernova Triggered Formation and Enrichment of Our Solar
  System}}.
\bjtitle{\apj}
\bvolume{745}(\bissue{1}),
\bfpage{22}
(\byear{2012})
{\href{https://arxiv.org/abs/1111.0012}{{arXiv:1111.0012}}}
{[astro-ph.SR]}.
\doiurl{10.1088/0004-637X/745/1/22}
\end{barticle}
\endbibitem

\bibitem{Diehl2010}
\begin{barticle}
\bauthor{\bsnm{{Diehl}}, \binits{R.}},
\bauthor{\bsnm{{Lang}}, \binits{M.G.}},
\bauthor{\bsnm{{Martin}}, \binits{P.}},
\bauthor{\bsnm{{Ohlendorf}}, \binits{H.}},
\bauthor{\bsnm{{Preibisch}}, \binits{T.}},
\bauthor{\bsnm{{Voss}}, \binits{R.}},
\bauthor{\bsnm{{Jean}}, \binits{P.}},
\bauthor{\bsnm{{Roques}}, \binits{J.-P.}},
\bauthor{\bsnm{{von Ballmoos}}, \binits{P.}},
\bauthor{\bsnm{{Wang}}, \binits{W.}}:
\batitle{{Radioactive $^{26}$Al from the Scorpius-Centaurus association}}.
\bjtitle{\aap}
\bvolume{522},
\bfpage{51}
(\byear{2010})
{\href{https://arxiv.org/abs/1007.4462}{{arXiv:1007.4462}}}
{[astro-ph.HE]}.
\doiurl{10.1051/0004-6361/201014302}
\end{barticle}
\endbibitem

\bibitem{Reiter2020}
\begin{barticle}
\bauthor{\bsnm{{Reiter}}, \binits{M.}}:
\batitle{{Observational constraints on the likelihood of $^{26}$Al in
  planet-forming environments}}.
\bjtitle{\aap}
\bvolume{644},
\bfpage{1}
(\byear{2020})
{\href{https://arxiv.org/abs/2011.09971}{{arXiv:2011.09971}}}
{[astro-ph.EP]}.
\doiurl{10.1051/0004-6361/202039334}
\end{barticle}
\endbibitem

\bibitem{Forbes2021}
\begin{barticle}
\bauthor{\bsnm{{Forbes}}, \binits{J.C.}},
\bauthor{\bsnm{{Alves}}, \binits{J.}},
\bauthor{\bsnm{{Lin}}, \binits{D.N.C.}}:
\batitle{{A Solar System formation analogue in the Ophiuchus star-forming
  complex}}.
\bjtitle{Nature Astronomy}
\bvolume{5},
\bfpage{1009}--\blpage{1016}
(\byear{2021})
{\href{https://arxiv.org/abs/2108.09326}{{arXiv:2108.09326}}}
{[astro-ph.EP]}.
\doiurl{10.1038/s41550-021-01442-9}
\end{barticle}
\endbibitem

\bibitem{Zdziarski2014}
\begin{barticle}
\bauthor{\bsnm{{Zdziarski}}, \binits{A.A.}},
\bauthor{\bsnm{{Pjanka}}, \binits{P.}},
\bauthor{\bsnm{{Sikora}}, \binits{M.}},
\bauthor{\bsnm{{Stawarz}}, \binits{{\L}.}}:
\batitle{{Jet contributions to the broad-band spectrum of Cyg X-1 in the hard
  state}}.
\bjtitle{\mnras}
\bvolume{442}(\bissue{4}),
\bfpage{3243}--\blpage{3255}
(\byear{2014})
{\href{https://arxiv.org/abs/1403.4768}{{arXiv:1403.4768}}}
{[astro-ph.HE]}.
\doiurl{10.1093/mnras/stu1009}
\end{barticle}
\endbibitem

\bibitem{Vadawale2018}
\begin{barticle}
\bauthor{\bsnm{{Vadawale}}, \binits{S.V.}},
\bauthor{\bsnm{{Chattopadhyay}}, \binits{T.}},
\bauthor{\bsnm{{Mithun}}, \binits{N.P.S.}},
\bauthor{\bsnm{{Rao}}, \binits{A.R.}},
\bauthor{\bsnm{{Bhattacharya}}, \binits{D.}},
\bauthor{\bsnm{{Vibhute}}, \binits{A.}},
\bauthor{\bsnm{{Bhalerao}}, \binits{V.B.}},
\bauthor{\bsnm{{Dewangan}}, \binits{G.C.}},
\bauthor{\bsnm{{Misra}}, \binits{R.}},
\bauthor{\bsnm{{Paul}}, \binits{B.}},
\bauthor{\bsnm{{Basu}}, \binits{A.}},
\bauthor{\bsnm{{Joshi}}, \binits{B.C.}},
\bauthor{\bsnm{{Sreekumar}}, \binits{S.}},
\bauthor{\bsnm{{Samuel}}, \binits{E.}},
\bauthor{\bsnm{{Priya}}, \binits{P.}},
\bauthor{\bsnm{{Vinod}}, \binits{P.}},
\bauthor{\bsnm{{Seetha}}, \binits{S.}}:
\batitle{{Phase-resolved X-ray polarimetry of the Crab pulsar with the AstroSat
  CZT Imager}}.
\bjtitle{Nature Astronomy}
\bvolume{2},
\bfpage{50}--\blpage{55}
(\byear{2018}).
\doiurl{10.1038/s41550-017-0293-z}
\end{barticle}
\endbibitem

\bibitem{Chauvin2018}
\begin{barticle}
\bauthor{\bsnm{{Chauvin}}, \binits{M.}},
\bauthor{\bsnm{{Flor{\'e}n}}, \binits{H.-G.}},
\bauthor{\bsnm{{Friis}}, \binits{M.}},
\bauthor{\bsnm{{Jackson}}, \binits{M.}},
\bauthor{\bsnm{{Kamae}}, \binits{T.}},
\bauthor{\bsnm{{Kataoka}}, \binits{J.}},
\bauthor{\bsnm{{Kawano}}, \binits{T.}},
\bauthor{\bsnm{{Kiss}}, \binits{M.}},
\bauthor{\bsnm{{Mikhalev}}, \binits{V.}},
\bauthor{\bsnm{{Mizuno}}, \binits{T.}},
\bauthor{\bsnm{{Tajima}}, \binits{H.}},
\bauthor{\bsnm{{Takahashi}}, \binits{H.}},
\bauthor{\bsnm{{Uchida}}, \binits{N.}},
\bauthor{\bsnm{{Pearce}}, \binits{M.}}:
\batitle{{The PoGO+ view on Crab off-pulse hard X-ray polarization}}.
\bjtitle{\mnras}
\bvolume{477}(\bissue{1}),
\bfpage{45}--\blpage{49}
(\byear{2018})
{\href{https://arxiv.org/abs/1802.07775}{{arXiv:1802.07775}}}
{[astro-ph.HE]}.
\doiurl{10.1093/mnrasl/sly027}
\end{barticle}
\endbibitem

\bibitem{Chattopadhyay2023}
\begin{botherref}
\oauthor{\bsnm{{Chattopadhyay}}, \binits{T.}},
\oauthor{\bsnm{{Kumar}}, \binits{A.}},
\oauthor{\bsnm{{Rao}}, \binits{A.R.}},
\oauthor{\bsnm{{Bhargava}}, \binits{Y.}},
\oauthor{\bsnm{{Vadawale}}, \binits{S.V.}},
\oauthor{\bsnm{{Ratheesh}}, \binits{A.}},
\oauthor{\bsnm{{Dewangan}}, \binits{G.}},
\oauthor{\bsnm{{Bhattacharyay}}, \binits{D.}},
\oauthor{\bsnm{{Mithun N.~P.}}, \binits{S.}},
\oauthor{\bsnm{{Bhalerao}}, \binits{V.}}:
{High hard X-ray polarization in Cygnus X-1 confined to the intermediate hard
  state: evidence for a variable jet component}.
arXiv e-prints,
2306--04057
(2023)
{\href{https://arxiv.org/abs/2306.04057}{{arXiv:2306.04057}}}
{[astro-ph.HE]}.
\doiurl{10.48550/arXiv.2306.04057}
\end{botherref}
\endbibitem

\bibitem{Laurent2016}
\begin{bchapter}
\bauthor{\bsnm{{Laurent}}, \binits{P.}},
\bauthor{\bsnm{{Gouiffes}}, \binits{C.}},
\bauthor{\bsnm{{Rodriguez}}, \binits{J.}},
\bauthor{\bsnm{{Chambouleyron}}, \binits{V.}}:
\bctitle{{INTEGRAL/IBIS observations of V404 Cygni polarimetric properties
  during its 2015 giant flares}}.
In: \bbtitle{11th INTEGRAL Conference Gamma-Ray Astrophysics in
  Multi-Wavelength Perspective},
p. \bfpage{22}
(\byear{2016}).
\doiurl{10.22323/1.285.0022}
\end{bchapter}
\endbibitem

\bibitem{Chattopadhyay2022}
\begin{barticle}
\bauthor{\bsnm{{Chattopadhyay}}, \binits{T.}},
\bauthor{\bsnm{{Gupta}}, \binits{S.}},
\bauthor{\bsnm{{Iyyani}}, \binits{S.}},
\bauthor{\bsnm{{Saraogi}}, \binits{D.}},
\bauthor{\bsnm{{Sharma}}, \binits{V.}},
\bauthor{\bsnm{{Tsvetkova}}, \binits{A.}},
\bauthor{\bsnm{{Ratheesh}}, \binits{A.}},
\bauthor{\bsnm{{Gupta}}, \binits{R.}},
\bauthor{\bsnm{{Mithun}}, \binits{N.P.S.}},
\bauthor{\bsnm{{Vaishnava}}, \binits{C.S.}},
\bauthor{\bsnm{{Prasad}}, \binits{V.}},
\bauthor{\bsnm{{Aarthy}}, \binits{E.}},
\bauthor{\bsnm{{Kumar}}, \binits{A.}},
\bauthor{\bsnm{{Rao}}, \binits{A.R.}},
\bauthor{\bsnm{{Vadawale}}, \binits{S.}},
\bauthor{\bsnm{{Bhalerao}}, \binits{V.}},
\bauthor{\bsnm{{Bhattacharya}}, \binits{D.}},
\bauthor{\bsnm{{Vibhute}}, \binits{A.}},
\bauthor{\bsnm{{Frederiks}}, \binits{D.}}:
\batitle{{Hard X-Ray Polarization Catalog for a Five-year Sample of Gamma-Ray
  Bursts Using AstroSat CZT Imager}}.
\bjtitle{\apj}
\bvolume{936}(\bissue{1}),
\bfpage{12}
(\byear{2022})
{\href{https://arxiv.org/abs/2207.09605}{{arXiv:2207.09605}}}
{[astro-ph.HE]}.
\doiurl{10.3847/1538-4357/ac82ef}
\end{barticle}
\endbibitem

\bibitem{Takata2007}
\begin{barticle}
\bauthor{\bsnm{{Takata}}, \binits{J.}},
\bauthor{\bsnm{{Chang}}, \binits{H.-K.}}:
\batitle{{Pulse Profiles, Spectra, and Polarization Characteristics of
  Nonthermal Emissions from the Crab-like Pulsars}}.
\bjtitle{\apj}
\bvolume{670}(\bissue{1}),
\bfpage{677}--\blpage{692}
(\byear{2007})
{\href{https://arxiv.org/abs/0707.3301}{{arXiv:0707.3301}}}
{[astro-ph]}.
\doiurl{10.1086/521785}
\end{barticle}
\endbibitem

\bibitem{Petri2013}
\begin{barticle}
\bauthor{\bsnm{{P{\'e}tri}}, \binits{J.}}:
\batitle{{Phase-resolved polarization properties of the pulsar striped wind
  synchrotron emission}}.
\bjtitle{\mnras}
\bvolume{434}(\bissue{3}),
\bfpage{2636}--\blpage{2644}
(\byear{2013})
{\href{https://arxiv.org/abs/1308.0973}{{arXiv:1308.0973}}}
{[astro-ph.HE]}.
\doiurl{10.1093/mnras/stt1214}
\end{barticle}
\endbibitem

\bibitem{Harding2017}
\begin{barticle}
\bauthor{\bsnm{{Harding}}, \binits{A.K.}},
\bauthor{\bsnm{{Kalapotharakos}}, \binits{C.}}:
\batitle{{Multiwavelength Polarization of Rotation-powered Pulsars}}.
\bjtitle{\apj}
\bvolume{840}(\bissue{2}),
\bfpage{73}
(\byear{2017})
{\href{https://arxiv.org/abs/1704.06183}{{arXiv:1704.06183}}}
{[astro-ph.HE]}.
\doiurl{10.3847/1538-4357/aa6ead}
\end{barticle}
\endbibitem

\bibitem{Barrett1995}
\begin{barticle}
\bauthor{\bsnm{{Barrett}}, \binits{H.H.}},
\bauthor{\bsnm{{Eskin}}, \binits{J.D.}},
\bauthor{\bsnm{{Barber}}, \binits{H.B.}}:
\batitle{{Charge Transport in Arrays of Semiconductor Gamma-Ray Detectors}}.
\bjtitle{\prl}
\bvolume{75}(\bissue{1}),
\bfpage{156}--\blpage{159}
(\byear{1995}).
\doiurl{10.1103/PhysRevLett.75.156}
\end{barticle}
\endbibitem

\bibitem{He1996}
\begin{barticle}
\bauthor{\bsnm{{He}}, \binits{Z.}},
\bauthor{\bsnm{{Knoll}}, \binits{G.F.}},
\bauthor{\bsnm{{Wehe}}, \binits{D.K.}},
\bauthor{\bsnm{{Rojeski}}, \binits{R.}},
\bauthor{\bsnm{{Mastrangelo}}, \binits{C.H.}},
\bauthor{\bsnm{{Hammig}}, \binits{M.}},
\bauthor{\bsnm{{Barrett}}, \binits{C.}},
\bauthor{\bsnm{{Uritani}}, \binits{A.}}:
\batitle{{1-D position sensitive single carrier semiconductor detectors}}.
\bjtitle{Nuclear Instruments and Methods in Physics Research A}
\bvolume{380}(\bissue{1-2}),
\bfpage{228}--\blpage{231}
(\byear{1996}).
\doiurl{10.1016/S0168-9002(96)00352-X}
\end{barticle}
\endbibitem

\bibitem{He1997}
\begin{barticle}
\bauthor{\bsnm{{He}}, \binits{Z.}},
\bauthor{\bsnm{{Knoll}}, \binits{G.F.}},
\bauthor{\bsnm{{Wehe}}, \binits{D.K.}},
\bauthor{\bsnm{{Miyamoto}}, \binits{J.}}:
\batitle{{Position-sensitive single carrier CdZnTe detectors}}.
\bjtitle{Nuclear Instruments and Methods in Physics Research A}
\bvolume{388}(\bissue{1-2}),
\bfpage{180}--\blpage{185}
(\byear{1997}).
\doiurl{10.1016/S0168-9002(97)00318-5}
\end{barticle}
\endbibitem

\bibitem{He1999}
\begin{barticle}
\bauthor{\bsnm{{He}}, \binits{Z.}},
\bauthor{\bsnm{{Li}}, \binits{W.}},
\bauthor{\bsnm{{Knoll}}, \binits{G.F.}},
\bauthor{\bsnm{{Wehe}}, \binits{D.K.}},
\bauthor{\bsnm{{Berry}}, \binits{J.}},
\bauthor{\bsnm{{Stahle}}, \binits{C.M.}}:
\batitle{{3-D position sensitive CdZnTe gamma-ray spectrometers}}.
\bjtitle{Nuclear Instruments and Methods in Physics Research A}
\bvolume{422}(\bissue{1-3}),
\bfpage{173}--\blpage{178}
(\byear{1999}).
\doiurl{10.1016/S0168-9002(98)00950-4}
\end{barticle}
\endbibitem

\bibitem{Du2001}
\begin{barticle}
\bauthor{\bsnm{{Du}}, \binits{Y.F.}},
\bauthor{\bsnm{{He}}, \binits{Z.}},
\bauthor{\bsnm{{Knoll}}, \binits{G.F.}},
\bauthor{\bsnm{{Wehe}}, \binits{D.K.}},
\bauthor{\bsnm{{Li}}, \binits{W.}}:
\batitle{{Evaluation of a Compton scattering camera using 3-D position
  sensitive CdZnTe detectors}}.
\bjtitle{Nuclear Instruments and Methods in Physics Research A}
\bvolume{457}(\bissue{1-2}),
\bfpage{203}--\blpage{211}
(\byear{2001}).
\doiurl{10.1016/S0168-9002(00)00669-0}
\end{barticle}
\endbibitem

\bibitem{Zhang2006}
\begin{barticle}
\bauthor{\bsnm{{Zhang}}, \binits{F.}},
\bauthor{\bsnm{{He}}, \binits{Z.}}:
\batitle{{New Readout Electronics for 3-D Position Sensitive
  CdZnTe/HgI\_2Detector Arrays}}.
\bjtitle{IEEE Transactions on Nuclear Science}
\bvolume{53}(\bissue{5}),
\bfpage{3021}--\blpage{3027}
(\byear{2006}).
\doiurl{10.1109/TNS.2006.879761}
\end{barticle}
\endbibitem

\bibitem{Zhang2007}
\begin{barticle}
\bauthor{\bsnm{{Zhang}}, \binits{F.}},
\bauthor{\bsnm{{He}}, \binits{Z.}},
\bauthor{\bsnm{{Seifert}}, \binits{C.E.}}:
\batitle{{A Prototype Three-Dimensional Position Sensitive CdZnTe Detector
  Array}}.
\bjtitle{IEEE Transactions on Nuclear Science}
\bvolume{54}(\bissue{4}),
\bfpage{843}--\blpage{848}
(\byear{2007}).
\doiurl{10.1109/TNS.2007.902354}
\end{barticle}
\endbibitem

\bibitem{Kim2012}
\begin{barticle}
\bauthor{\bsnm{{Kim}}, \binits{J.C.}},
\bauthor{\bsnm{{Kaye}}, \binits{W.R.}},
\bauthor{\bsnm{{Wang}}, \binits{W.}},
\bauthor{\bsnm{{Zhang}}, \binits{F.}},
\bauthor{\bsnm{{He}}, \binits{Z.}}:
\batitle{{Impact of drift time variation on the Compton image from large-volume
  CdZnTe crystals}}.
\bjtitle{Nuclear Instruments and Methods in Physics Research A}
\bvolume{683},
\bfpage{53}--\blpage{62}
(\byear{2012}).
\doiurl{10.1016/j.nima.2012.04.057}
\end{barticle}
\endbibitem

\bibitem{Yang2020}
\begin{barticle}
\bauthor{\bsnm{{Yang}}, \binits{J.}},
\bauthor{\bsnm{{Li}}, \binits{Y.L.}},
\bauthor{\bsnm{{Tian}}, \binits{Y.}},
\bauthor{\bsnm{{Fu}}, \binits{Y.D.}},
\bauthor{\bsnm{{Xu}}, \binits{L.}},
\bauthor{\bsnm{{Cai}}, \binits{Y.M.}},
\bauthor{\bsnm{{Li}}, \binits{Y.J.}}:
\batitle{{Performance optimization of pixelated CdZnTe detector readout by
  analog ASIC using cathode waveform}}.
\bjtitle{Journal of Instrumentation}
\bvolume{15}(\bissue{4}),
\bfpage{04005}
(\byear{2020}).
\doiurl{10.1088/1748-0221/15/04/P04005}
\end{barticle}
\endbibitem

\bibitem{Tomsick2021}
\begin{bchapter}
\bauthor{\bsnm{{Tomsick}}, \binits{J.}},
\bauthor{\bsnm{{Boggs}}, \binits{S.}},
\bauthor{\bsnm{{Zoglauer}}, \binits{A.}},
\bauthor{\bsnm{{Lazar}}, \binits{H.}},
\bauthor{\bsnm{{Beechert}}, \binits{J.}},
\bauthor{\bsnm{{Gulick}}, \binits{H.}},
\bauthor{\bsnm{{Roberts}}, \binits{J.}},
\bauthor{\bsnm{{Siegert}}, \binits{T.}},
\bauthor{\bsnm{{Wulf}}, \binits{E.}},
\bauthor{\bsnm{{Sleator}}, \binits{C.}},
\bauthor{\bsnm{{Grove}}, \binits{J.}},
\bauthor{\bsnm{{Phlips}}, \binits{B.}},
\bauthor{\bsnm{{Brandt}}, \binits{T.}},
\bauthor{\bsnm{{Smale}}, \binits{A.}},
\bauthor{\bsnm{{Kierans}}, \binits{C.}},
\bauthor{\bsnm{{Hartmann}}, \binits{D.}},
\bauthor{\bsnm{{Leising}}, \binits{M.}},
\bauthor{\bsnm{{Ajello}}, \binits{M.}},
\bauthor{\bsnm{{Jean}}, \binits{P.}},
\bauthor{\bsnm{{von Ballmoos}}, \binits{P.}},
\bauthor{\bsnm{{Malzac}}, \binits{J.}},
\bauthor{\bsnm{{Burns}}, \binits{E.}},
\bauthor{\bsnm{{Fryer}}, \binits{C.}},
\bauthor{\bsnm{{Chang}}, \binits{H.}},
\bauthor{\bsnm{{Tavecchio}}, \binits{F.}},
\bauthor{\bsnm{{Takahashi}}, \binits{T.}}:
\bctitle{{The Compton Spectrometer and Imager Project for MeV Astronomy}}.
In: \bbtitle{American Astronomical Society Meeting Abstracts}.
\bsertitle{American Astronomical Society Meeting Abstracts},
vol. \bseriesno{53},
pp. \bfpage{315}--\blpage{01}
(\byear{2021})
\end{bchapter}
\endbibitem

\bibitem{Feng2019}
\begin{barticle}
\bauthor{\bsnm{{Feng}}, \binits{H.}},
\bauthor{\bsnm{{Jiang}}, \binits{W.}},
\bauthor{\bsnm{{Minuti}}, \binits{M.}},
\bauthor{\bsnm{{Wu}}, \binits{Q.}},
\bauthor{\bsnm{{Jung}}, \binits{A.}},
\bauthor{\bsnm{{Yang}}, \binits{D.}},
\bauthor{\bsnm{{Citraro}}, \binits{S.}},
\bauthor{\bsnm{{Nasimi}}, \binits{H.}},
\bauthor{\bsnm{{Yu}}, \binits{J.}},
\bauthor{\bsnm{{Jin}}, \binits{G.}},
\bauthor{\bsnm{{Huang}}, \binits{J.}},
\bauthor{\bsnm{{Zeng}}, \binits{M.}},
\bauthor{\bsnm{{An}}, \binits{P.}},
\bauthor{\bsnm{{Baldini}}, \binits{L.}},
\bauthor{\bsnm{{Bellazzini}}, \binits{R.}},
\bauthor{\bsnm{{Brez}}, \binits{A.}},
\bauthor{\bsnm{{Latronico}}, \binits{L.}},
\bauthor{\bsnm{{Sgr{\`o}}}, \binits{C.}},
\bauthor{\bsnm{{Spandre}}, \binits{G.}},
\bauthor{\bsnm{{Pinchera}}, \binits{M.}},
\bauthor{\bsnm{{Muleri}}, \binits{F.}},
\bauthor{\bsnm{{Soffitta}}, \binits{P.}},
\bauthor{\bsnm{{Costa}}, \binits{E.}}:
\batitle{{PolarLight: a CubeSat X-ray polarimeter based on the gas pixel
  detector}}.
\bjtitle{Experimental Astronomy}
\bvolume{47}(\bissue{1-2}),
\bfpage{225}--\blpage{243}
(\byear{2019})
{\href{https://arxiv.org/abs/1903.01619}{{arXiv:1903.01619}}}
{[astro-ph.IM]}.
\doiurl{10.1007/s10686-019-09625-z}
\end{barticle}
\endbibitem

\bibitem{Wen2019}
\begin{barticle}
\bauthor{\bsnm{{Wen}}, \binits{J.}},
\bauthor{\bsnm{{Long}}, \binits{X.}},
\bauthor{\bsnm{{Zheng}}, \binits{X.}},
\bauthor{\bsnm{{An}}, \binits{Y.}},
\bauthor{\bsnm{{Cai}}, \binits{Z.}},
\bauthor{\bsnm{{Cang}}, \binits{J.}},
\bauthor{\bsnm{{Che}}, \binits{Y.}},
\bauthor{\bsnm{{Chen}}, \binits{C.}},
\bauthor{\bsnm{{Chen}}, \binits{L.}},
\bauthor{\bsnm{{Chen}}, \binits{Q.}},
\bauthor{\bsnm{{Chen}}, \binits{Z.}},
\bauthor{\bsnm{{Cheng}}, \binits{Y.}},
\bauthor{\bsnm{{Deng}}, \binits{L.}},
\bauthor{\bsnm{{Deng}}, \binits{W.}},
\bauthor{\bsnm{{Ding}}, \binits{W.}},
\bauthor{\bsnm{{Du}}, \binits{H.}},
\bauthor{\bsnm{{Duan}}, \binits{L.}},
\bauthor{\bsnm{{Gan}}, \binits{Q.}},
\bauthor{\bsnm{{Gao}}, \binits{T.}},
\bauthor{\bsnm{{Gao}}, \binits{Z.}},
\bauthor{\bsnm{{Han}}, \binits{W.}},
\bauthor{\bsnm{{Han}}, \binits{Y.}},
\bauthor{\bsnm{{He}}, \binits{X.}},
\bauthor{\bsnm{{He}}, \binits{X.}},
\bauthor{\bsnm{{Hou}}, \binits{L.}},
\bauthor{\bsnm{{Hu}}, \binits{F.}},
\bauthor{\bsnm{{Hu}}, \binits{J.}},
\bauthor{\bsnm{{Huang}}, \binits{B.}},
\bauthor{\bsnm{{Huang}}, \binits{D.}},
\bauthor{\bsnm{{Huang}}, \binits{X.}},
\bauthor{\bsnm{{Jia}}, \binits{S.}},
\bauthor{\bsnm{{Jiang}}, \binits{Y.}},
\bauthor{\bsnm{{Jin}}, \binits{Y.}},
\bauthor{\bsnm{{Li}}, \binits{K.}},
\bauthor{\bsnm{{Li}}, \binits{S.}},
\bauthor{\bsnm{{Li}}, \binits{Y.}},
\bauthor{\bsnm{{Liang}}, \binits{J.}},
\bauthor{\bsnm{{Liang}}, \binits{Y.}},
\bauthor{\bsnm{{Lin}}, \binits{W.}},
\bauthor{\bsnm{{Liu}}, \binits{C.}},
\bauthor{\bsnm{{Liu}}, \binits{G.}},
\bauthor{\bsnm{{Liu}}, \binits{M.}},
\bauthor{\bsnm{{Liu}}, \binits{R.}},
\bauthor{\bsnm{{Liu}}, \binits{T.}},
\bauthor{\bsnm{{Liu}}, \binits{W.}},
\bauthor{\bsnm{{Lu}}, \binits{D.}},
\bauthor{\bsnm{{Lu}}, \binits{P.}},
\bauthor{\bsnm{{Lu}}, \binits{Z.}},
\bauthor{\bsnm{{Luo}}, \binits{X.}},
\bauthor{\bsnm{{Ma}}, \binits{S.}},
\bauthor{\bsnm{{Ma}}, \binits{Y.}},
\bauthor{\bsnm{{Mao}}, \binits{X.}},
\bauthor{\bsnm{{Mo}}, \binits{Y.}},
\bauthor{\bsnm{{Nie}}, \binits{Q.}},
\bauthor{\bsnm{{Qu}}, \binits{S.}},
\bauthor{\bsnm{{Shan}}, \binits{X.}},
\bauthor{\bsnm{{Shi}}, \binits{G.}},
\bauthor{\bsnm{{Song}}, \binits{W.}},
\bauthor{\bsnm{{Sun}}, \binits{Z.}},
\bauthor{\bsnm{{Tan}}, \binits{X.}},
\bauthor{\bsnm{{Tang}}, \binits{S.}},
\bauthor{\bsnm{{Tao}}, \binits{M.}},
\bauthor{\bsnm{{Wang}}, \binits{B.}},
\bauthor{\bsnm{{Wang}}, \binits{Y.}},
\bauthor{\bsnm{{Wang}}, \binits{Z.}},
\bauthor{\bsnm{{Wu}}, \binits{Q.}},
\bauthor{\bsnm{{Wu}}, \binits{X.}},
\bauthor{\bsnm{{Xia}}, \binits{Y.}},
\bauthor{\bsnm{{Xiao}}, \binits{H.}},
\bauthor{\bsnm{{Xie}}, \binits{W.}},
\bauthor{\bsnm{{Xu}}, \binits{D.}},
\bauthor{\bsnm{{Xu}}, \binits{R.}},
\bauthor{\bsnm{{Xu}}, \binits{W.}},
\bauthor{\bsnm{{Yan}}, \binits{L.}},
\bauthor{\bsnm{{Yan}}, \binits{S.}},
\bauthor{\bsnm{{Yang}}, \binits{D.}},
\bauthor{\bsnm{{Yang}}, \binits{H.}},
\bauthor{\bsnm{{Yang}}, \binits{H.}},
\bauthor{\bsnm{{Yang}}, \binits{Y.-S.}},
\bauthor{\bsnm{{Yang}}, \binits{Y.}},
\bauthor{\bsnm{{Yao}}, \binits{L.}},
\bauthor{\bsnm{{Yu}}, \binits{H.}},
\bauthor{\bsnm{{Yu}}, \binits{Y.}},
\bauthor{\bsnm{{Zhang}}, \binits{A.}},
\bauthor{\bsnm{{Zhang}}, \binits{B.}},
\bauthor{\bsnm{{Zhang}}, \binits{L.}},
\bauthor{\bsnm{{Zhang}}, \binits{M.}},
\bauthor{\bsnm{{Zhang}}, \binits{S.}},
\bauthor{\bsnm{{Zhang}}, \binits{T.}},
\bauthor{\bsnm{{Zhang}}, \binits{Y.}},
\bauthor{\bsnm{{Zhao}}, \binits{Q.}},
\bauthor{\bsnm{{Zhao}}, \binits{R.}},
\bauthor{\bsnm{{Zheng}}, \binits{S.}},
\bauthor{\bsnm{{Zhou}}, \binits{X.}},
\bauthor{\bsnm{{Zhu}}, \binits{R.}},
\bauthor{\bsnm{{Zou}}, \binits{Y.}},
\bauthor{\bsnm{{An}}, \binits{P.}},
\bauthor{\bsnm{{Cai}}, \binits{Y.}},
\bauthor{\bsnm{{Chen}}, \binits{H.}},
\bauthor{\bsnm{{Dai}}, \binits{Z.}},
\bauthor{\bsnm{{Fan}}, \binits{Y.}},
\bauthor{\bsnm{{Feng}}, \binits{C.}},
\bauthor{\bsnm{{Feng}}, \binits{H.}},
\bauthor{\bsnm{{Gao}}, \binits{H.}},
\bauthor{\bsnm{{Huang}}, \binits{L.}},
\bauthor{\bsnm{{Kang}}, \binits{M.}},
\bauthor{\bsnm{{Li}}, \binits{L.}},
\bauthor{\bsnm{{Li}}, \binits{Z.}},
\bauthor{\bsnm{{Liang}}, \binits{E.}},
\bauthor{\bsnm{{Lin}}, \binits{L.}},
\bauthor{\bsnm{{Lin}}, \binits{Q.}},
\bauthor{\bsnm{{Liu}}, \binits{C.}},
\bauthor{\bsnm{{Liu}}, \binits{H.}},
\bauthor{\bsnm{{Liu}}, \binits{X.}},
\bauthor{\bsnm{{Liu}}, \binits{Y.}},
\bauthor{\bsnm{{Lu}}, \binits{X.}},
\bauthor{\bsnm{{Mao}}, \binits{S.}},
\bauthor{\bsnm{{Shen}}, \binits{R.}},
\bauthor{\bsnm{{Shu}}, \binits{J.}},
\bauthor{\bsnm{{Su}}, \binits{M.}},
\bauthor{\bsnm{{Sun}}, \binits{H.}},
\bauthor{\bsnm{{Tam}}, \binits{P.-H.}},
\bauthor{\bsnm{{Tang}}, \binits{C.-P.}},
\bauthor{\bsnm{{Tian}}, \binits{Y.}},
\bauthor{\bsnm{{Wang}}, \binits{F.}},
\bauthor{\bsnm{{Wang}}, \binits{J.}},
\bauthor{\bsnm{{Wang}}, \binits{W.}},
\bauthor{\bsnm{{Wang}}, \binits{Z.}},
\bauthor{\bsnm{{Wu}}, \binits{J.}},
\bauthor{\bsnm{{Wu}}, \binits{X.}},
\bauthor{\bsnm{{Xiong}}, \binits{S.}},
\bauthor{\bsnm{{Xu}}, \binits{C.}},
\bauthor{\bsnm{{Yu}}, \binits{J.}},
\bauthor{\bsnm{{Yu}}, \binits{W.}},
\bauthor{\bsnm{{Yu}}, \binits{Y.}},
\bauthor{\bsnm{{Zeng}}, \binits{M.}},
\bauthor{\bsnm{{Zeng}}, \binits{Z.}},
\bauthor{\bsnm{{Zhang}}, \binits{B.-B.}},
\bauthor{\bsnm{{Zhang}}, \binits{B.}},
\bauthor{\bsnm{{Zhao}}, \binits{Z.}},
\bauthor{\bsnm{{Zhou}}, \binits{R.}},
\bauthor{\bsnm{{Zhu}}, \binits{Z.}}:
\batitle{{GRID: a student project to monitor the transient gamma-ray sky in the
  multi-messenger astronomy era}}.
\bjtitle{Experimental Astronomy}
\bvolume{48}(\bissue{1}),
\bfpage{77}--\blpage{95}
(\byear{2019})
{\href{https://arxiv.org/abs/1907.06842}{{arXiv:1907.06842}}}
{[astro-ph.IM]}.
\doiurl{10.1007/s10686-019-09636-w}
\end{barticle}
\endbibitem

\bibitem{Beechert2022}
\begin{barticle}
\bauthor{\bsnm{{Beechert}}, \binits{J.}},
\bauthor{\bsnm{{Lazar}}, \binits{H.}},
\bauthor{\bsnm{{Boggs}}, \binits{S.E.}},
\bauthor{\bsnm{{Brandt}}, \binits{T.J.}},
\bauthor{\bsnm{{Chang}}, \binits{Y.-C.}},
\bauthor{\bsnm{{Chu}}, \binits{C.-Y.}},
\bauthor{\bsnm{{Gulick}}, \binits{H.}},
\bauthor{\bsnm{{Kierans}}, \binits{C.}},
\bauthor{\bsnm{{Lowell}}, \binits{A.}},
\bauthor{\bsnm{{Pellegrini}}, \binits{N.}},
\bauthor{\bsnm{{Roberts}}, \binits{J.M.}},
\bauthor{\bsnm{{Siegert}}, \binits{T.}},
\bauthor{\bsnm{{Sleator}}, \binits{C.}},
\bauthor{\bsnm{{Tomsick}}, \binits{J.A.}},
\bauthor{\bsnm{{Zoglauer}}, \binits{A.}}:
\batitle{{Calibrations of the Compton Spectrometer and Imager}}.
\bjtitle{Nuclear Instruments and Methods in Physics Research A}
\bvolume{1031},
\bfpage{166510}
(\byear{2022})
{\href{https://arxiv.org/abs/2203.00695}{{arXiv:2203.00695}}}
{[astro-ph.IM]}.
\doiurl{10.1016/j.nima.2022.166510}
\end{barticle}
\endbibitem

\bibitem{Yoneda2023}
\begin{barticle}
\bauthor{\bsnm{{Yoneda}}, \binits{H.}},
\bauthor{\bsnm{{Odaka}}, \binits{H.}},
\bauthor{\bsnm{{Ichinohe}}, \binits{Y.}},
\bauthor{\bsnm{{Takashima}}, \binits{S.}},
\bauthor{\bsnm{{Aramaki}}, \binits{T.}},
\bauthor{\bsnm{{Aoyama}}, \binits{K.}},
\bauthor{\bsnm{{Asaadi}}, \binits{J.}},
\bauthor{\bsnm{{Fabris}}, \binits{L.}},
\bauthor{\bsnm{{Inoue}}, \binits{Y.}},
\bauthor{\bsnm{{Karagiorgi}}, \binits{G.}},
\bauthor{\bsnm{{Khangulyan}}, \binits{D.}},
\bauthor{\bsnm{{Kimura}}, \binits{M.}},
\bauthor{\bsnm{{Leyva}}, \binits{J.}},
\bauthor{\bsnm{{Mukherjee}}, \binits{R.}},
\bauthor{\bsnm{{Nakasone}}, \binits{T.}},
\bauthor{\bsnm{{Perez}}, \binits{K.}},
\bauthor{\bsnm{{Sakurai}}, \binits{M.}},
\bauthor{\bsnm{{Seligman}}, \binits{W.}},
\bauthor{\bsnm{{Tanaka}}, \binits{M.}},
\bauthor{\bsnm{{Tsuji}}, \binits{N.}},
\bauthor{\bsnm{{Yorita}}, \binits{K.}},
\bauthor{\bsnm{{Zeng}}, \binits{J.}}:
\batitle{{Reconstruction of multiple Compton scattering events in MeV gamma-ray
  Compton telescopes towards GRAMS: The physics-based probabilistic model}}.
\bjtitle{Astroparticle Physics}
\bvolume{144},
\bfpage{102765}
(\byear{2023}).
\doiurl{10.1016/j.astropartphys.2022.102765}
\end{barticle}
\endbibitem

\bibitem{Takashima2022}
\begin{barticle}
\bauthor{\bsnm{{Takashima}}, \binits{S.}},
\bauthor{\bsnm{{Odaka}}, \binits{H.}},
\bauthor{\bsnm{{Yoneda}}, \binits{H.}},
\bauthor{\bsnm{{Ichinohe}}, \binits{Y.}},
\bauthor{\bsnm{{Bamba}}, \binits{A.}},
\bauthor{\bsnm{{Aramaki}}, \binits{T.}},
\bauthor{\bsnm{{Inoue}}, \binits{Y.}}:
\batitle{{Event reconstruction of Compton telescopes using a multi-task neural
  network}}.
\bjtitle{Nuclear Instruments and Methods in Physics Research A}
\bvolume{1038},
\bfpage{166897}
(\byear{2022})
{\href{https://arxiv.org/abs/2205.08082}{{arXiv:2205.08082}}}
{[astro-ph.IM]}.
\doiurl{10.1016/j.nima.2022.166897}
\end{barticle}
\endbibitem

\bibitem{Gruber1999}
\begin{barticle}
\bauthor{\bsnm{{Gruber}}, \binits{D.E.}},
\bauthor{\bsnm{{Matteson}}, \binits{J.L.}},
\bauthor{\bsnm{{Peterson}}, \binits{L.E.}},
\bauthor{\bsnm{{Jung}}, \binits{G.V.}}:
\batitle{{The Spectrum of Diffuse Cosmic Hard X-Rays Measured with HEAO 1}}.
\bjtitle{\apj}
\bvolume{520}(\bissue{1}),
\bfpage{124}--\blpage{129}
(\byear{1999})
{\href{https://arxiv.org/abs/astro-ph/9903492}{{arXiv:astro-ph/9903492}}}
{[astro-ph]}.
\doiurl{10.1086/307450}
\end{barticle}
\endbibitem

\bibitem{Mizuno2004}
\begin{barticle}
\bauthor{\bsnm{{Mizuno}}, \binits{T.}},
\bauthor{\bsnm{{Kamae}}, \binits{T.}},
\bauthor{\bsnm{{Godfrey}}, \binits{G.}},
\bauthor{\bsnm{{Handa}}, \binits{T.}},
\bauthor{\bsnm{{Thompson}}, \binits{D.J.}},
\bauthor{\bsnm{{Lauben}}, \binits{D.}},
\bauthor{\bsnm{{Fukazawa}}, \binits{Y.}},
\bauthor{\bsnm{{Ozaki}}, \binits{M.}}:
\batitle{{Cosmic-Ray Background Flux Model Based on a Gamma-Ray Large Area
  Space Telescope Balloon Flight Engineering Model}}.
\bjtitle{\apj}
\bvolume{614}(\bissue{2}),
\bfpage{1113}--\blpage{1123}
(\byear{2004})
{\href{https://arxiv.org/abs/astro-ph/0406684}{{arXiv:astro-ph/0406684}}}
{[astro-ph]}.
\doiurl{10.1086/423801}
\end{barticle}
\endbibitem

\bibitem{Tuerler2010}
\begin{barticle}
\bauthor{\bsnm{{T{\"u}rler}}, \binits{M.}},
\bauthor{\bsnm{{Chernyakova}}, \binits{M.}},
\bauthor{\bsnm{{Courvoisier}}, \binits{T.J.-L.}},
\bauthor{\bsnm{{Lubi{\'n}ski}}, \binits{P.}},
\bauthor{\bsnm{{Neronov}}, \binits{A.}},
\bauthor{\bsnm{{Produit}}, \binits{N.}},
\bauthor{\bsnm{{Walter}}, \binits{R.}}:
\batitle{{INTEGRAL hard X-ray spectra of the cosmic X-ray background and
  Galactic ridge emission}}.
\bjtitle{\aap}
\bvolume{512},
\bfpage{49}
(\byear{2010})
{\href{https://arxiv.org/abs/1001.2110}{{arXiv:1001.2110}}}
{[astro-ph.CO]}.
\doiurl{10.1051/0004-6361/200913072}
\end{barticle}
\endbibitem

\bibitem{Abdo2010}
\begin{barticle}
\bauthor{\bsnm{{Abdo}}, \binits{A.A.}},
\bauthor{\bsnm{{Ackermann}}, \binits{M.}},
\bauthor{\bsnm{{Ajello}}, \binits{M.}},
\bauthor{\bsnm{{Atwood}}, \binits{W.B.}},
\bauthor{\bsnm{{Baldini}}, \binits{L.}},
\bauthor{\bsnm{{Ballet}}, \binits{J.}},
\bauthor{\bsnm{{Barbiellini}}, \binits{G.}},
\bauthor{\bsnm{{Bastieri}}, \binits{D.}},
\bauthor{\bsnm{{Baughman}}, \binits{B.M.}},
\bauthor{\bsnm{{Bechtol}}, \binits{K.}},
\bauthor{\bsnm{{Bellazzini}}, \binits{R.}},
\bauthor{\bsnm{{Berenji}}, \binits{B.}},
\bauthor{\bsnm{{Blandford}}, \binits{R.D.}},
\bauthor{\bsnm{{Bloom}}, \binits{E.D.}},
\bauthor{\bsnm{{Bonamente}}, \binits{E.}},
\bauthor{\bsnm{{Borgland}}, \binits{A.W.}},
\bauthor{\bsnm{{Bregeon}}, \binits{J.}},
\bauthor{\bsnm{{Brez}}, \binits{A.}},
\bauthor{\bsnm{{Brigida}}, \binits{M.}},
\bauthor{\bsnm{{Bruel}}, \binits{P.}},
\bauthor{\bsnm{{Burnett}}, \binits{T.H.}},
\bauthor{\bsnm{{Buson}}, \binits{S.}},
\bauthor{\bsnm{{Caliandro}}, \binits{G.A.}},
\bauthor{\bsnm{{Cameron}}, \binits{R.A.}},
\bauthor{\bsnm{{Caraveo}}, \binits{P.A.}},
\bauthor{\bsnm{{Casandjian}}, \binits{J.M.}},
\bauthor{\bsnm{{Cavazzuti}}, \binits{E.}},
\bauthor{\bsnm{{Cecchi}}, \binits{C.}},
\bauthor{\bsnm{{{\c{C}}elik}}, \binits{{\"O}.}},
\bauthor{\bsnm{{Charles}}, \binits{E.}},
\bauthor{\bsnm{{Chekhtman}}, \binits{A.}},
\bauthor{\bsnm{{Cheung}}, \binits{C.C.}},
\bauthor{\bsnm{{Chiang}}, \binits{J.}},
\bauthor{\bsnm{{Ciprini}}, \binits{S.}},
\bauthor{\bsnm{{Claus}}, \binits{R.}},
\bauthor{\bsnm{{Cohen-Tanugi}}, \binits{J.}},
\bauthor{\bsnm{{Cominsky}}, \binits{L.R.}},
\bauthor{\bsnm{{Conrad}}, \binits{J.}},
\bauthor{\bsnm{{Cutini}}, \binits{S.}},
\bauthor{\bsnm{{Dermer}}, \binits{C.D.}},
\bauthor{\bsnm{{de Angelis}}, \binits{A.}},
\bauthor{\bsnm{{de Palma}}, \binits{F.}},
\bauthor{\bsnm{{Digel}}, \binits{S.W.}},
\bauthor{\bsnm{{di Bernardo}}, \binits{G.}},
\bauthor{\bsnm{{do Couto e Silva}}, \binits{E.}},
\bauthor{\bsnm{{Drell}}, \binits{P.S.}},
\bauthor{\bsnm{{Drlica-Wagner}}, \binits{A.}},
\bauthor{\bsnm{{Dubois}}, \binits{R.}},
\bauthor{\bsnm{{Dumora}}, \binits{D.}},
\bauthor{\bsnm{{Farnier}}, \binits{C.}},
\bauthor{\bsnm{{Favuzzi}}, \binits{C.}},
\bauthor{\bsnm{{Fegan}}, \binits{S.J.}},
\bauthor{\bsnm{{Focke}}, \binits{W.B.}},
\bauthor{\bsnm{{Fortin}}, \binits{P.}},
\bauthor{\bsnm{{Frailis}}, \binits{M.}},
\bauthor{\bsnm{{Fukazawa}}, \binits{Y.}},
\bauthor{\bsnm{{Funk}}, \binits{S.}},
\bauthor{\bsnm{{Fusco}}, \binits{P.}},
\bauthor{\bsnm{{Gaggero}}, \binits{D.}},
\bauthor{\bsnm{{Gargano}}, \binits{F.}},
\bauthor{\bsnm{{Gasparrini}}, \binits{D.}},
\bauthor{\bsnm{{Gehrels}}, \binits{N.}},
\bauthor{\bsnm{{Germani}}, \binits{S.}},
\bauthor{\bsnm{{Giebels}}, \binits{B.}},
\bauthor{\bsnm{{Giglietto}}, \binits{N.}},
\bauthor{\bsnm{{Giommi}}, \binits{P.}},
\bauthor{\bsnm{{Giordano}}, \binits{F.}},
\bauthor{\bsnm{{Glanzman}}, \binits{T.}},
\bauthor{\bsnm{{Godfrey}}, \binits{G.}},
\bauthor{\bsnm{{Grenier}}, \binits{I.A.}},
\bauthor{\bsnm{{Grondin}}, \binits{M.-H.}},
\bauthor{\bsnm{{Grove}}, \binits{J.E.}},
\bauthor{\bsnm{{Guillemot}}, \binits{L.}},
\bauthor{\bsnm{{Guiriec}}, \binits{S.}},
\bauthor{\bsnm{{Gustafsson}}, \binits{M.}},
\bauthor{\bsnm{{Hanabata}}, \binits{Y.}},
\bauthor{\bsnm{{Harding}}, \binits{A.K.}},
\bauthor{\bsnm{{Hayashida}}, \binits{M.}},
\bauthor{\bsnm{{Hughes}}, \binits{R.E.}},
\bauthor{\bsnm{{Itoh}}, \binits{R.}},
\bauthor{\bsnm{{Jackson}}, \binits{M.S.}},
\bauthor{\bsnm{{J{\'o}hannesson}}, \binits{G.}},
\bauthor{\bsnm{{Johnson}}, \binits{A.S.}},
\bauthor{\bsnm{{Johnson}}, \binits{R.P.}},
\bauthor{\bsnm{{Johnson}}, \binits{T.J.}},
\bauthor{\bsnm{{Johnson}}, \binits{W.N.}},
\bauthor{\bsnm{{Kamae}}, \binits{T.}},
\bauthor{\bsnm{{Katagiri}}, \binits{H.}},
\bauthor{\bsnm{{Kataoka}}, \binits{J.}},
\bauthor{\bsnm{{Kawai}}, \binits{N.}},
\bauthor{\bsnm{{Kerr}}, \binits{M.}},
\bauthor{\bsnm{{Kn{\"o}dlseder}}, \binits{J.}},
\bauthor{\bsnm{{Kocian}}, \binits{M.L.}},
\bauthor{\bsnm{{Kuehn}}, \binits{F.}},
\bauthor{\bsnm{{Kuss}}, \binits{M.}},
\bauthor{\bsnm{{Lande}}, \binits{J.}},
\bauthor{\bsnm{{Latronico}}, \binits{L.}},
\bauthor{\bsnm{{Lemoine-Goumard}}, \binits{M.}},
\bauthor{\bsnm{{Longo}}, \binits{F.}},
\bauthor{\bsnm{{Loparco}}, \binits{F.}},
\bauthor{\bsnm{{Lott}}, \binits{B.}},
\bauthor{\bsnm{{Lovellette}}, \binits{M.N.}},
\bauthor{\bsnm{{Lubrano}}, \binits{P.}},
\bauthor{\bsnm{{Madejski}}, \binits{G.M.}},
\bauthor{\bsnm{{Makeev}}, \binits{A.}},
\bauthor{\bsnm{{Mazziotta}}, \binits{M.N.}},
\bauthor{\bsnm{{McConville}}, \binits{W.}},
\bauthor{\bsnm{{McEnery}}, \binits{J.E.}},
\bauthor{\bsnm{{Meurer}}, \binits{C.}},
\bauthor{\bsnm{{Michelson}}, \binits{P.F.}},
\bauthor{\bsnm{{Mitthumsiri}}, \binits{W.}},
\bauthor{\bsnm{{Mizuno}}, \binits{T.}},
\bauthor{\bsnm{{Moiseev}}, \binits{A.A.}},
\bauthor{\bsnm{{Monte}}, \binits{C.}},
\bauthor{\bsnm{{Monzani}}, \binits{M.E.}},
\bauthor{\bsnm{{Morselli}}, \binits{A.}},
\bauthor{\bsnm{{Moskalenko}}, \binits{I.V.}},
\bauthor{\bsnm{{Murgia}}, \binits{S.}},
\bauthor{\bsnm{{Nolan}}, \binits{P.L.}},
\bauthor{\bsnm{{Norris}}, \binits{J.P.}},
\bauthor{\bsnm{{Nuss}}, \binits{E.}},
\bauthor{\bsnm{{Ohsugi}}, \binits{T.}},
\bauthor{\bsnm{{Omodei}}, \binits{N.}},
\bauthor{\bsnm{{Orlando}}, \binits{E.}},
\bauthor{\bsnm{{Ormes}}, \binits{J.F.}},
\bauthor{\bsnm{{Paneque}}, \binits{D.}},
\bauthor{\bsnm{{Panetta}}, \binits{J.H.}},
\bauthor{\bsnm{{Parent}}, \binits{D.}},
\bauthor{\bsnm{{Pelassa}}, \binits{V.}},
\bauthor{\bsnm{{Pepe}}, \binits{M.}},
\bauthor{\bsnm{{Pesce-Rollins}}, \binits{M.}},
\bauthor{\bsnm{{Piron}}, \binits{F.}},
\bauthor{\bsnm{{Porter}}, \binits{T.A.}},
\bauthor{\bsnm{{Rain{\`o}}}, \binits{S.}},
\bauthor{\bsnm{{Rando}}, \binits{R.}},
\bauthor{\bsnm{{Razzano}}, \binits{M.}},
\bauthor{\bsnm{{Reimer}}, \binits{A.}},
\bauthor{\bsnm{{Reimer}}, \binits{O.}},
\bauthor{\bsnm{{Reposeur}}, \binits{T.}},
\bauthor{\bsnm{{Ritz}}, \binits{S.}},
\bauthor{\bsnm{{Rochester}}, \binits{L.S.}},
\bauthor{\bsnm{{Rodriguez}}, \binits{A.Y.}},
\bauthor{\bsnm{{Roth}}, \binits{M.}},
\bauthor{\bsnm{{Ryde}}, \binits{F.}},
\bauthor{\bsnm{{Sadrozinski}}, \binits{H.F.-W.}},
\bauthor{\bsnm{{Sanchez}}, \binits{D.}},
\bauthor{\bsnm{{Sander}}, \binits{A.}},
\bauthor{\bsnm{{Parkinson}}, \binits{P.M.S.}},
\bauthor{\bsnm{{Scargle}}, \binits{J.D.}},
\bauthor{\bsnm{{Sellerholm}}, \binits{A.}},
\bauthor{\bsnm{{Sgr{\`o}}}, \binits{C.}},
\bauthor{\bsnm{{Shaw}}, \binits{M.S.}},
\bauthor{\bsnm{{Siskind}}, \binits{E.J.}},
\bauthor{\bsnm{{Smith}}, \binits{D.A.}},
\bauthor{\bsnm{{Smith}}, \binits{P.D.}},
\bauthor{\bsnm{{Spandre}}, \binits{G.}},
\bauthor{\bsnm{{Spinelli}}, \binits{P.}},
\bauthor{\bsnm{{Starck}}, \binits{J.-L.}},
\bauthor{\bsnm{{Strickman}}, \binits{M.S.}},
\bauthor{\bsnm{{Strong}}, \binits{A.W.}},
\bauthor{\bsnm{{Suson}}, \binits{D.J.}},
\bauthor{\bsnm{{Tajima}}, \binits{H.}},
\bauthor{\bsnm{{Takahashi}}, \binits{H.}},
\bauthor{\bsnm{{Takahashi}}, \binits{T.}},
\bauthor{\bsnm{{Tanaka}}, \binits{T.}},
\bauthor{\bsnm{{Thayer}}, \binits{J.B.}},
\bauthor{\bsnm{{Thayer}}, \binits{J.G.}},
\bauthor{\bsnm{{Thompson}}, \binits{D.J.}},
\bauthor{\bsnm{{Tibaldo}}, \binits{L.}},
\bauthor{\bsnm{{Torres}}, \binits{D.F.}},
\bauthor{\bsnm{{Tosti}}, \binits{G.}},
\bauthor{\bsnm{{Tramacere}}, \binits{A.}},
\bauthor{\bsnm{{Uchiyama}}, \binits{Y.}},
\bauthor{\bsnm{{Usher}}, \binits{T.L.}},
\bauthor{\bsnm{{Vasileiou}}, \binits{V.}},
\bauthor{\bsnm{{Vilchez}}, \binits{N.}},
\bauthor{\bsnm{{Vitale}}, \binits{V.}},
\bauthor{\bsnm{{Waite}}, \binits{A.P.}},
\bauthor{\bsnm{{Wang}}, \binits{P.}},
\bauthor{\bsnm{{Winer}}, \binits{B.L.}},
\bauthor{\bsnm{{Wood}}, \binits{K.S.}},
\bauthor{\bsnm{{Ylinen}}, \binits{T.}},
\bauthor{\bsnm{{Ziegler}}, \binits{M.}},
\bauthor{\bsnm{{Fermi LAT Collaboration}}}:
\batitle{{Spectrum of the Isotropic Diffuse Gamma-Ray Emission Derived from
  First-Year Fermi Large Area Telescope Data}}.
\bjtitle{\prl}
\bvolume{104}(\bissue{10}),
\bfpage{101101}
(\byear{2010})
{\href{https://arxiv.org/abs/1002.3603}{{arXiv:1002.3603}}}
{[astro-ph.HE]}.
\doiurl{10.1103/PhysRevLett.104.101101}
\end{barticle}
\endbibitem

\bibitem{Alcaraz2000}
\begin{barticle}
\bauthor{\bsnm{{Alcaraz}}, \binits{J.}},
\bauthor{\bsnm{{Alpat}}, \binits{B.}},
\bauthor{\bsnm{{Ambrosi}}, \binits{G.}},
\bauthor{\bsnm{{Anderhub}}, \binits{H.}},
\bauthor{\bsnm{{Ao}}, \binits{L.}},
\bauthor{\bsnm{{Arefiev}}, \binits{A.}},
\bauthor{\bsnm{{Azzarello}}, \binits{P.}},
\bauthor{\bsnm{{Babucci}}, \binits{E.}},
\bauthor{\bsnm{{Baldini}}, \binits{L.}},
\bauthor{\bsnm{{Basile}}, \binits{M.}},
\bauthor{\bsnm{{Barancourt}}, \binits{D.}},
\bauthor{\bsnm{{Barao}}, \binits{F.}},
\bauthor{\bsnm{{Barbier}}, \binits{G.}},
\bauthor{\bsnm{{Barreira}}, \binits{G.}},
\bauthor{\bsnm{{Battiston}}, \binits{R.}},
\bauthor{\bsnm{{Becker}}, \binits{R.}},
\bauthor{\bsnm{{Becker}}, \binits{U.}},
\bauthor{\bsnm{{Bellagamba}}, \binits{L.}},
\bauthor{\bsnm{{B{\'e}n{\'e}}}, \binits{P.}},
\bauthor{\bsnm{{Berdugo}}, \binits{J.}},
\bauthor{\bsnm{{Berges}}, \binits{P.}},
\bauthor{\bsnm{{Bertucci}}, \binits{B.}},
\bauthor{\bsnm{{Biland}}, \binits{A.}},
\bauthor{\bsnm{{Bizzaglia}}, \binits{S.}},
\bauthor{\bsnm{{Blasko}}, \binits{S.}},
\bauthor{\bsnm{{Boella}}, \binits{G.}},
\bauthor{\bsnm{{Boschini}}, \binits{M.}},
\bauthor{\bsnm{{Bourquin}}, \binits{M.}},
\bauthor{\bsnm{{Brocco}}, \binits{L.}},
\bauthor{\bsnm{{Bruni}}, \binits{G.}},
\bauthor{\bsnm{{Buenerd}}, \binits{M.}},
\bauthor{\bsnm{{Burger}}, \binits{J.D.}},
\bauthor{\bsnm{{Burger}}, \binits{W.J.}},
\bauthor{\bsnm{{Cai}}, \binits{X.D.}},
\bauthor{\bsnm{{Camps}}, \binits{C.}},
\bauthor{\bsnm{{Cannarsa}}, \binits{P.}},
\bauthor{\bsnm{{Capell}}, \binits{M.}},
\bauthor{\bsnm{{Casadei}}, \binits{D.}},
\bauthor{\bsnm{{Casaus}}, \binits{J.}},
\bauthor{\bsnm{{Castellini}}, \binits{G.}},
\bauthor{\bsnm{{Cecchi}}, \binits{C.}},
\bauthor{\bsnm{{Chang}}, \binits{Y.H.}},
\bauthor{\bsnm{{Chen}}, \binits{H.F.}},
\bauthor{\bsnm{{Chen}}, \binits{H.S.}},
\bauthor{\bsnm{{Chen}}, \binits{Z.G.}},
\bauthor{\bsnm{{Chernoplekov}}, \binits{N.A.}},
\bauthor{\bsnm{{Chiueh}}, \binits{T.H.}},
\bauthor{\bsnm{{Chuang}}, \binits{Y.L.}},
\bauthor{\bsnm{{Cindolo}}, \binits{F.}},
\bauthor{\bsnm{{Commichau}}, \binits{V.}},
\bauthor{\bsnm{{Contin}}, \binits{A.}},
\bauthor{\bsnm{{Crespo}}, \binits{P.}},
\bauthor{\bsnm{{Cristinziani}}, \binits{M.}},
\bauthor{\bsnm{{da Cunha}}, \binits{J.P.}},
\bauthor{\bsnm{{Dai}}, \binits{T.S.}},
\bauthor{\bsnm{{Deus}}, \binits{J.D.}},
\bauthor{\bsnm{{Dinu}}, \binits{N.}},
\bauthor{\bsnm{{Djambazov}}, \binits{L.}},
\bauthor{\bsnm{{D'Antone}}, \binits{I.}},
\bauthor{\bsnm{{Dong}}, \binits{Z.R.}},
\bauthor{\bsnm{{Emonet}}, \binits{P.}},
\bauthor{\bsnm{{Engelberg}}, \binits{J.}},
\bauthor{\bsnm{{Eppling}}, \binits{F.J.}},
\bauthor{\bsnm{{Eronen}}, \binits{T.}},
\bauthor{\bsnm{{Esposito}}, \binits{G.}},
\bauthor{\bsnm{{Extermann}}, \binits{P.}},
\bauthor{\bsnm{{Favier}}, \binits{J.}},
\bauthor{\bsnm{{Fiandrini}}, \binits{E.}},
\bauthor{\bsnm{{Fisher}}, \binits{P.H.}},
\bauthor{\bsnm{{Fluegge}}, \binits{G.}},
\bauthor{\bsnm{{Fouque}}, \binits{N.}},
\bauthor{\bsnm{{Galaktionov}}, \binits{Y.}},
\bauthor{\bsnm{{Gervasi}}, \binits{M.}},
\bauthor{\bsnm{{Giusti}}, \binits{P.}},
\bauthor{\bsnm{{Grandi}}, \binits{D.}},
\bauthor{\bsnm{{Grimm}}, \binits{O.}},
\bauthor{\bsnm{{Gu}}, \binits{W.Q.}},
\bauthor{\bsnm{{Hangarter}}, \binits{K.}},
\bauthor{\bsnm{{Hasan}}, \binits{A.}},
\bauthor{\bsnm{{Hermel}}, \binits{V.}},
\bauthor{\bsnm{{Hofer}}, \binits{H.}},
\bauthor{\bsnm{{Huang}}, \binits{M.A.}},
\bauthor{\bsnm{{Hungerford}}, \binits{W.}},
\bauthor{\bsnm{{Ionica}}, \binits{M.}},
\bauthor{\bsnm{{Ionica}}, \binits{R.}},
\bauthor{\bsnm{{Jongmanns}}, \binits{M.}},
\bauthor{\bsnm{{Karlamaa}}, \binits{K.}},
\bauthor{\bsnm{{Karpinski}}, \binits{W.}},
\bauthor{\bsnm{{Kenney}}, \binits{G.}},
\bauthor{\bsnm{{Kenny}}, \binits{J.}},
\bauthor{\bsnm{{Kim}}, \binits{W.}},
\bauthor{\bsnm{{Klimentov}}, \binits{A.}},
\bauthor{\bsnm{{Kossakowski}}, \binits{R.}},
\bauthor{\bsnm{{Koutsenko}}, \binits{V.}},
\bauthor{\bsnm{{Kraeber}}, \binits{M.}},
\bauthor{\bsnm{{Laborie}}, \binits{G.}},
\bauthor{\bsnm{{Laitinen}}, \binits{T.}},
\bauthor{\bsnm{{Lamanna}}, \binits{G.}},
\bauthor{\bsnm{{Laurenti}}, \binits{G.}},
\bauthor{\bsnm{{Lebedev}}, \binits{A.}},
\bauthor{\bsnm{{Lee}}, \binits{S.C.}},
\bauthor{\bsnm{{Levi}}, \binits{G.}},
\bauthor{\bsnm{{Levtchenko}}, \binits{P.}},
\bauthor{\bsnm{{Liu}}, \binits{C.L.}},
\bauthor{\bsnm{{Liu}}, \binits{H.T.}},
\bauthor{\bsnm{{Lopes}}, \binits{I.}},
\bauthor{\bsnm{{Lu}}, \binits{G.}},
\bauthor{\bsnm{{Lu}}, \binits{Y.S.}},
\bauthor{\bsnm{{L{\"u}belsmeyer}}, \binits{K.}},
\bauthor{\bsnm{{Luckey}}, \binits{D.}},
\bauthor{\bsnm{{Lustermann}}, \binits{W.}},
\bauthor{\bsnm{{Ma{\~n}a}}, \binits{C.}},
\bauthor{\bsnm{{Margotti}}, \binits{A.}},
\bauthor{\bsnm{{Mayet}}, \binits{F.}},
\bauthor{\bsnm{{McNeil}}, \binits{R.R.}},
\bauthor{\bsnm{{Meillon}}, \binits{B.}},
\bauthor{\bsnm{{Menichelli}}, \binits{M.}},
\bauthor{\bsnm{{Mihul}}, \binits{A.}},
\bauthor{\bsnm{{Mourao}}, \binits{A.}},
\bauthor{\bsnm{{Mujunen}}, \binits{A.}},
\bauthor{\bsnm{{Palmonari}}, \binits{F.}},
\bauthor{\bsnm{{Papi}}, \binits{A.}},
\bauthor{\bsnm{{Park}}, \binits{I.H.}},
\bauthor{\bsnm{{Pauluzzi}}, \binits{M.}},
\bauthor{\bsnm{{Pauss}}, \binits{F.}},
\bauthor{\bsnm{{Perrin}}, \binits{E.}},
\bauthor{\bsnm{{Pesci}}, \binits{A.}},
\bauthor{\bsnm{{Pevsner}}, \binits{A.}},
\bauthor{\bsnm{{Pimenta}}, \binits{M.}},
\bauthor{\bsnm{{Plyaskin}}, \binits{V.}},
\bauthor{\bsnm{{Pojidaev}}, \binits{V.}},
\bauthor{\bsnm{{Postolache}}, \binits{V.}},
\bauthor{\bsnm{{Produit}}, \binits{N.}},
\bauthor{\bsnm{{Rancoita}}, \binits{P.G.}},
\bauthor{\bsnm{{Rapin}}, \binits{D.}},
\bauthor{\bsnm{{Raupach}}, \binits{F.}},
\bauthor{\bsnm{{Ren}}, \binits{D.}},
\bauthor{\bsnm{{Ren}}, \binits{Z.}},
\bauthor{\bsnm{{Ribordy}}, \binits{M.}},
\bauthor{\bsnm{{Richeux}}, \binits{J.P.}},
\bauthor{\bsnm{{Riihonen}}, \binits{E.}},
\bauthor{\bsnm{{Ritakari}}, \binits{J.}},
\bauthor{\bsnm{{Roeser}}, \binits{U.}},
\bauthor{\bsnm{{Roissin}}, \binits{C.}},
\bauthor{\bsnm{{Sagdeev}}, \binits{R.}},
\bauthor{\bsnm{{Sartorelli}}, \binits{G.}},
\bauthor{\bsnm{{Schultz von Dratzig}}, \binits{A.}},
\bauthor{\bsnm{{Schwering}}, \binits{G.}},
\bauthor{\bsnm{{Scolieri}}, \binits{G.}},
\bauthor{\bsnm{{Seo}}, \binits{E.S.}},
\bauthor{\bsnm{{Shoutko}}, \binits{V.}},
\bauthor{\bsnm{{Shoumilov}}, \binits{E.}},
\bauthor{\bsnm{{Siedling}}, \binits{R.}},
\bauthor{\bsnm{{Son}}, \binits{D.}},
\bauthor{\bsnm{{Song}}, \binits{T.}},
\bauthor{\bsnm{{Steuer}}, \binits{M.}},
\bauthor{\bsnm{{Sun}}, \binits{G.S.}},
\bauthor{\bsnm{{Suter}}, \binits{H.}},
\bauthor{\bsnm{{Tang}}, \binits{X.W.}},
\bauthor{\bsnm{{Ting}}, \binits{S.C.C.}},
\bauthor{\bsnm{{Ting}}, \binits{S.M.}},
\bauthor{\bsnm{{Tornikoski}}, \binits{M.}},
\bauthor{\bsnm{{Torsti}}, \binits{J.}},
\bauthor{\bsnm{{Tr{\"u}mper}}, \binits{J.}},
\bauthor{\bsnm{{Ulbricht}}, \binits{J.}},
\bauthor{\bsnm{{Urpo}}, \binits{S.}},
\bauthor{\bsnm{{Usoskin}}, \binits{I.}},
\bauthor{\bsnm{{Valtonen}}, \binits{E.}},
\bauthor{\bsnm{{Vandenhirtz}}, \binits{J.}},
\bauthor{\bsnm{{Velcea}}, \binits{F.}},
\bauthor{\bsnm{{Velikhov}}, \binits{E.}},
\bauthor{\bsnm{{Verlaat}}, \binits{B.}},
\bauthor{\bsnm{{Vetlitsky}}, \binits{I.}},
\bauthor{\bsnm{{Vezzu}}, \binits{F.}},
\bauthor{\bsnm{{Vialle}}, \binits{J.P.}},
\bauthor{\bsnm{{Viertel}}, \binits{G.}},
\bauthor{\bsnm{{Vit{\'e}}}, \binits{D.}},
\bauthor{\bsnm{{Von Gunten}}, \binits{H.}},
\bauthor{\bsnm{{Waldmeier Wicki}}, \binits{S.}},
\bauthor{\bsnm{{Wallraff}}, \binits{W.}},
\bauthor{\bsnm{{Wang}}, \binits{B.C.}},
\bauthor{\bsnm{{Wang}}, \binits{J.Z.}},
\bauthor{\bsnm{{Wang}}, \binits{Y.H.}},
\bauthor{\bsnm{{Wiik}}, \binits{K.}},
\bauthor{\bsnm{{Williams}}, \binits{C.}},
\bauthor{\bsnm{{Wu}}, \binits{S.X.}},
\bauthor{\bsnm{{Xia}}, \binits{P.C.}},
\bauthor{\bsnm{{Yan}}, \binits{J.L.}},
\bauthor{\bsnm{{Yan}}, \binits{L.G.}},
\bauthor{\bsnm{{Yang}}, \binits{C.G.}},
\bauthor{\bsnm{{Yang}}, \binits{M.}},
\bauthor{\bsnm{{Ye}}, \binits{S.W.}},
\bauthor{\bsnm{{Yeh}}, \binits{P.}},
\bauthor{\bsnm{{Xu}}, \binits{Z.Z.}},
\bauthor{\bsnm{{Zhang}}, \binits{H.Y.}},
\bauthor{\bsnm{{Zhang}}, \binits{Z.P.}},
\bauthor{\bsnm{{Zhao}}, \binits{D.X.}},
\bauthor{\bsnm{{Zhu}}, \binits{G.Y.}},
\bauthor{\bsnm{{Zhu}}, \binits{W.Z.}},
\bauthor{\bsnm{{Zhuang}}, \binits{H.L.}},
\bauthor{\bsnm{{Zichichi}}, \binits{A.}},
\bauthor{\bsnm{{Zimmermann}}, \binits{B.}}:
\batitle{{Leptons in near earth orbit}}.
\bjtitle{Physics Letters B}
\bvolume{484}(\bissue{1-2}),
\bfpage{10}--\blpage{22}
(\byear{2000}).
\doiurl{10.1016/S0370-2693(00)00588-8}
\end{barticle}
\endbibitem

\bibitem{Kole2015}
\begin{barticle}
\bauthor{\bsnm{{Kole}}, \binits{M.}},
\bauthor{\bsnm{{Pearce}}, \binits{M.}},
\bauthor{\bsnm{{Mu{\~n}oz Salinas}}, \binits{M.}}:
\batitle{{A model of the cosmic ray induced atmospheric neutron environment}}.
\bjtitle{Astroparticle Physics}
\bvolume{62},
\bfpage{230}--\blpage{240}
(\byear{2015})
{\href{https://arxiv.org/abs/1410.1364}{{arXiv:1410.1364}}}
{[astro-ph.IM]}.
\doiurl{10.1016/j.astropartphys.2014.10.002}
\end{barticle}
\endbibitem

\bibitem{Schoenfelder1993}
\begin{barticle}
\bauthor{\bsnm{{Schoenfelder}}, \binits{V.}},
\bauthor{\bsnm{{Aarts}}, \binits{H.}},
\bauthor{\bsnm{{Bennett}}, \binits{K.}},
\bauthor{\bsnm{{de Boer}}, \binits{H.}},
\bauthor{\bsnm{{Clear}}, \binits{J.}},
\bauthor{\bsnm{{Collmar}}, \binits{W.}},
\bauthor{\bsnm{{Connors}}, \binits{A.}},
\bauthor{\bsnm{{Deerenberg}}, \binits{A.}},
\bauthor{\bsnm{{Diehl}}, \binits{R.}},
\bauthor{\bsnm{{von Dordrecht}}, \binits{A.}},
\bauthor{\bsnm{{den Herder}}, \binits{J.W.}},
\bauthor{\bsnm{{Hermsen}}, \binits{W.}},
\bauthor{\bsnm{{Kippen}}, \binits{M.}},
\bauthor{\bsnm{{Kuiper}}, \binits{L.}},
\bauthor{\bsnm{{Lichti}}, \binits{G.}},
\bauthor{\bsnm{{Lockwood}}, \binits{J.}},
\bauthor{\bsnm{{Macri}}, \binits{J.}},
\bauthor{\bsnm{{McConnell}}, \binits{M.}},
\bauthor{\bsnm{{Morris}}, \binits{D.}},
\bauthor{\bsnm{{Much}}, \binits{R.}},
\bauthor{\bsnm{{Ryan}}, \binits{J.}},
\bauthor{\bsnm{{Simpson}}, \binits{G.}},
\bauthor{\bsnm{{Snelling}}, \binits{M.}},
\bauthor{\bsnm{{Stacy}}, \binits{G.}},
\bauthor{\bsnm{{Steinle}}, \binits{H.}},
\bauthor{\bsnm{{Strong}}, \binits{A.}},
\bauthor{\bsnm{{Swanenburg}}, \binits{B.N.}},
\bauthor{\bsnm{{Taylor}}, \binits{B.}},
\bauthor{\bsnm{{de Vries}}, \binits{C.}},
\bauthor{\bsnm{{Winkler}}, \binits{C.}}:
\batitle{{Instrument Description and Performance of the Imaging Gamma-Ray
  Telescope COMPTEL aboard the Compton Gamma-Ray Observatory}}.
\bjtitle{\apjs}
\bvolume{86},
\bfpage{657}
(\byear{1993}).
\doiurl{10.1086/191794}
\end{barticle}
\endbibitem

\bibitem{Wang2009}
\begin{barticle}
\bauthor{\bsnm{{Wang}}, \binits{W.}},
\bauthor{\bsnm{{Lang}}, \binits{M.G.}},
\bauthor{\bsnm{{Diehl}}, \binits{R.}},
\bauthor{\bsnm{{Halloin}}, \binits{H.}},
\bauthor{\bsnm{{Jean}}, \binits{P.}},
\bauthor{\bsnm{{Kn{\"o}dlseder}}, \binits{J.}},
\bauthor{\bsnm{{Kretschmer}}, \binits{K.}},
\bauthor{\bsnm{{Martin}}, \binits{P.}},
\bauthor{\bsnm{{Roques}}, \binits{J.P.}},
\bauthor{\bsnm{{Strong}}, \binits{A.W.}},
\bauthor{\bsnm{{Winkler}}, \binits{C.}},
\bauthor{\bsnm{{Zhang}}, \binits{X.L.}}:
\batitle{{Spectral and intensity variations of Galactic $^{26}$Al emission}}.
\bjtitle{\aap}
\bvolume{496}(\bissue{3}),
\bfpage{713}--\blpage{724}
(\byear{2009})
{\href{https://arxiv.org/abs/0902.0211}{{arXiv:0902.0211}}}
{[astro-ph.HE]}.
\doiurl{10.1051/0004-6361/200811175}
\end{barticle}
\endbibitem

\bibitem{Diehl2015}
\begin{barticle}
\bauthor{\bsnm{{Diehl}}, \binits{R.}},
\bauthor{\bsnm{{Siegert}}, \binits{T.}},
\bauthor{\bsnm{{Hillebrandt}}, \binits{W.}},
\bauthor{\bsnm{{Krause}}, \binits{M.}},
\bauthor{\bsnm{{Greiner}}, \binits{J.}},
\bauthor{\bsnm{{Maeda}}, \binits{K.}},
\bauthor{\bsnm{{R{\"o}pke}}, \binits{F.K.}},
\bauthor{\bsnm{{Sim}}, \binits{S.A.}},
\bauthor{\bsnm{{Wang}}, \binits{W.}},
\bauthor{\bsnm{{Zhang}}, \binits{X.}}:
\batitle{{SN2014J gamma rays from the $^{56}$Ni decay chain}}.
\bjtitle{\aap}
\bvolume{574},
\bfpage{72}
(\byear{2015})
{\href{https://arxiv.org/abs/1409.5477}{{arXiv:1409.5477}}}
{[astro-ph.HE]}.
\doiurl{10.1051/0004-6361/201424991}
\end{barticle}
\endbibitem

\bibitem{Siegert2015}
\begin{barticle}
\bauthor{\bsnm{{Siegert}}, \binits{T.}},
\bauthor{\bsnm{{Diehl}}, \binits{R.}},
\bauthor{\bsnm{{Krause}}, \binits{M.G.H.}},
\bauthor{\bsnm{{Greiner}}, \binits{J.}}:
\batitle{{Revisiting INTEGRAL/SPI observations of $^{44}$Ti from Cassiopeia
  A}}.
\bjtitle{\aap}
\bvolume{579},
\bfpage{124}
(\byear{2015})
{\href{https://arxiv.org/abs/1505.05999}{{arXiv:1505.05999}}}
{[astro-ph.HE]}.
\doiurl{10.1051/0004-6361/201525877}
\end{barticle}
\endbibitem

\bibitem{Winkler2003}
\begin{barticle}
\bauthor{\bsnm{{Winkler}}, \binits{C.}},
\bauthor{\bsnm{{Courvoisier}}, \binits{T.J.-L.}},
\bauthor{\bsnm{{Di Cocco}}, \binits{G.}},
\bauthor{\bsnm{{Gehrels}}, \binits{N.}},
\bauthor{\bsnm{{Gim{\'e}nez}}, \binits{A.}},
\bauthor{\bsnm{{Grebenev}}, \binits{S.}},
\bauthor{\bsnm{{Hermsen}}, \binits{W.}},
\bauthor{\bsnm{{Mas-Hesse}}, \binits{J.M.}},
\bauthor{\bsnm{{Lebrun}}, \binits{F.}},
\bauthor{\bsnm{{Lund}}, \binits{N.}},
\bauthor{\bsnm{{Palumbo}}, \binits{G.G.C.}},
\bauthor{\bsnm{{Paul}}, \binits{J.}},
\bauthor{\bsnm{{Roques}}, \binits{J.-P.}},
\bauthor{\bsnm{{Schnopper}}, \binits{H.}},
\bauthor{\bsnm{{Sch{\"o}nfelder}}, \binits{V.}},
\bauthor{\bsnm{{Sunyaev}}, \binits{R.}},
\bauthor{\bsnm{{Teegarden}}, \binits{B.}},
\bauthor{\bsnm{{Ubertini}}, \binits{P.}},
\bauthor{\bsnm{{Vedrenne}}, \binits{G.}},
\bauthor{\bsnm{{Dean}}, \binits{A.J.}}:
\batitle{{The INTEGRAL mission}}.
\bjtitle{\aap}
\bvolume{411},
\bfpage{1}--\blpage{6}
(\byear{2003}).
\doiurl{10.1051/0004-6361:20031288}
\end{barticle}
\endbibitem

\bibitem{Harrison2013}
\begin{barticle}
\bauthor{\bsnm{{Harrison}}, \binits{F.A.}},
\bauthor{\bsnm{{Craig}}, \binits{W.W.}},
\bauthor{\bsnm{{Christensen}}, \binits{F.E.}},
\bauthor{\bsnm{{Hailey}}, \binits{C.J.}},
\bauthor{\bsnm{{Zhang}}, \binits{W.W.}},
\bauthor{\bsnm{{Boggs}}, \binits{S.E.}},
\bauthor{\bsnm{{Stern}}, \binits{D.}},
\bauthor{\bsnm{{Cook}}, \binits{W.R.}},
\bauthor{\bsnm{{Forster}}, \binits{K.}},
\bauthor{\bsnm{{Giommi}}, \binits{P.}},
\bauthor{\bsnm{{Grefenstette}}, \binits{B.W.}},
\bauthor{\bsnm{{Kim}}, \binits{Y.}},
\bauthor{\bsnm{{Kitaguchi}}, \binits{T.}},
\bauthor{\bsnm{{Koglin}}, \binits{J.E.}},
\bauthor{\bsnm{{Madsen}}, \binits{K.K.}},
\bauthor{\bsnm{{Mao}}, \binits{P.H.}},
\bauthor{\bsnm{{Miyasaka}}, \binits{H.}},
\bauthor{\bsnm{{Mori}}, \binits{K.}},
\bauthor{\bsnm{{Perri}}, \binits{M.}},
\bauthor{\bsnm{{Pivovaroff}}, \binits{M.J.}},
\bauthor{\bsnm{{Puccetti}}, \binits{S.}},
\bauthor{\bsnm{{Rana}}, \binits{V.R.}},
\bauthor{\bsnm{{Westergaard}}, \binits{N.J.}},
\bauthor{\bsnm{{Willis}}, \binits{J.}},
\bauthor{\bsnm{{Zoglauer}}, \binits{A.}},
\bauthor{\bsnm{{An}}, \binits{H.}},
\bauthor{\bsnm{{Bachetti}}, \binits{M.}},
\bauthor{\bsnm{{Barri{\`e}re}}, \binits{N.M.}},
\bauthor{\bsnm{{Bellm}}, \binits{E.C.}},
\bauthor{\bsnm{{Bhalerao}}, \binits{V.}},
\bauthor{\bsnm{{Brejnholt}}, \binits{N.F.}},
\bauthor{\bsnm{{Fuerst}}, \binits{F.}},
\bauthor{\bsnm{{Liebe}}, \binits{C.C.}},
\bauthor{\bsnm{{Markwardt}}, \binits{C.B.}},
\bauthor{\bsnm{{Nynka}}, \binits{M.}},
\bauthor{\bsnm{{Vogel}}, \binits{J.K.}},
\bauthor{\bsnm{{Walton}}, \binits{D.J.}},
\bauthor{\bsnm{{Wik}}, \binits{D.R.}},
\bauthor{\bsnm{{Alexander}}, \binits{D.M.}},
\bauthor{\bsnm{{Cominsky}}, \binits{L.R.}},
\bauthor{\bsnm{{Hornschemeier}}, \binits{A.E.}},
\bauthor{\bsnm{{Hornstrup}}, \binits{A.}},
\bauthor{\bsnm{{Kaspi}}, \binits{V.M.}},
\bauthor{\bsnm{{Madejski}}, \binits{G.M.}},
\bauthor{\bsnm{{Matt}}, \binits{G.}},
\bauthor{\bsnm{{Molendi}}, \binits{S.}},
\bauthor{\bsnm{{Smith}}, \binits{D.M.}},
\bauthor{\bsnm{{Tomsick}}, \binits{J.A.}},
\bauthor{\bsnm{{Ajello}}, \binits{M.}},
\bauthor{\bsnm{{Ballantyne}}, \binits{D.R.}},
\bauthor{\bsnm{{Balokovi{\'c}}}, \binits{M.}},
\bauthor{\bsnm{{Barret}}, \binits{D.}},
\bauthor{\bsnm{{Bauer}}, \binits{F.E.}},
\bauthor{\bsnm{{Blandford}}, \binits{R.D.}},
\bauthor{\bsnm{{Brandt}}, \binits{W.N.}},
\bauthor{\bsnm{{Brenneman}}, \binits{L.W.}},
\bauthor{\bsnm{{Chiang}}, \binits{J.}},
\bauthor{\bsnm{{Chakrabarty}}, \binits{D.}},
\bauthor{\bsnm{{Chenevez}}, \binits{J.}},
\bauthor{\bsnm{{Comastri}}, \binits{A.}},
\bauthor{\bsnm{{Dufour}}, \binits{F.}},
\bauthor{\bsnm{{Elvis}}, \binits{M.}},
\bauthor{\bsnm{{Fabian}}, \binits{A.C.}},
\bauthor{\bsnm{{Farrah}}, \binits{D.}},
\bauthor{\bsnm{{Fryer}}, \binits{C.L.}},
\bauthor{\bsnm{{Gotthelf}}, \binits{E.V.}},
\bauthor{\bsnm{{Grindlay}}, \binits{J.E.}},
\bauthor{\bsnm{{Helfand}}, \binits{D.J.}},
\bauthor{\bsnm{{Krivonos}}, \binits{R.}},
\bauthor{\bsnm{{Meier}}, \binits{D.L.}},
\bauthor{\bsnm{{Miller}}, \binits{J.M.}},
\bauthor{\bsnm{{Natalucci}}, \binits{L.}},
\bauthor{\bsnm{{Ogle}}, \binits{P.}},
\bauthor{\bsnm{{Ofek}}, \binits{E.O.}},
\bauthor{\bsnm{{Ptak}}, \binits{A.}},
\bauthor{\bsnm{{Reynolds}}, \binits{S.P.}},
\bauthor{\bsnm{{Rigby}}, \binits{J.R.}},
\bauthor{\bsnm{{Tagliaferri}}, \binits{G.}},
\bauthor{\bsnm{{Thorsett}}, \binits{S.E.}},
\bauthor{\bsnm{{Treister}}, \binits{E.}},
\bauthor{\bsnm{{Urry}}, \binits{C.M.}}:
\batitle{{The Nuclear Spectroscopic Telescope Array (NuSTAR) High-energy X-Ray
  Mission}}.
\bjtitle{\apj}
\bvolume{770}(\bissue{2}),
\bfpage{103}
(\byear{2013})
{\href{https://arxiv.org/abs/1301.7307}{{arXiv:1301.7307}}}
{[astro-ph.IM]}.
\doiurl{10.1088/0004-637X/770/2/103}
\end{barticle}
\endbibitem

\bibitem{Winkler1995}
\begin{barticle}
\bauthor{\bsnm{{Winkler}}, \binits{C.}}:
\batitle{{The INTEGRAL mission}}.
\bjtitle{Experimental Astronomy}
\bvolume{6}(\bissue{4}),
\bfpage{71}--\blpage{76}
(\byear{1995}).
\doiurl{10.1007/BF00419260}
\end{barticle}
\endbibitem

\bibitem{Lebrun2003}
\begin{barticle}
\bauthor{\bsnm{{Lebrun}}, \binits{F.}},
\bauthor{\bsnm{{Leray}}, \binits{J.P.}},
\bauthor{\bsnm{{Lavocat}}, \binits{P.}},
\bauthor{\bsnm{{Cr{\'e}tolle}}, \binits{J.}},
\bauthor{\bsnm{{Arqu{\`e}s}}, \binits{M.}},
\bauthor{\bsnm{{Blondel}}, \binits{C.}},
\bauthor{\bsnm{{Bonnin}}, \binits{C.}},
\bauthor{\bsnm{{Bou{\`e}re}}, \binits{A.}},
\bauthor{\bsnm{{Cara}}, \binits{C.}},
\bauthor{\bsnm{{Chaleil}}, \binits{T.}},
\bauthor{\bsnm{{Daly}}, \binits{F.}},
\bauthor{\bsnm{{Desages}}, \binits{F.}},
\bauthor{\bsnm{{Dzitko}}, \binits{H.}},
\bauthor{\bsnm{{Horeau}}, \binits{B.}},
\bauthor{\bsnm{{Laurent}}, \binits{P.}},
\bauthor{\bsnm{{Limousin}}, \binits{O.}},
\bauthor{\bsnm{{Mathy}}, \binits{F.}},
\bauthor{\bsnm{{Mauguen}}, \binits{V.}},
\bauthor{\bsnm{{Meignier}}, \binits{F.}},
\bauthor{\bsnm{{Molini{\'e}}}, \binits{F.}},
\bauthor{\bsnm{{Poindron}}, \binits{E.}},
\bauthor{\bsnm{{Rouger}}, \binits{M.}},
\bauthor{\bsnm{{Sauvageon}}, \binits{A.}},
\bauthor{\bsnm{{Tourrette}}, \binits{T.}}:
\batitle{{ISGRI: The INTEGRAL Soft Gamma-Ray Imager}}.
\bjtitle{\aap}
\bvolume{411},
\bfpage{141}--\blpage{148}
(\byear{2003})
{\href{https://arxiv.org/abs/astro-ph/0310362}{{arXiv:astro-ph/0310362}}}
{[astro-ph]}.
\doiurl{10.1051/0004-6361:20031367}
\end{barticle}
\endbibitem

\bibitem{Atwood2009}
\begin{barticle}
\bauthor{\bsnm{{Atwood}}, \binits{W.B.}},
\bauthor{\bsnm{{Abdo}}, \binits{A.A.}},
\bauthor{\bsnm{{Ackermann}}, \binits{M.}},
\bauthor{\bsnm{{Althouse}}, \binits{W.}},
\bauthor{\bsnm{{Anderson}}, \binits{B.}},
\bauthor{\bsnm{{Axelsson}}, \binits{M.}},
\bauthor{\bsnm{{Baldini}}, \binits{L.}},
\bauthor{\bsnm{{Ballet}}, \binits{J.}},
\bauthor{\bsnm{{Band}}, \binits{D.L.}},
\bauthor{\bsnm{{Barbiellini}}, \binits{G.}},
\bauthor{\bsnm{{Bartelt}}, \binits{J.}},
\bauthor{\bsnm{{Bastieri}}, \binits{D.}},
\bauthor{\bsnm{{Baughman}}, \binits{B.M.}},
\bauthor{\bsnm{{Bechtol}}, \binits{K.}},
\bauthor{\bsnm{{B{\'e}d{\'e}r{\`e}de}}, \binits{D.}},
\bauthor{\bsnm{{Bellardi}}, \binits{F.}},
\bauthor{\bsnm{{Bellazzini}}, \binits{R.}},
\bauthor{\bsnm{{Berenji}}, \binits{B.}},
\bauthor{\bsnm{{Bignami}}, \binits{G.F.}},
\bauthor{\bsnm{{Bisello}}, \binits{D.}},
\bauthor{\bsnm{{Bissaldi}}, \binits{E.}},
\bauthor{\bsnm{{Blandford}}, \binits{R.D.}},
\bauthor{\bsnm{{Bloom}}, \binits{E.D.}},
\bauthor{\bsnm{{Bogart}}, \binits{J.R.}},
\bauthor{\bsnm{{Bonamente}}, \binits{E.}},
\bauthor{\bsnm{{Bonnell}}, \binits{J.}},
\bauthor{\bsnm{{Borgland}}, \binits{A.W.}},
\bauthor{\bsnm{{Bouvier}}, \binits{A.}},
\bauthor{\bsnm{{Bregeon}}, \binits{J.}},
\bauthor{\bsnm{{Brez}}, \binits{A.}},
\bauthor{\bsnm{{Brigida}}, \binits{M.}},
\bauthor{\bsnm{{Bruel}}, \binits{P.}},
\bauthor{\bsnm{{Burnett}}, \binits{T.H.}},
\bauthor{\bsnm{{Busetto}}, \binits{G.}},
\bauthor{\bsnm{{Caliandro}}, \binits{G.A.}},
\bauthor{\bsnm{{Cameron}}, \binits{R.A.}},
\bauthor{\bsnm{{Caraveo}}, \binits{P.A.}},
\bauthor{\bsnm{{Carius}}, \binits{S.}},
\bauthor{\bsnm{{Carlson}}, \binits{P.}},
\bauthor{\bsnm{{Casandjian}}, \binits{J.M.}},
\bauthor{\bsnm{{Cavazzuti}}, \binits{E.}},
\bauthor{\bsnm{{Ceccanti}}, \binits{M.}},
\bauthor{\bsnm{{Cecchi}}, \binits{C.}},
\bauthor{\bsnm{{Charles}}, \binits{E.}},
\bauthor{\bsnm{{Chekhtman}}, \binits{A.}},
\bauthor{\bsnm{{Cheung}}, \binits{C.C.}},
\bauthor{\bsnm{{Chiang}}, \binits{J.}},
\bauthor{\bsnm{{Chipaux}}, \binits{R.}},
\bauthor{\bsnm{{Cillis}}, \binits{A.N.}},
\bauthor{\bsnm{{Ciprini}}, \binits{S.}},
\bauthor{\bsnm{{Claus}}, \binits{R.}},
\bauthor{\bsnm{{Cohen-Tanugi}}, \binits{J.}},
\bauthor{\bsnm{{Condamoor}}, \binits{S.}},
\bauthor{\bsnm{{Conrad}}, \binits{J.}},
\bauthor{\bsnm{{Corbet}}, \binits{R.}},
\bauthor{\bsnm{{Corucci}}, \binits{L.}},
\bauthor{\bsnm{{Costamante}}, \binits{L.}},
\bauthor{\bsnm{{Cutini}}, \binits{S.}},
\bauthor{\bsnm{{Davis}}, \binits{D.S.}},
\bauthor{\bsnm{{Decotigny}}, \binits{D.}},
\bauthor{\bsnm{{DeKlotz}}, \binits{M.}},
\bauthor{\bsnm{{Dermer}}, \binits{C.D.}},
\bauthor{\bsnm{{de Angelis}}, \binits{A.}},
\bauthor{\bsnm{{Digel}}, \binits{S.W.}},
\bauthor{\bsnm{{do Couto e Silva}}, \binits{E.}},
\bauthor{\bsnm{{Drell}}, \binits{P.S.}},
\bauthor{\bsnm{{Dubois}}, \binits{R.}},
\bauthor{\bsnm{{Dumora}}, \binits{D.}},
\bauthor{\bsnm{{Edmonds}}, \binits{Y.}},
\bauthor{\bsnm{{Fabiani}}, \binits{D.}},
\bauthor{\bsnm{{Farnier}}, \binits{C.}},
\bauthor{\bsnm{{Favuzzi}}, \binits{C.}},
\bauthor{\bsnm{{Flath}}, \binits{D.L.}},
\bauthor{\bsnm{{Fleury}}, \binits{P.}},
\bauthor{\bsnm{{Focke}}, \binits{W.B.}},
\bauthor{\bsnm{{Funk}}, \binits{S.}},
\bauthor{\bsnm{{Fusco}}, \binits{P.}},
\bauthor{\bsnm{{Gargano}}, \binits{F.}},
\bauthor{\bsnm{{Gasparrini}}, \binits{D.}},
\bauthor{\bsnm{{Gehrels}}, \binits{N.}},
\bauthor{\bsnm{{Gentit}}, \binits{F.-X.}},
\bauthor{\bsnm{{Germani}}, \binits{S.}},
\bauthor{\bsnm{{Giebels}}, \binits{B.}},
\bauthor{\bsnm{{Giglietto}}, \binits{N.}},
\bauthor{\bsnm{{Giommi}}, \binits{P.}},
\bauthor{\bsnm{{Giordano}}, \binits{F.}},
\bauthor{\bsnm{{Glanzman}}, \binits{T.}},
\bauthor{\bsnm{{Godfrey}}, \binits{G.}},
\bauthor{\bsnm{{Grenier}}, \binits{I.A.}},
\bauthor{\bsnm{{Grondin}}, \binits{M.-H.}},
\bauthor{\bsnm{{Grove}}, \binits{J.E.}},
\bauthor{\bsnm{{Guillemot}}, \binits{L.}},
\bauthor{\bsnm{{Guiriec}}, \binits{S.}},
\bauthor{\bsnm{{Haller}}, \binits{G.}},
\bauthor{\bsnm{{Harding}}, \binits{A.K.}},
\bauthor{\bsnm{{Hart}}, \binits{P.A.}},
\bauthor{\bsnm{{Hays}}, \binits{E.}},
\bauthor{\bsnm{{Healey}}, \binits{S.E.}},
\bauthor{\bsnm{{Hirayama}}, \binits{M.}},
\bauthor{\bsnm{{Hjalmarsdotter}}, \binits{L.}},
\bauthor{\bsnm{{Horn}}, \binits{R.}},
\bauthor{\bsnm{{Hughes}}, \binits{R.E.}},
\bauthor{\bsnm{{J{\'o}hannesson}}, \binits{G.}},
\bauthor{\bsnm{{Johansson}}, \binits{G.}},
\bauthor{\bsnm{{Johnson}}, \binits{A.S.}},
\bauthor{\bsnm{{Johnson}}, \binits{R.P.}},
\bauthor{\bsnm{{Johnson}}, \binits{T.J.}},
\bauthor{\bsnm{{Johnson}}, \binits{W.N.}},
\bauthor{\bsnm{{Kamae}}, \binits{T.}},
\bauthor{\bsnm{{Katagiri}}, \binits{H.}},
\bauthor{\bsnm{{Kataoka}}, \binits{J.}},
\bauthor{\bsnm{{Kavelaars}}, \binits{A.}},
\bauthor{\bsnm{{Kawai}}, \binits{N.}},
\bauthor{\bsnm{{Kelly}}, \binits{H.}},
\bauthor{\bsnm{{Kerr}}, \binits{M.}},
\bauthor{\bsnm{{Klamra}}, \binits{W.}},
\bauthor{\bsnm{{Kn{\"o}dlseder}}, \binits{J.}},
\bauthor{\bsnm{{Kocian}}, \binits{M.L.}},
\bauthor{\bsnm{{Komin}}, \binits{N.}},
\bauthor{\bsnm{{Kuehn}}, \binits{F.}},
\bauthor{\bsnm{{Kuss}}, \binits{M.}},
\bauthor{\bsnm{{Landriu}}, \binits{D.}},
\bauthor{\bsnm{{Latronico}}, \binits{L.}},
\bauthor{\bsnm{{Lee}}, \binits{B.}},
\bauthor{\bsnm{{Lee}}, \binits{S.-H.}},
\bauthor{\bsnm{{Lemoine-Goumard}}, \binits{M.}},
\bauthor{\bsnm{{Lionetto}}, \binits{A.M.}},
\bauthor{\bsnm{{Longo}}, \binits{F.}},
\bauthor{\bsnm{{Loparco}}, \binits{F.}},
\bauthor{\bsnm{{Lott}}, \binits{B.}},
\bauthor{\bsnm{{Lovellette}}, \binits{M.N.}},
\bauthor{\bsnm{{Lubrano}}, \binits{P.}},
\bauthor{\bsnm{{Madejski}}, \binits{G.M.}},
\bauthor{\bsnm{{Makeev}}, \binits{A.}},
\bauthor{\bsnm{{Marangelli}}, \binits{B.}},
\bauthor{\bsnm{{Massai}}, \binits{M.M.}},
\bauthor{\bsnm{{Mazziotta}}, \binits{M.N.}},
\bauthor{\bsnm{{McEnery}}, \binits{J.E.}},
\bauthor{\bsnm{{Menon}}, \binits{N.}},
\bauthor{\bsnm{{Meurer}}, \binits{C.}},
\bauthor{\bsnm{{Michelson}}, \binits{P.F.}},
\bauthor{\bsnm{{Minuti}}, \binits{M.}},
\bauthor{\bsnm{{Mirizzi}}, \binits{N.}},
\bauthor{\bsnm{{Mitthumsiri}}, \binits{W.}},
\bauthor{\bsnm{{Mizuno}}, \binits{T.}},
\bauthor{\bsnm{{Moiseev}}, \binits{A.A.}},
\bauthor{\bsnm{{Monte}}, \binits{C.}},
\bauthor{\bsnm{{Monzani}}, \binits{M.E.}},
\bauthor{\bsnm{{Moretti}}, \binits{E.}},
\bauthor{\bsnm{{Morselli}}, \binits{A.}},
\bauthor{\bsnm{{Moskalenko}}, \binits{I.V.}},
\bauthor{\bsnm{{Murgia}}, \binits{S.}},
\bauthor{\bsnm{{Nakamori}}, \binits{T.}},
\bauthor{\bsnm{{Nishino}}, \binits{S.}},
\bauthor{\bsnm{{Nolan}}, \binits{P.L.}},
\bauthor{\bsnm{{Norris}}, \binits{J.P.}},
\bauthor{\bsnm{{Nuss}}, \binits{E.}},
\bauthor{\bsnm{{Ohno}}, \binits{M.}},
\bauthor{\bsnm{{Ohsugi}}, \binits{T.}},
\bauthor{\bsnm{{Omodei}}, \binits{N.}},
\bauthor{\bsnm{{Orlando}}, \binits{E.}},
\bauthor{\bsnm{{Ormes}}, \binits{J.F.}},
\bauthor{\bsnm{{Paccagnella}}, \binits{A.}},
\bauthor{\bsnm{{Paneque}}, \binits{D.}},
\bauthor{\bsnm{{Panetta}}, \binits{J.H.}},
\bauthor{\bsnm{{Parent}}, \binits{D.}},
\bauthor{\bsnm{{Pearce}}, \binits{M.}},
\bauthor{\bsnm{{Pepe}}, \binits{M.}},
\bauthor{\bsnm{{Perazzo}}, \binits{A.}},
\bauthor{\bsnm{{Pesce-Rollins}}, \binits{M.}},
\bauthor{\bsnm{{Picozza}}, \binits{P.}},
\bauthor{\bsnm{{Pieri}}, \binits{L.}},
\bauthor{\bsnm{{Pinchera}}, \binits{M.}},
\bauthor{\bsnm{{Piron}}, \binits{F.}},
\bauthor{\bsnm{{Porter}}, \binits{T.A.}},
\bauthor{\bsnm{{Poupard}}, \binits{L.}},
\bauthor{\bsnm{{Rain{\`o}}}, \binits{S.}},
\bauthor{\bsnm{{Rando}}, \binits{R.}},
\bauthor{\bsnm{{Rapposelli}}, \binits{E.}},
\bauthor{\bsnm{{Razzano}}, \binits{M.}},
\bauthor{\bsnm{{Reimer}}, \binits{A.}},
\bauthor{\bsnm{{Reimer}}, \binits{O.}},
\bauthor{\bsnm{{Reposeur}}, \binits{T.}},
\bauthor{\bsnm{{Reyes}}, \binits{L.C.}},
\bauthor{\bsnm{{Ritz}}, \binits{S.}},
\bauthor{\bsnm{{Rochester}}, \binits{L.S.}},
\bauthor{\bsnm{{Rodriguez}}, \binits{A.Y.}},
\bauthor{\bsnm{{Romani}}, \binits{R.W.}},
\bauthor{\bsnm{{Roth}}, \binits{M.}},
\bauthor{\bsnm{{Russell}}, \binits{J.J.}},
\bauthor{\bsnm{{Ryde}}, \binits{F.}},
\bauthor{\bsnm{{Sabatini}}, \binits{S.}},
\bauthor{\bsnm{{Sadrozinski}}, \binits{H.F.-W.}},
\bauthor{\bsnm{{Sanchez}}, \binits{D.}},
\bauthor{\bsnm{{Sander}}, \binits{A.}},
\bauthor{\bsnm{{Sapozhnikov}}, \binits{L.}},
\bauthor{\bsnm{{Parkinson}}, \binits{P.M.S.}},
\bauthor{\bsnm{{Scargle}}, \binits{J.D.}},
\bauthor{\bsnm{{Schalk}}, \binits{T.L.}},
\bauthor{\bsnm{{Scolieri}}, \binits{G.}},
\bauthor{\bsnm{{Sgr{\`o}}}, \binits{C.}},
\bauthor{\bsnm{{Share}}, \binits{G.H.}},
\bauthor{\bsnm{{Shaw}}, \binits{M.}},
\bauthor{\bsnm{{Shimokawabe}}, \binits{T.}},
\bauthor{\bsnm{{Shrader}}, \binits{C.}},
\bauthor{\bsnm{{Sierpowska-Bartosik}}, \binits{A.}},
\bauthor{\bsnm{{Siskind}}, \binits{E.J.}},
\bauthor{\bsnm{{Smith}}, \binits{D.A.}},
\bauthor{\bsnm{{Smith}}, \binits{P.D.}},
\bauthor{\bsnm{{Spandre}}, \binits{G.}},
\bauthor{\bsnm{{Spinelli}}, \binits{P.}},
\bauthor{\bsnm{{Starck}}, \binits{J.-L.}},
\bauthor{\bsnm{{Stephens}}, \binits{T.E.}},
\bauthor{\bsnm{{Strickman}}, \binits{M.S.}},
\bauthor{\bsnm{{Strong}}, \binits{A.W.}},
\bauthor{\bsnm{{Suson}}, \binits{D.J.}},
\bauthor{\bsnm{{Tajima}}, \binits{H.}},
\bauthor{\bsnm{{Takahashi}}, \binits{H.}},
\bauthor{\bsnm{{Takahashi}}, \binits{T.}},
\bauthor{\bsnm{{Tanaka}}, \binits{T.}},
\bauthor{\bsnm{{Tenze}}, \binits{A.}},
\bauthor{\bsnm{{Tether}}, \binits{S.}},
\bauthor{\bsnm{{Thayer}}, \binits{J.B.}},
\bauthor{\bsnm{{Thayer}}, \binits{J.G.}},
\bauthor{\bsnm{{Thompson}}, \binits{D.J.}},
\bauthor{\bsnm{{Tibaldo}}, \binits{L.}},
\bauthor{\bsnm{{Tibolla}}, \binits{O.}},
\bauthor{\bsnm{{Torres}}, \binits{D.F.}},
\bauthor{\bsnm{{Tosti}}, \binits{G.}},
\bauthor{\bsnm{{Tramacere}}, \binits{A.}},
\bauthor{\bsnm{{Turri}}, \binits{M.}},
\bauthor{\bsnm{{Usher}}, \binits{T.L.}},
\bauthor{\bsnm{{Vilchez}}, \binits{N.}},
\bauthor{\bsnm{{Vitale}}, \binits{V.}},
\bauthor{\bsnm{{Wang}}, \binits{P.}},
\bauthor{\bsnm{{Watters}}, \binits{K.}},
\bauthor{\bsnm{{Winer}}, \binits{B.L.}},
\bauthor{\bsnm{{Wood}}, \binits{K.S.}},
\bauthor{\bsnm{{Ylinen}}, \binits{T.}},
\bauthor{\bsnm{{Ziegler}}, \binits{M.}}:
\batitle{{The Large Area Telescope on the Fermi Gamma-Ray Space Telescope
  Mission}}.
\bjtitle{\apj}
\bvolume{697}(\bissue{2}),
\bfpage{1071}--\blpage{1102}
(\byear{2009})
{\href{https://arxiv.org/abs/0902.1089}{{arXiv:0902.1089}}}
{[astro-ph.IM]}.
\doiurl{10.1088/0004-637X/697/2/1071}
\end{barticle}
\endbibitem

\bibitem{Thompson1993}
\begin{barticle}
\bauthor{\bsnm{{Thompson}}, \binits{D.J.}},
\bauthor{\bsnm{{Bertsch}}, \binits{D.L.}},
\bauthor{\bsnm{{Fichtel}}, \binits{C.E.}},
\bauthor{\bsnm{{Hartman}}, \binits{R.C.}},
\bauthor{\bsnm{{Hofstadter}}, \binits{R.}},
\bauthor{\bsnm{{Hughes}}, \binits{E.B.}},
\bauthor{\bsnm{{Hunter}}, \binits{S.D.}},
\bauthor{\bsnm{{Hughlock}}, \binits{B.W.}},
\bauthor{\bsnm{{Kanbach}}, \binits{G.}},
\bauthor{\bsnm{{Kniffen}}, \binits{D.A.}},
\bauthor{\bsnm{{Lin}}, \binits{Y.C.}},
\bauthor{\bsnm{{Mattox}}, \binits{J.R.}},
\bauthor{\bsnm{{Mayer-Hasselwander}}, \binits{H.A.}},
\bauthor{\bsnm{{von Montigny}}, \binits{C.}},
\bauthor{\bsnm{{Nolan}}, \binits{P.L.}},
\bauthor{\bsnm{{Nel}}, \binits{H.I.}},
\bauthor{\bsnm{{Pinkau}}, \binits{K.}},
\bauthor{\bsnm{{Rothermel}}, \binits{H.}},
\bauthor{\bsnm{{Schneid}}, \binits{E.J.}},
\bauthor{\bsnm{{Sommer}}, \binits{M.}},
\bauthor{\bsnm{{Sreekumar}}, \binits{P.}},
\bauthor{\bsnm{{Tieger}}, \binits{D.}},
\bauthor{\bsnm{{Walker}}, \binits{A.H.}}:
\batitle{{Calibration of the Energetic Gamma-Ray Experiment Telescope (EGRET)
  for the Compton Gamma-Ray Observatory}}.
\bjtitle{\apjs}
\bvolume{86},
\bfpage{629}
(\byear{1993}).
\doiurl{10.1086/191793}
\end{barticle}
\endbibitem

\bibitem{deJager1996}
\begin{barticle}
\bauthor{\bsnm{{de Jager}}, \binits{O.C.}},
\bauthor{\bsnm{{Harding}}, \binits{A.K.}},
\bauthor{\bsnm{{Michelson}}, \binits{P.F.}},
\bauthor{\bsnm{{Nel}}, \binits{H.I.}},
\bauthor{\bsnm{{Nolan}}, \binits{P.L.}},
\bauthor{\bsnm{{Sreekumar}}, \binits{P.}},
\bauthor{\bsnm{{Thompson}}, \binits{D.J.}}:
\batitle{{Gamma-Ray Observations of the Crab Nebula: A Study of the
  Synchro-Compton Spectrum}}.
\bjtitle{\apj}
\bvolume{457},
\bfpage{253}
(\byear{1996}).
\doiurl{10.1086/176726}
\end{barticle}
\endbibitem

\bibitem{McConnell2000}
\begin{barticle}
\bauthor{\bsnm{{McConnell}}, \binits{M.L.}},
\bauthor{\bsnm{{Ryan}}, \binits{J.M.}},
\bauthor{\bsnm{{Collmar}}, \binits{W.}},
\bauthor{\bsnm{{Sch{\"o}nfelder}}, \binits{V.}},
\bauthor{\bsnm{{Steinle}}, \binits{H.}},
\bauthor{\bsnm{{Strong}}, \binits{A.W.}},
\bauthor{\bsnm{{Bloemen}}, \binits{H.}},
\bauthor{\bsnm{{Hermsen}}, \binits{W.}},
\bauthor{\bsnm{{Kuiper}}, \binits{L.}},
\bauthor{\bsnm{{Bennett}}, \binits{K.}},
\bauthor{\bsnm{{Phlips}}, \binits{B.F.}},
\bauthor{\bsnm{{Ling}}, \binits{J.C.}}:
\batitle{{A High-Sensitivity Measurement of the MeV Gamma-Ray Spectrum of
  Cygnus X-1}}.
\bjtitle{\apj}
\bvolume{543}(\bissue{2}),
\bfpage{928}--\blpage{937}
(\byear{2000})
{\href{https://arxiv.org/abs/astro-ph/0001484}{{arXiv:astro-ph/0001484}}}
{[astro-ph]}.
\doiurl{10.1086/317128}
\end{barticle}
\endbibitem

\bibitem{Abraham2023}
\begin{barticle}
\bauthor{\bsnm{Abraham}, \binits{S.}},
\bauthor{\bsnm{Zhu}, \binits{Y.}},
\bauthor{\bsnm{Nowicki}, \binits{S.}},
\bauthor{\bsnm{Bloser}, \binits{P.}},
\bauthor{\bsnm{Berry}, \binits{J.}},
\bauthor{\bsnm{Sandoval}, \binits{B.}},
\bauthor{\bsnm{Lanctot}, \binits{S.}},
\bauthor{\bsnm{Petryk}, \binits{M.}},
\bauthor{\bsnm{Deming}, \binits{J.}},
\bauthor{\bsnm{Klimenko}, \binits{A.}},
\bauthor{\bsnm{He}, \binits{Z.}}:
\batitle{Capability demonstration of a 3d cdznte detector on a high-altitude
  balloon flight}.
\bjtitle{Nuclear Instruments and Methods in Physics Research Section A:
  Accelerators, Spectrometers, Detectors and Associated Equipment}
\bvolume{1054},
\bfpage{168413}
(\byear{2023}).
\doiurl{10.1016/j.nima.2023.168413}
\end{barticle}
\endbibitem

\bibitem{Shy2023}
\begin{barticle}
\bauthor{\bsnm{{Shy}}, \binits{D.}},
\bauthor{\bsnm{{Goodman}}, \binits{D.}},
\bauthor{\bsnm{{Parsons}}, \binits{R.}},
\bauthor{\bsnm{{Streicher}}, \binits{M.}},
\bauthor{\bsnm{{Kaye}}, \binits{W.}},
\bauthor{\bsnm{{Mitchell}}, \binits{L.}},
\bauthor{\bsnm{{He}}, \binits{Z.}},
\bauthor{\bsnm{{Phlips}}, \binits{B.}}:
\batitle{{Radiation damage of 2 {\texttimes} 2 {\texttimes} 1cm$^{3}$ pixelated
  CdZnTe due to high-energy protons}}.
\bjtitle{Nuclear Instruments and Methods in Physics Research A}
\bvolume{1056},
\bfpage{168622}
(\byear{2023})
{\href{https://arxiv.org/abs/2308.02858}{{arXiv:2308.02858}}}
{[physics.ins-det]}.
\doiurl{10.1016/j.nima.2023.168622}
\end{barticle}
\endbibitem

\bibitem{Kuvvetli2003}
\begin{barticle}
\bauthor{\bsnm{{Kuvvetli}}, \binits{I.}},
\bauthor{\bsnm{{Budtz-J{\o}rgensen}}, \binits{C.}},
\bauthor{\bsnm{{Korsbech}}, \binits{U.}},
\bauthor{\bsnm{{Jensen}}, \binits{H.J.}}:
\batitle{{Radiation damage measurements on CZT drift strip detectors}}.
\bjtitle{Nuclear Instruments and Methods in Physics Research A}
\bvolume{512}(\bissue{1-2}),
\bfpage{98}--\blpage{105}
(\byear{2003}).
\doiurl{10.1016/S0168-9002(03)01881-3}
\end{barticle}
\endbibitem

\bibitem{Fraboni2004}
\begin{bchapter}
\bauthor{\bsnm{Fraboni}, \binits{B.}},
\bauthor{\bsnm{Cavallini}, \binits{A.}},
\bauthor{\bsnm{Auricchio}, \binits{N.}},
\bauthor{\bsnm{Dusi}, \binits{W.}},
\bauthor{\bsnm{Zanarini}, \binits{M.}},
\bauthor{\bsnm{Siffert}, \binits{P.}}:
\bctitle{Recovery of radiation damage in cdte and cdznte detectors}.
In: \bbtitle{IEEE Symposium Conference Record Nuclear Science 2004.},
vol. \bseriesno{7},
pp. \bfpage{4312}--\blpage{4317}
(\byear{2004}).
\doiurl{10.1109/NSSMIC.2004.1466842}
\end{bchapter}
\endbibitem

\bibitem{Wang2014}
\begin{barticle}
\bauthor{\bsnm{{Wang}}, \binits{W.}},
\bauthor{\bsnm{{Li}}, \binits{Z.}}:
\batitle{{Hard X-Ray Emission and $^{44}$Ti Line Features of the Tycho
  Supernova Remnant}}.
\bjtitle{\apj}
\bvolume{789}(\bissue{2}),
\bfpage{123}
(\byear{2014})
{\href{https://arxiv.org/abs/1405.6463}{{arXiv:1405.6463}}}
{[astro-ph.HE]}.
\doiurl{10.1088/0004-637X/789/2/123}
\end{barticle}
\endbibitem

\bibitem{Wang2016}
\begin{barticle}
\bauthor{\bsnm{{Wang}}, \binits{W.}},
\bauthor{\bsnm{{Li}}, \binits{Z.}}:
\batitle{{Hard X-Ray Emissions from Cassiopeia A Observed by INTEGRAL}}.
\bjtitle{\apj}
\bvolume{825}(\bissue{2}),
\bfpage{102}
(\byear{2016})
{\href{https://arxiv.org/abs/1605.00360}{{arXiv:1605.00360}}}
{[astro-ph.HE]}.
\doiurl{10.3847/0004-637X/825/2/102}
\end{barticle}
\endbibitem

\bibitem{Ruderman1975}
\begin{barticle}
\bauthor{\bsnm{{Ruderman}}, \binits{M.A.}},
\bauthor{\bsnm{{Sutherland}}, \binits{P.G.}}:
\batitle{{Theory of pulsars: polar gaps, sparks, and coherent microwave
  radiation.}}
\bjtitle{\apj}
\bvolume{196},
\bfpage{51}--\blpage{72}
(\byear{1975}).
\doiurl{10.1086/153393}
\end{barticle}
\endbibitem

\bibitem{Arons1981}
\begin{barticle}
\bauthor{\bsnm{{Arons}}, \binits{J.}}:
\batitle{{Pair creation above pulsar polar caps - Steady flow in the surface
  acceleration zone and polar CAP X-ray emission}}.
\bjtitle{\apj}
\bvolume{248},
\bfpage{1099}--\blpage{1116}
(\byear{1981}).
\doiurl{10.1086/159239}
\end{barticle}
\endbibitem

\bibitem{Cheng1986}
\begin{barticle}
\bauthor{\bsnm{{Cheng}}, \binits{K.S.}},
\bauthor{\bsnm{{Ho}}, \binits{C.}},
\bauthor{\bsnm{{Ruderman}}, \binits{M.}}:
\batitle{{Energetic Radiation from Rapidly Spinning Pulsars. I. Outer
  Magnetosphere Gaps}}.
\bjtitle{\apj}
\bvolume{300},
\bfpage{500}
(\byear{1986}).
\doiurl{10.1086/163829}
\end{barticle}
\endbibitem

\bibitem{Harding2005}
\begin{barticle}
\bauthor{\bsnm{{Harding}}, \binits{A.K.}},
\bauthor{\bsnm{{Usov}}, \binits{V.V.}},
\bauthor{\bsnm{{Muslimov}}, \binits{A.G.}}:
\batitle{{High-Energy Emission from Millisecond Pulsars}}.
\bjtitle{\apj}
\bvolume{622}(\bissue{1}),
\bfpage{531}--\blpage{543}
(\byear{2005})
{\href{https://arxiv.org/abs/astro-ph/0411805}{{arXiv:astro-ph/0411805}}}
{[astro-ph]}.
\doiurl{10.1086/427840}
\end{barticle}
\endbibitem

\bibitem{Harding2006}
\begin{barticle}
\bauthor{\bsnm{{Harding}}, \binits{A.K.}},
\bauthor{\bsnm{{Lai}}, \binits{D.}}:
\batitle{{Physics of strongly magnetized neutron stars}}.
\bjtitle{Reports on Progress in Physics}
\bvolume{69}(\bissue{9}),
\bfpage{2631}--\blpage{2708}
(\byear{2006})
{\href{https://arxiv.org/abs/astro-ph/0606674}{{arXiv:astro-ph/0606674}}}
{[astro-ph]}.
\doiurl{10.1088/0034-4885/69/9/R03}
\end{barticle}
\endbibitem

\bibitem{Kuiper2018}
\begin{barticle}
\bauthor{\bsnm{{Kuiper}}, \binits{L.}},
\bauthor{\bsnm{{Hermsen}}, \binits{W.}},
\bauthor{\bsnm{{Dekker}}, \binits{A.}}:
\batitle{{The Fermi-LAT detection of magnetar-like pulsar PSR J1846-0258 at
  high-energy gamma-rays}}.
\bjtitle{\mnras}
\bvolume{475}(\bissue{1}),
\bfpage{1238}--\blpage{1250}
(\byear{2018})
{\href{https://arxiv.org/abs/1709.00899}{{arXiv:1709.00899}}}
{[astro-ph.HE]}.
\doiurl{10.1093/mnras/stx3128}
\end{barticle}
\endbibitem

\bibitem{Petri2015}
\begin{barticle}
\bauthor{\bsnm{{P{\'e}tri}}, \binits{J.}}:
\batitle{{Effect of geodetic precession on the evolution of pulsar high-energy
  pulse profiles as derived with the striped-wind model}}.
\bjtitle{\aap}
\bvolume{574},
\bfpage{51}
(\byear{2015})
{\href{https://arxiv.org/abs/1410.7618}{{arXiv:1410.7618}}}
{[astro-ph.HE]}.
\doiurl{10.1051/0004-6361/201424289}
\end{barticle}
\endbibitem

\bibitem{Takata2017}
\begin{barticle}
\bauthor{\bsnm{{Takata}}, \binits{J.}},
\bauthor{\bsnm{{Cheng}}, \binits{K.S.}}:
\batitle{{X-Ray/GeV Emissions from Crab-like Pulsars in the LMC}}.
\bjtitle{\apj}
\bvolume{834}(\bissue{1}),
\bfpage{4}
(\byear{2017})
{\href{https://arxiv.org/abs/1612.00158}{{arXiv:1612.00158}}}
{[astro-ph.HE]}.
\doiurl{10.3847/1538-4357/834/1/4}
\end{barticle}
\endbibitem

\bibitem{Barnard2022}
\begin{barticle}
\bauthor{\bsnm{{Barnard}}, \binits{M.}},
\bauthor{\bsnm{{Venter}}, \binits{C.}},
\bauthor{\bsnm{{Harding}}, \binits{A.K.}},
\bauthor{\bsnm{{Kalapotharakos}}, \binits{C.}},
\bauthor{\bsnm{{Johnson}}, \binits{T.J.}}:
\batitle{{Probing the High-energy Gamma-Ray Emission Mechanism in the Vela
  Pulsar via Phase-resolved Spectral and Energy-dependent Light-curve
  Modeling}}.
\bjtitle{\apj}
\bvolume{925}(\bissue{2}),
\bfpage{184}
(\byear{2022})
{\href{https://arxiv.org/abs/2111.03405}{{arXiv:2111.03405}}}
{[astro-ph.HE]}.
\doiurl{10.3847/1538-4357/ac2a3d}
\end{barticle}
\endbibitem

\bibitem{Iniguez-Pascual2022}
\begin{barticle}
\bauthor{\bsnm{{{\'I}{\~n}iguez-Pascual}}, \binits{D.}},
\bauthor{\bsnm{{Vigan{\`o}}}, \binits{D.}},
\bauthor{\bsnm{{Torres}}, \binits{D.F.}}:
\batitle{{Synchro-curvature emitting regions in high-energy pulsar models}}.
\bjtitle{\mnras}
\bvolume{516}(\bissue{2}),
\bfpage{2475}--\blpage{2485}
(\byear{2022})
{\href{https://arxiv.org/abs/2208.05549}{{arXiv:2208.05549}}}
{[astro-ph.HE]}.
\doiurl{10.1093/mnras/stac2275}
\end{barticle}
\endbibitem

\bibitem{Torres2018}
\begin{barticle}
\bauthor{\bsnm{{Torres}}, \binits{D.F.}}:
\batitle{{Order parameters for the high-energy spectra of pulsars}}.
\bjtitle{Nature Astronomy}
\bvolume{2},
\bfpage{247}--\blpage{256}
(\byear{2018})
{\href{https://arxiv.org/abs/1802.04177}{{arXiv:1802.04177}}}
{[astro-ph.HE]}.
\doiurl{10.1038/s41550-018-0384-5}
\end{barticle}
\endbibitem

\bibitem{Torres2019}
\begin{barticle}
\bauthor{\bsnm{{Torres}}, \binits{D.F.}},
\bauthor{\bsnm{{Vigan{\`o}}}, \binits{D.}},
\bauthor{\bsnm{{Coti Zelati}}, \binits{F.}},
\bauthor{\bsnm{{Li}}, \binits{J.}}:
\batitle{{Synchrocurvature modelling of the multifrequency non-thermal emission
  of pulsars}}.
\bjtitle{\mnras}
\bvolume{489}(\bissue{4}),
\bfpage{5494}--\blpage{5512}
(\byear{2019})
{\href{https://arxiv.org/abs/1908.11574}{{arXiv:1908.11574}}}
{[astro-ph.HE]}.
\doiurl{10.1093/mnras/stz2403}
\end{barticle}
\endbibitem

\bibitem{Acciari2021}
\begin{barticle}
\bauthor{\bsnm{{Acciari}}, \binits{V.A.}},
\bauthor{\bsnm{{Ansoldi}}, \binits{S.}},
\bauthor{\bsnm{{Antonelli}}, \binits{L.A.}},
\bauthor{\bsnm{{Arbet Engels}}, \binits{A.}},
\bauthor{\bsnm{{Artero}}, \binits{M.}},
\bauthor{\bsnm{{Asano}}, \binits{K.}},
\bauthor{\bsnm{{Baack}}, \binits{D.}},
\bauthor{\bsnm{{Babi{\'c}}}, \binits{A.}},
\bauthor{\bsnm{{Baquero}}, \binits{A.}},
\bauthor{\bsnm{{Barres de Almeida}}, \binits{U.}},
\bauthor{\bsnm{{Barrio}}, \binits{J.A.}},
\bauthor{\bsnm{{Batkovi{\'c}}}, \binits{I.}},
\bauthor{\bsnm{{Becerra Gonz{\'a}lez}}, \binits{J.}},
\bauthor{\bsnm{{Bednarek}}, \binits{W.}},
\bauthor{\bsnm{{Bellizzi}}, \binits{L.}},
\bauthor{\bsnm{{Bernardini}}, \binits{E.}},
\bauthor{\bsnm{{Bernardos}}, \binits{M.}},
\bauthor{\bsnm{{Berti}}, \binits{A.}},
\bauthor{\bsnm{{Besenrieder}}, \binits{J.}},
\bauthor{\bsnm{{Bhattacharyya}}, \binits{W.}},
\bauthor{\bsnm{{Bigongiari}}, \binits{C.}},
\bauthor{\bsnm{{Biland}}, \binits{A.}},
\bauthor{\bsnm{{Blanch}}, \binits{O.}},
\bauthor{\bsnm{{Bonnoli}}, \binits{G.}},
\bauthor{\bsnm{{Bo{\v{s}}njak}}, \binits{{\v{Z}}.}},
\bauthor{\bsnm{{Busetto}}, \binits{G.}},
\bauthor{\bsnm{{Carosi}}, \binits{R.}},
\bauthor{\bsnm{{Ceribella}}, \binits{G.}},
\bauthor{\bsnm{{Cerruti}}, \binits{M.}},
\bauthor{\bsnm{{Chai}}, \binits{Y.}},
\bauthor{\bsnm{{Chilingarian}}, \binits{A.}},
\bauthor{\bsnm{{Cikota}}, \binits{S.}},
\bauthor{\bsnm{{Colak}}, \binits{S.M.}},
\bauthor{\bsnm{{Colombo}}, \binits{E.}},
\bauthor{\bsnm{{Contreras}}, \binits{J.L.}},
\bauthor{\bsnm{{Cortina}}, \binits{J.}},
\bauthor{\bsnm{{Covino}}, \binits{S.}},
\bauthor{\bsnm{{D'Amico}}, \binits{G.}},
\bauthor{\bsnm{{D'Elia}}, \binits{V.}},
\bauthor{\bsnm{{Da Vela}}, \binits{P.}},
\bauthor{\bsnm{{Dazzi}}, \binits{F.}},
\bauthor{\bsnm{{De Angelis}}, \binits{A.}},
\bauthor{\bsnm{{De Lotto}}, \binits{B.}},
\bauthor{\bsnm{{Delfino}}, \binits{M.}},
\bauthor{\bsnm{{Delgado}}, \binits{J.}},
\bauthor{\bsnm{{Delgado Mendez}}, \binits{C.}},
\bauthor{\bsnm{{Depaoli}}, \binits{D.}},
\bauthor{\bsnm{{Di Pierro}}, \binits{F.}},
\bauthor{\bsnm{{Di Venere}}, \binits{L.}},
\bauthor{\bsnm{{Do Souto Espi{\~n}eira}}, \binits{E.}},
\bauthor{\bsnm{{Dominis Prester}}, \binits{D.}},
\bauthor{\bsnm{{Donini}}, \binits{A.}},
\bauthor{\bsnm{{Dorner}}, \binits{D.}},
\bauthor{\bsnm{{Doro}}, \binits{M.}},
\bauthor{\bsnm{{Elsaesser}}, \binits{D.}},
\bauthor{\bsnm{{Fallah Ramazani}}, \binits{V.}},
\bauthor{\bsnm{{Fattorini}}, \binits{A.}},
\bauthor{\bsnm{{Ferrara}}, \binits{G.}},
\bauthor{\bsnm{{Fonseca}}, \binits{M.V.}},
\bauthor{\bsnm{{Font}}, \binits{L.}},
\bauthor{\bsnm{{Fruck}}, \binits{C.}},
\bauthor{\bsnm{{Fukami}}, \binits{S.}},
\bauthor{\bsnm{{Garc{\'\i}a L{\'o}pez}}, \binits{R.J.}},
\bauthor{\bsnm{{Garczarczyk}}, \binits{M.}},
\bauthor{\bsnm{{Gasparyan}}, \binits{S.}},
\bauthor{\bsnm{{Gaug}}, \binits{M.}},
\bauthor{\bsnm{{Giglietto}}, \binits{N.}},
\bauthor{\bsnm{{Giordano}}, \binits{F.}},
\bauthor{\bsnm{{Gliwny}}, \binits{P.}},
\bauthor{\bsnm{{Godinovi{\'c}}}, \binits{N.}},
\bauthor{\bsnm{{Green}}, \binits{J.G.}},
\bauthor{\bsnm{{Green}}, \binits{D.}},
\bauthor{\bsnm{{Hadasch}}, \binits{D.}},
\bauthor{\bsnm{{Hahn}}, \binits{A.}},
\bauthor{\bsnm{{Heckmann}}, \binits{L.}},
\bauthor{\bsnm{{Herrera}}, \binits{J.}},
\bauthor{\bsnm{{Hoang}}, \binits{J.}},
\bauthor{\bsnm{{Hrupec}}, \binits{D.}},
\bauthor{\bsnm{{H{\"u}tten}}, \binits{M.}},
\bauthor{\bsnm{{Inada}}, \binits{T.}},
\bauthor{\bsnm{{Inoue}}, \binits{S.}},
\bauthor{\bsnm{{Ishio}}, \binits{K.}},
\bauthor{\bsnm{{Iwamura}}, \binits{Y.}},
\bauthor{\bsnm{{Jim{\'e}nez}}, \binits{I.}},
\bauthor{\bsnm{{Jormanainen}}, \binits{J.}},
\bauthor{\bsnm{{Jouvin}}, \binits{L.}},
\bauthor{\bsnm{{Kajiwara}}, \binits{Y.}},
\bauthor{\bsnm{{Karjalainen}}, \binits{M.}},
\bauthor{\bsnm{{Kerszberg}}, \binits{D.}},
\bauthor{\bsnm{{Kobayashi}}, \binits{Y.}},
\bauthor{\bsnm{{Kubo}}, \binits{H.}},
\bauthor{\bsnm{{Kushida}}, \binits{J.}},
\bauthor{\bsnm{{Lamastra}}, \binits{A.}},
\bauthor{\bsnm{{Lelas}}, \binits{D.}},
\bauthor{\bsnm{{Leone}}, \binits{F.}},
\bauthor{\bsnm{{Lindfors}}, \binits{E.}},
\bauthor{\bsnm{{Lombardi}}, \binits{S.}},
\bauthor{\bsnm{{Longo}}, \binits{F.}},
\bauthor{\bsnm{{L{\'o}pez-Coto}}, \binits{R.}},
\bauthor{\bsnm{{L{\'o}pez-Moya}}, \binits{M.}},
\bauthor{\bsnm{{L{\'o}pez-Oramas}}, \binits{A.}},
\bauthor{\bsnm{{Loporchio}}, \binits{S.}},
\bauthor{\bsnm{{Machado de Oliveira Fraga}}, \binits{B.}},
\bauthor{\bsnm{{Maggio}}, \binits{C.}},
\bauthor{\bsnm{{Majumdar}}, \binits{P.}},
\bauthor{\bsnm{{Makariev}}, \binits{M.}},
\bauthor{\bsnm{{Mallamaci}}, \binits{M.}},
\bauthor{\bsnm{{Maneva}}, \binits{G.}},
\bauthor{\bsnm{{Manganaro}}, \binits{M.}},
\bauthor{\bsnm{{Mannheim}}, \binits{K.}},
\bauthor{\bsnm{{Maraschi}}, \binits{L.}},
\bauthor{\bsnm{{Mariotti}}, \binits{M.}},
\bauthor{\bsnm{{Mart{\'\i}nez}}, \binits{M.}},
\bauthor{\bsnm{{Mazin}}, \binits{D.}},
\bauthor{\bsnm{{Menchiari}}, \binits{S.}},
\bauthor{\bsnm{{Mender}}, \binits{S.}},
\bauthor{\bsnm{{Mi{\'c}anovi{\'c}}}, \binits{S.}},
\bauthor{\bsnm{{Miceli}}, \binits{D.}},
\bauthor{\bsnm{{Miener}}, \binits{T.}},
\bauthor{\bsnm{{Minev}}, \binits{M.}},
\bauthor{\bsnm{{Miranda}}, \binits{J.M.}},
\bauthor{\bsnm{{Mirzoyan}}, \binits{R.}},
\bauthor{\bsnm{{Molina}}, \binits{E.}},
\bauthor{\bsnm{{Moralejo}}, \binits{A.}},
\bauthor{\bsnm{{Morcuende}}, \binits{D.}},
\bauthor{\bsnm{{Moreno}}, \binits{V.}},
\bauthor{\bsnm{{Moretti}}, \binits{E.}},
\bauthor{\bsnm{{Neustroev}}, \binits{V.}},
\bauthor{\bsnm{{Nigro}}, \binits{C.}},
\bauthor{\bsnm{{Nilsson}}, \binits{K.}},
\bauthor{\bsnm{{Nishijima}}, \binits{K.}},
\bauthor{\bsnm{{Noda}}, \binits{K.}},
\bauthor{\bsnm{{Nozaki}}, \binits{S.}},
\bauthor{\bsnm{{Ohtani}}, \binits{Y.}},
\bauthor{\bsnm{{Oka}}, \binits{T.}},
\bauthor{\bsnm{{Otero-Santos}}, \binits{J.}},
\bauthor{\bsnm{{Paiano}}, \binits{S.}},
\bauthor{\bsnm{{Palatiello}}, \binits{M.}},
\bauthor{\bsnm{{Paneque}}, \binits{D.}},
\bauthor{\bsnm{{Paoletti}}, \binits{R.}},
\bauthor{\bsnm{{Paredes}}, \binits{J.M.}},
\bauthor{\bsnm{{Pavleti{\'c}}}, \binits{L.}},
\bauthor{\bsnm{{Pe{\~n}il}}, \binits{P.}},
\bauthor{\bsnm{{Perennes}}, \binits{C.}},
\bauthor{\bsnm{{Persic}}, \binits{M.}},
\bauthor{\bsnm{{Prada Moroni}}, \binits{P.G.}},
\bauthor{\bsnm{{Prandini}}, \binits{E.}},
\bauthor{\bsnm{{Priyadarshi}}, \binits{C.}},
\bauthor{\bsnm{{Puljak}}, \binits{I.}},
\bauthor{\bsnm{{Rhode}}, \binits{W.}},
\bauthor{\bsnm{{Rib{\'o}}}, \binits{M.}},
\bauthor{\bsnm{{Rico}}, \binits{J.}},
\bauthor{\bsnm{{Righi}}, \binits{C.}},
\bauthor{\bsnm{{Rugliancich}}, \binits{A.}},
\bauthor{\bsnm{{Saha}}, \binits{L.}},
\bauthor{\bsnm{{Sahakyan}}, \binits{N.}},
\bauthor{\bsnm{{Saito}}, \binits{T.}},
\bauthor{\bsnm{{Sakurai}}, \binits{S.}},
\bauthor{\bsnm{{Satalecka}}, \binits{K.}},
\bauthor{\bsnm{{Saturni}}, \binits{F.G.}},
\bauthor{\bsnm{{Schleicher}}, \binits{B.}},
\bauthor{\bsnm{{Schmidt}}, \binits{K.}},
\bauthor{\bsnm{{Schweizer}}, \binits{T.}},
\bauthor{\bsnm{{Sitarek}}, \binits{J.}},
\bauthor{\bsnm{{{\v{S}}nidari{\'c}}}, \binits{I.}},
\bauthor{\bsnm{{Sobczynska}}, \binits{D.}},
\bauthor{\bsnm{{Spolon}}, \binits{A.}},
\bauthor{\bsnm{{Stamerra}}, \binits{A.}},
\bauthor{\bsnm{{Strom}}, \binits{D.}},
\bauthor{\bsnm{{Strzys}}, \binits{M.}},
\bauthor{\bsnm{{Suda}}, \binits{Y.}},
\bauthor{\bsnm{{Suri{\'c}}}, \binits{T.}},
\bauthor{\bsnm{{Takahashi}}, \binits{M.}},
\bauthor{\bsnm{{Tavecchio}}, \binits{F.}},
\bauthor{\bsnm{{Temnikov}}, \binits{P.}},
\bauthor{\bsnm{{Terzi{\'c}}}, \binits{T.}},
\bauthor{\bsnm{{Teshima}}, \binits{M.}},
\bauthor{\bsnm{{Tosti}}, \binits{L.}},
\bauthor{\bsnm{{Truzzi}}, \binits{S.}},
\bauthor{\bsnm{{Tutone}}, \binits{A.}},
\bauthor{\bsnm{{Ubach}}, \binits{S.}},
\bauthor{\bsnm{{van Scherpenberg}}, \binits{J.}},
\bauthor{\bsnm{{Vanzo}}, \binits{G.}},
\bauthor{\bsnm{{Vazquez Acosta}}, \binits{M.}},
\bauthor{\bsnm{{Ventura}}, \binits{S.}},
\bauthor{\bsnm{{Verguilov}}, \binits{V.}},
\bauthor{\bsnm{{Vigorito}}, \binits{C.F.}},
\bauthor{\bsnm{{Vitale}}, \binits{V.}},
\bauthor{\bsnm{{Vovk}}, \binits{I.}},
\bauthor{\bsnm{{Will}}, \binits{M.}},
\bauthor{\bsnm{{Wunderlich}}, \binits{C.}},
\bauthor{\bsnm{{Zari{\'c}}}, \binits{D.}},
\bauthor{\bsnm{{Zari{\'c}}}, \binits{D.}},
\bauthor{\bsnm{{Caraveo}}, \binits{P.A.}},
\bauthor{\bsnm{{Cognard}}, \binits{I.}},
\bauthor{\bsnm{{Guillemot}}, \binits{L.}},
\bauthor{\bsnm{{Harding}}, \binits{A.K.}},
\bauthor{\bsnm{{Li}}, \binits{J.}},
\bauthor{\bsnm{{Limyansky}}, \binits{B.}},
\bauthor{\bsnm{{Ng}}, \binits{C.Y.}},
\bauthor{\bsnm{{Torres}}, \binits{D.F.}},
\bauthor{\bsnm{{Saz Parkinson}}, \binits{P.M.}}:
\batitle{{Search for Very High-energy Emission from the Millisecond Pulsar PSR
  J0218+4232}}.
\bjtitle{\apj}
\bvolume{922}(\bissue{2}),
\bfpage{251}
(\byear{2021})
{\href{https://arxiv.org/abs/2108.11373}{{arXiv:2108.11373}}}
{[astro-ph.HE]}.
\doiurl{10.3847/1538-4357/ac20d7}
\end{barticle}
\endbibitem

\bibitem{Kuiper2015}
\begin{barticle}
\bauthor{\bsnm{{Kuiper}}, \binits{L.}},
\bauthor{\bsnm{{Hermsen}}, \binits{W.}}:
\batitle{{The soft {\ensuremath{\gamma}}-ray pulsar population: a high-energy
  overview}}.
\bjtitle{\mnras}
\bvolume{449}(\bissue{4}),
\bfpage{3827}--\blpage{3866}
(\byear{2015})
{\href{https://arxiv.org/abs/1502.06769}{{arXiv:1502.06769}}}
{[astro-ph.HE]}.
\doiurl{10.1093/mnras/stv426}
\end{barticle}
\endbibitem

\bibitem{CotiZelati2020}
\begin{barticle}
\bauthor{\bsnm{{Coti Zelati}}, \binits{F.}},
\bauthor{\bsnm{{Torres}}, \binits{D.F.}},
\bauthor{\bsnm{{Li}}, \binits{J.}},
\bauthor{\bsnm{{Vigan{\`o}}}, \binits{D.}}:
\batitle{{Spectral characterization of the non-thermal X-ray emission of
  gamma-ray pulsars}}.
\bjtitle{\mnras}
\bvolume{492}(\bissue{1}),
\bfpage{1025}--\blpage{1043}
(\byear{2020})
{\href{https://arxiv.org/abs/1912.03953}{{arXiv:1912.03953}}}
{[astro-ph.HE]}.
\doiurl{10.1093/mnras/stz3485}
\end{barticle}
\endbibitem

\bibitem{Hare2021}
\begin{barticle}
\bauthor{\bsnm{{Hare}}, \binits{J.}},
\bauthor{\bsnm{{Volkov}}, \binits{I.}},
\bauthor{\bsnm{{Pavlov}}, \binits{G.G.}},
\bauthor{\bsnm{{Kargaltsev}}, \binits{O.}},
\bauthor{\bsnm{{Johnston}}, \binits{S.}}:
\batitle{{Precise Timing and Phase-resolved Spectroscopy of the Young Pulsar
  J1617-5055 with NuSTAR}}.
\bjtitle{\apj}
\bvolume{923}(\bissue{2}),
\bfpage{249}
(\byear{2021})
{\href{https://arxiv.org/abs/2110.08077}{{arXiv:2110.08077}}}
{[astro-ph.HE]}.
\doiurl{10.3847/1538-4357/ac30e2}
\end{barticle}
\endbibitem

\bibitem{Zhang2013}
\begin{barticle}
\bauthor{\bsnm{{Zhang}}, \binits{H.}},
\bauthor{\bsnm{{B{\"o}ttcher}}, \binits{M.}}:
\batitle{{X-Ray and Gamma-Ray Polarization in Leptonic and Hadronic Jet Models
  of Blazars}}.
\bjtitle{\apj}
\bvolume{774}(\bissue{1}),
\bfpage{18}
(\byear{2013})
{\href{https://arxiv.org/abs/1307.4187}{{arXiv:1307.4187}}}
{[astro-ph.HE]}.
\doiurl{10.1088/0004-637X/774/1/18}
\end{barticle}
\endbibitem

\bibitem{Petropoulou2015}
\begin{barticle}
\bauthor{\bsnm{{Petropoulou}}, \binits{M.}},
\bauthor{\bsnm{{Dimitrakoudis}}, \binits{S.}},
\bauthor{\bsnm{{Padovani}}, \binits{P.}},
\bauthor{\bsnm{{Mastichiadis}}, \binits{A.}},
\bauthor{\bsnm{{Resconi}}, \binits{E.}}:
\batitle{{Photohadronic origin of {\ensuremath{\gamma}} -ray BL Lac emission:
  implications for IceCube neutrinos}}.
\bjtitle{\mnras}
\bvolume{448}(\bissue{3}),
\bfpage{2412}--\blpage{2429}
(\byear{2015})
{\href{https://arxiv.org/abs/1501.07115}{{arXiv:1501.07115}}}
{[astro-ph.HE]}.
\doiurl{10.1093/mnras/stv179}
\end{barticle}
\endbibitem

\bibitem{Baldi2019}
\begin{barticle}
\bauthor{\bsnm{{Baldi}}, \binits{R.D.}},
\bauthor{\bsnm{{Torresi}}, \binits{E.}},
\bauthor{\bsnm{{Migliori}}, \binits{G.}},
\bauthor{\bsnm{{Balmaverde}}, \binits{B.}}:
\batitle{{The High Energy View of FR0 Radio Galaxies}}.
\bjtitle{Galaxies}
\bvolume{7}(\bissue{3}),
\bfpage{76}
(\byear{2019})
{\href{https://arxiv.org/abs/1909.04113}{{arXiv:1909.04113}}}
{[astro-ph.HE]}.
\doiurl{10.3390/galaxies7030076}
\end{barticle}
\endbibitem

\end{thebibliography}



\end{document}